\documentclass[preprint]{aastex63}
\usepackage{shortvrb}
\usepackage{graphicx}                    
\usepackage{amssymb}                    
\usepackage{color}                       
\usepackage{url}                         

\begin{document}
\title{Temporal evolutions and quasi-periodic variations present in the sunspot number and group sunspot area data measured at Kodaikanal Observatory for solar cycles 14 to 24}

\author{Belur Ravindra}
\affil{Indian Institute of Astrophysics, Koramangala, Bengaluru-560034. INDIA.}
\author{Partha Chowdhury}
\affil{University College of Science and Technology, Department of Chemical Technology, University of Calcutta, 92, A.P.C. Road, Kolkata, 700009, West Bengal, INDIA.}
\author{Pratap Chandra Ray}
\affil{Post Graduate Department of Mathematics, Bethune College, University of Calcutta, 700006 Kolkata, India}
\author{Kumaravel Pichamani}
\affil{Indian Institute of Astrophysics, Koramangala, Bengaluru-560034. INDIA.}


\correspondingauthor{Belur Ravindra}
\email{ravindra@iiap.res.in}

\begin{abstract}

The Kodaikanal Observatory has made synoptic observations of the Sun in white light since 1904, and these images are sketched on the Stonyhurst grids called sun charts. These continuous hand-drawn data sets are used for long-term studies of the Sun. This article investigates temporal and periodic variations of the monthly hemispheric sunspot number and sunspot group area for 1905--2016, covering solar cycles 14 to 24. We find that the temporal variations of the sunspot number and group area are different in each hemisphere and peak at different times of the solar cycle in the opposite hemisphere. For both the data sets, Cycle 19 shows maximum amplitude. For the sunspot number time series, Cycle 24 was the weakest, and Cycle 15 for the group area. The existence of double peaks and violation of the ``odd-even rule'' was found in both data sets. We have studied the periodic and quasi-periodic variations in both the time series by wavelet technique. We noticed that along with the fundamental mode of the $\sim$ 11~year cycle and polarity reversal period of 22~years, the sunspot activity data also exhibited several mid-term periodicities in the opposite hemispheres, specifically the Riger group and quasi-biennial periodicities. The temporal evolution of these detected quasi-periodicities also differs in the northern and southern hemispheres. We analyzed the data set statistically to understand the bulk properties and coupling between the opposite hemispheres. The study indicates that the two hemispheric data sets differ, but some dependency could be present.

\end{abstract}


\section{Introduction}
The sunspots are the most prominent features visible even with a small-sized telescope from the Earth. The first record of these features started 400~years ago. The sunspots number count began at the starting of the 18th century \citep{1861MNRAS..21...77W} with or without the sunspot drawings made on the charts. Later, refined the sunspot counts with correction factors to accommodate all the observatory data taken with different sized telescopes \citep{1851AN.....32..193W, 1902PA.....10..449W}. The refined sunspot number records were used to study various properties and characteristics of sunspots and activity cycle \citep{1939Obs....62..158G, 1940Obs....63..215G, 1943Obs....65...24G, 1944ApJ...100..219G}. In the late 19th century, several observatories in the world have measured the sunspot group area along with the sunspot number \citep[e.g.,][]{1983ApJ...275..878H}. Among them, the Greenwich data has the longest record of the sunspot number and sunspot group area \citep{2015LRSP...12....4H}.

With the recorded sunspot number, it was noticed that sunspots have an 11-year periodicity. On using several years of data, it was realized that the sunspot cycle range between 8.5 -- 13.5 years. At the beginning of the 19th century, the 11-year periodicity is noted as the appearance and disappearance of sunspots on the sun's disk. Later, in the 20th century, it was seen that this is also related to the reversal in the polarity of sunspot groups. It was also seen that over the 11 years, the polar field was observed to be out of phase with the solar cycle \citep{1959ApJ...130..364B}. In addition to the 11-year cycle, a shorter period such as 27~day periodicity was also noticed. This time interval is attributed to the rotation period of the low-latitude sunspots on the sun. 

Later part of the 20th century, a 154~days periodicity was observed in the flare data \citep{1984Natur.312..623R}, sunspot number and area data as well \citep{1989ApJ...337..568L, 1990A&A...238..377C}. Several mid-range periodicities were observed in between the 27-days and 11-year periods. Those are referred to as quasi or mid-range periodicities \citep[e.g.][and the references therein]{2003ApJ...591..406B, 2009MNRAS.392.1159C, 2010SoPh..266..173K}. The Rieger type periodicity and mid-range periodicities were observed in hemispheric sunspot area \citep{1990ApJ...363..718L} and plage area \citep{2022ApJ...925...81C}. However, it was reported only in a few cycles of hemispheric sunspot number data \citep{2006A&A...447..735T}. 

The hemispheric sunspot number data extracted from Kanzelh{\"o}he Solar Observatory (KSO), Austria is available from 1975 -- 2000 \citep{2002A&A...390..707T}. By combining the KSO data with the Skalnat{\'e} Pleso Observatory, Slovak Republic sunspot data \citet{2006A&A...447..735T} made the sunspot number data starting from 1945 till 2004. They have normalized the data with the International Sunspot Number full sphere sunspot number data and then compared it with the Sunspot Index Data Center (SIDC) hemispheric sunspot number data, which started in 1992 and continues till today. These data have reported the North-South asymmetry in hemispheric sunspot number and rotational period in both the hemisphere. Recently,  \citet{2021A&A...652A..56V} have extended this time series and made the hemispheric sunspot number catalog starting from 1874 till 2020 and studied some of the hemispheric properties of solar activity. 

At the Kodaikanal observatory, the sunspot observations started in 1905 using photographic plates and continue to use photographic films. The photographic record of white-light images is digitized at the Kodaikanal Observatory \citep{2013A&A...550A..19R, 2017A&A...601A.106M}. Along with the photographs, sunspots, plages, and filament drawings are also made on the Stonyhurst grid by projecting the recorded image onto the paper. The hemispheric sunspot number and area parameters were extracted from these data set \citep{2020Ap&SS.365...14R}. These data sets are useful for studying the sunspot cycles' different characteristic properties and mid-range periodicities.  

This paper presents the results on the temporal evolution of different solar cycles, coupling between the opposite hemispheres (statistical relationship), and mid-range periodicities in the hemispheric sunspot number and area data extracted from the KO sunspot drawings. In Section~\ref{sec:data}, we present the data and the analysis procedure. Section~\ref{sec:analysis} presents results on the Quasi-biennial oscillations in the hemispheric sunspot parameters. Section~\ref{sec:mod} outline the model fitting and their statistical accuracy. Section~5 delineate the non-linear, statistical relationship between the northern and southern hemispheric sunspot number and group sunspot area time series.
Section~\ref{sec:dis} summarizes the results, compares the asymmetry behaviour, and Quasi-Biennial oscillations found in sunspot number and area data. In the present work, we have utilized a number of analysis techniques, including complex non-linear wavelet method, Auto-regressive Integrated Moving Average Model (ARIMA) fitting with goodness of fit, Wavestrapping and Dynamic Time Warping to study the temporal variation and the statistical relationship between the long-term northern and southern hemispheric sunspot activities. We also compare them with the past observations.

\section{Data}
\label{sec:data}
Systematic observations of sunspots have been carried out at Kodaikanal Observatory (KO) since 1904 with an unchanged single white-light telescope having a 10-cm objective \citep{1993SoPh..146...27S}. From Jan 1976, these photographic plates were replaced by high-contrast films of size 25.4~cm $\times$ 30.5~cm. These photographic plates and the films have recently been digitized with the help of modern digitizer \citep[see:][for details]{2013A&A...550A..19R}. Utilizing this century-long data-set, several researchers investigated various aspects of the sunspots and their solar cycle characteristics \citep[e.g.,][]{2017A&A...601A.106M, 2021SoPh..296....2R}.

Along with this process, in parallel, drawings of the different observed features on the solar surface like sunspots, plages, filaments, etc., were made on the Stonyhurst grids, which are called sun charts. All those mentioned above solar magnetic indices were detected distinctly and optically marked with different colors in these sun charts. This was furnished by projecting the sunspot images onto the Stonyhurst latitude and longitude grid with a binning of 5$^{\circ}$ in both directions. Thus, KO provides a repository of more than 100-years of handwritten sunspot drawings covering nearly ten solar cycles. These sunspot drawings are preserved for scientific studies with due care at the KO library. The information of daily solar observations as well as the sky conditions were regularly maintained in a logbook. These drawings currently stand among one of the biggest and historical sunspot archives. Recently, \citet{2020Ap&SS.365...14R} have provided a detailed description about the sun charts, the process of counting sunspot numbers, and determining the sunspot area from these historical grids. 

\citet{2020Ap&SS.365...14R} determined the monthly average time series of sunspot numbers measured in both the opposite hemispheres using these KO sun charts from January 1905 to November 2016. The KO sunspot numbers are defined in the usual way as R$_{n}$= k(10.g + n) where g and n are the number of sunspot groups and a total number of individual sunspots observed on the visible solar disk, respectively. The correction factor k is considered as 1.

From these sun-charts, the monthly mean time series of group sunspot areas of both the northern and southern hemispheres were also extracted for the aforesaid time span. Sunspot group area represents the total areas covered on the solar disk by all sunspots. A larger sunspot group implies areas of larger magnetic flux, which in turn indicates larger magnetic energy content of an active region. Coupled with magnetic complexity, it would give a measure of flare productivity of the group. Here, the sunspot group area measured on the solar disk is defined as: $A_{M}$ = 2$A_{s}10^{6}/ \pi D^{2}cos(\rho)$. Here A$_{s}$ is the measured size of a sunspot group, $\rho$ is the angular distance of the center of the sunspot group to the center of the solar disk. D is the diameter of the solar image. The unit of the sunspot group area is the millionth of the solar hemisphere ($\mu$Hem), and the foreshortening effect has been corrected.

\begin{figure}[!h]
\centering
\includegraphics[width=0.43\textwidth]{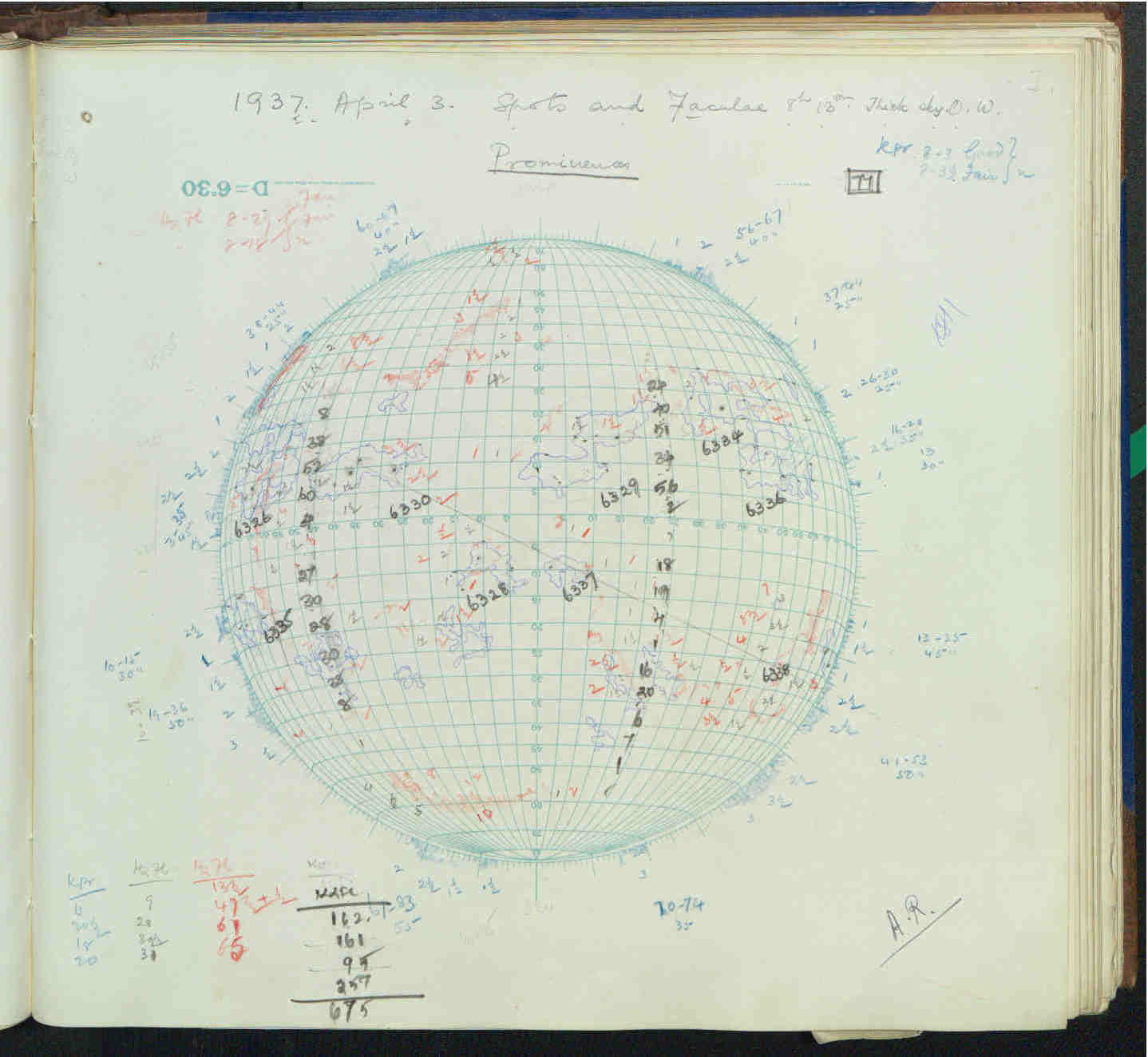}
\includegraphics[width=0.4\textwidth]{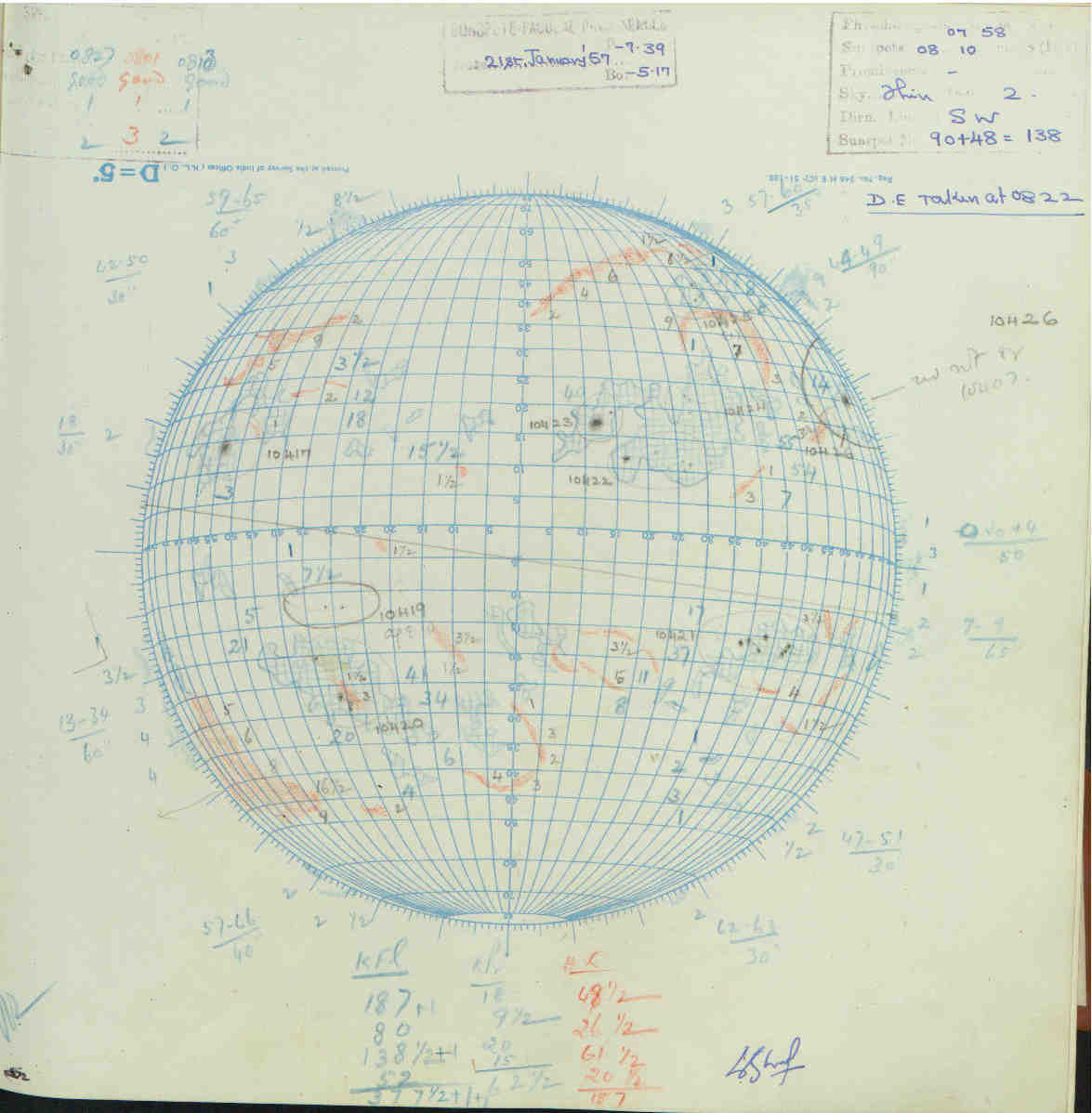} \\
\caption{The sketches of sunspots, plages, filaments, and prominences are made on the Stonyhurst grid for two different days of observations. A specific colors are used to draw different features.  The left-side image is for the date April 3, 1937 and the right side image is for the date January 21, 1957. The B$_{0}$ angle and p-angle are shown the chart.}
\label{fig:1}
\end{figure}

Figures~\ref{fig:1} represent typical images of these KO sun charts in which the sunspots, plages, and filaments are drawn to their size. Each of these features is shown in different colors. The area covered by the sunspots and plages is displayed next to those features and expressed in millimeters. These numbers are converted into millionth of hemisphere using the Tables available at KO. The Kodaikanal active region number (KKL number) is shown next to the sunspot group region in pencil. 

We have utilized the wavelet analysis tool \citep{1998BAMS..79..61T} to study the periodic and quasi-periodic variations in the monthly averaged sunspot number and group area time series for the period 1905 -- 2016, covering solar cycles 14 to 24. The wavelet analysis is a valuable tool that can reveal the presence of localized oscillations in two-dimensional time-frequency domains. This method provides information about the exact location of the detected periods and their temporal variations present in a time series. To study the periodic behavior of the KO sunspot number and group sunspot area, we have used complex Morlet wavelet \citep{1982Geop...47..203M} function,
\begin{equation}
\psi_n(\eta) = \pi^{-1/4}\mathrm{e}^{\mathrm{i}\omega_{0} \eta}\mathrm{e}^{-\eta^2}/2
\end{equation}

Here, $\omega_{0}$ is a non-dimensional frequency and we have adopted $\omega_{0}$ as 12 for mid-term frequency (low periodic zone; 4 months to 1.4 years) range and $\omega_{0}$ = 6 for the low-frequency regime (long periods; 1.4 years to 11 years) \citep{2019SoPh..294..142C, 2021SoPh..296....2R}. In addition to this, we have also investigated the presence of $\sim$~22 years Hale cycle which indicates the returning of the Sun's magnetic field in all the data sets considering $\omega_{0}$ = 6. The cone-of-influence (COI) indicates a reduction of the wavelet power due to edge effects \citep{1998BAMS..79..61T, 2004NPGeo..11..561G} is shown with a bold dashed line. In all wavelet spectrum, the thin black contours represented the periods above 95\% confidence level considering the red-noise background and detected using the recipe by \citet{2004NPGeo..11..561G}.

Under the condition of the red-noise background, the discrete Fourier power spectrum, after normalization, takes the form,
 
\begin{equation}
P_{k} = \frac{1 - \alpha^{2}}{1 + \alpha^{2} - 2\alpha cos(2\pi k/N)}
\end{equation}

Where k = 0, ..., N/2 is the frequency index, N is the number of data points in the time series, and $\alpha$ is the lag-1 serial correlation coefficient. We have considered a red-noise background and the lag-1 serial correlation coefficients of all the data sets of KO sunspot number and group area under study.

Global Wavelet Power Spectra (GWPS), which determines wavelet power at each period and is averaged over the time span, have been calculated for all data sets under investigation considering red-noise background. GWPS is formulated as,  

\begin{equation}
\bar{W^2}(s)=\frac{1}{N}\sum_{n=0}^{N-1}\left|W_{n}(s)\right|^{2}
\end{equation}

Here $W_{n}(s)$ is the wavelet power, and N is the number of local wavelet spectra. In this way, we can detect an unbiased and consistent estimation of the true power spectrum of any time series. The working nature of this GWPS has similarities with the computation of the Fourier power spectra of the time series. Here, GWPS plots with 95\% confidence level have been calculated using the method provided by \citet{1998BAMS..79..61T}.

\section{ANALYSIS AND RESULTS}
\label{sec:analysis}
\subsection{Temporal variations of sunspot numbers}

\begin{figure}[!h]
\centering
\includegraphics[width=0.45\textwidth]{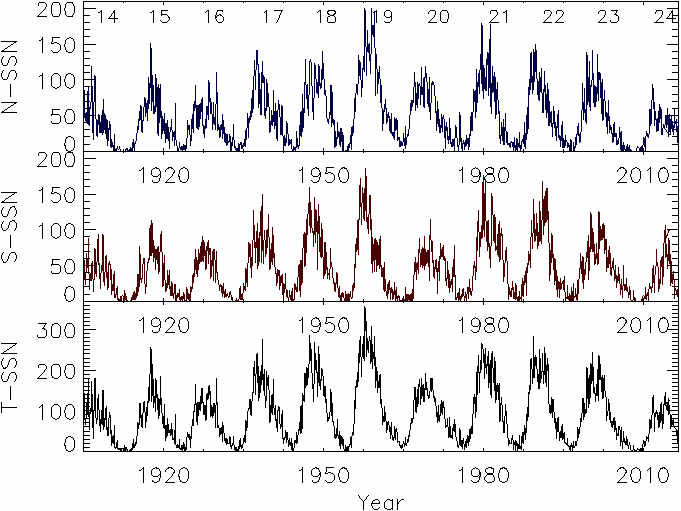}
\includegraphics[width=0.45\textwidth]{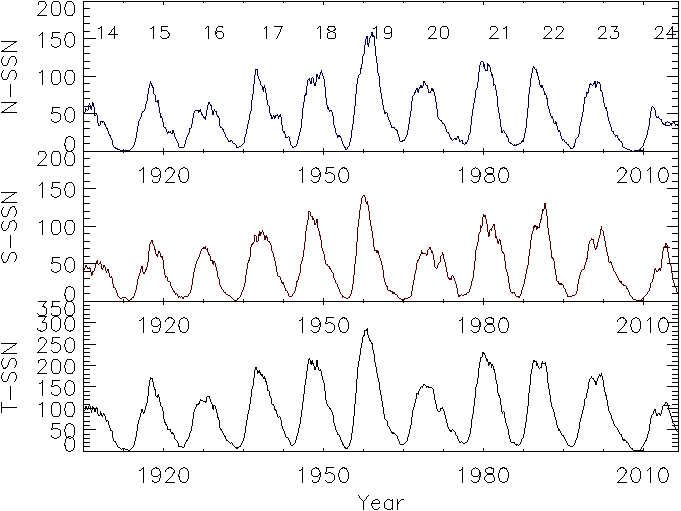} \\
\caption{Left: Monthly average values of northern (top), southern (middle),
and full-disk (bottom) sunspot numbers measured at KO from 1905 to 2016. Right: Same as left side plot but for the 13-month smoothed data.}
\label{fig:2}
\end{figure}

Figure~\ref{fig:2}(left) shows the monthly averaged sunspot numbers of both the hemispheres and the whole solar disk for cycles 14 to 24 measured at the KO. The 13-month smoothed monthly mean of these data sets is displayed in figure~\ref{fig:2}(right). These Figures indicate that the sunspot number time series in both hemispheres and the whole disk follows a regular cyclic pattern of about 11-years, with different dynamical behavior in different sunspot cycles. Solar cycle 19 exhibits the maximum amplitude in all data sets. However, cycle 21 had the second-highest amplitude for the northern hemisphere and full-disk data.
On the other hand, cycle 22 was the next powerful cycle for the southern hemisphere. Cycles 16, 20, and 24 are the weakest cycles for the northern, southern, and full-disk data. The strength of cycle 24 is distinctively weaker than cycles 23 and 22 in all cases. 

Figure~\ref{fig:2}(right) demonstrates that the even-numbered solar cycles have low amplitude compared to the following odd-numbered solar cycles in North, South and full-disk data. For example, cycles 16, 18, and 20 have smaller amplitude compared to following odd numbered cycles 17, 19, and 21. These behaviors satisfy the even-odd pair or G-O rule \citep{1948AZH...25..18G, 2012SoPh..281..827J}. However, we have noticed that the even numbered cycle 22 has larger amplitude than the following cycle 23 in the northern, southern, and full-disk time series. So, cycle 22--23 pair did not maintain the G-O rule.  Utilizing yearly group sunspot number and yearly sunspot numbers, \citet{2015Ge&Ae..55..902Z} showed that solar cycles 16-17 and 18-19 followed even-odd pair instead of odd-even. However, the concrete reason behind pairing of adjacent cycles in a manner of even-odd or odd-even is still unknown \citep{2015LRSP...12....4H}.

Several solar cycles exhibit double peaks during their maximum phase.
In Figure~\ref{fig:2}(right), we find that for cycles 16, 18, 21, 22, and 23, there are clear signatures of double peaks in both the hemispheres and the whole disk. For solar cycles 20 and 24, there is the signature of double peaks in the southern hemisphere and full-disk data. Cycle 20 shows multiple peaks and a step-like decrease during its descending phase, which is much more prominent in the case of the southern hemisphere. A step like a decrease is also detected only in the northern hemisphere during cycle 17. The amplitude of the sunspot number in the full disk shows a continuous decline from cycle 21 to 24. Recently, \citet{2021A&A...652A..56V} constructed sunspot number time series for both the opposite hemispheres from 1874 to 2020 considering the WDC-SILSO sunspot number database and Greenwich Royal Observatory sunspot area database and the properties mentioned above detected in the KO sunspot number are also observed in their data set. There is quite a similarity about the strength and temporal evolution of sunspot numbers in the opposite hemispheres as well as whole disk between these two data bases during different cycles \citep[see: Figure~7 of][]{2021A&A...652A..56V}.

\subsection{Temporal variations of sunspot group area}
\begin{figure}[!h]
\centering
\includegraphics[width=0.45\textwidth]{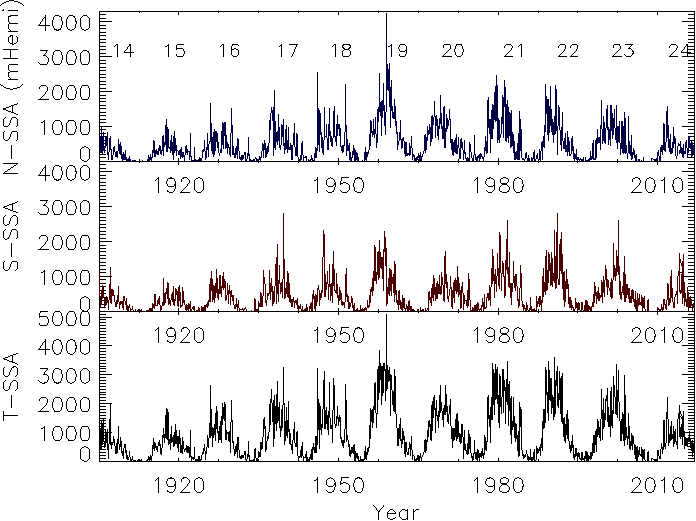}
\includegraphics[width=0.45\textwidth]{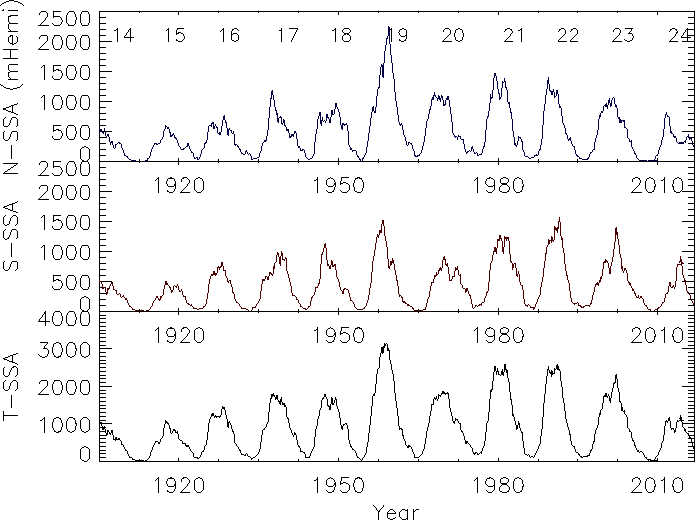} \\
\caption{Left: Monthly average values of northern (top), southern (middle),
and full-disk (bottom) sunspot group area ($\mu$Hem) measured at KO from 1905 to 2016. Right: Same as left side plot but for the 13-month smoothed data.}
\label{fig:3}
\end{figure}

Figure~\ref{fig:3}(left) shows the monthly averaged sunspot group area of the northern and southern hemisphere and the whole solar disk for cycles 14 to 24 measured at the KO. The monthly averaged and then 13-month data sets are displayed in Figure~\ref{fig:3}(right). These Figures indicate that sunspot number and sunspot group area exhibit a cyclic pattern of about 11 years. The solar cycle 19 shows the maximum amplitude and cycle 15 weakest in all data sets. However, for full-disk data, the amplitude of cycle 24 is very close to cycle 15. Cycles 21 and 22 possess the second-highest amplitude for the northern and southern hemispheres. Both these cycles exhibit nearly the same amplitude in the full-disk data set.

Cycle 16 had the second weak strength in the last century after cycle 15. In the northern, southern hemisphere and full-disk data shows a pair of cycles 16-17, 18-19, and 20-21 followed the G-O rule. The pair of cycles 22-23 violated the G-O rule in both the hemispheres and full-disk data. The double peaks around the maximum epoch were prominent for cycles 16, 18, 21, 22, and 24. The existence of double peaks in sunspot number and area in opposite hemispheres suggests that the origin is related to the presence of sub photospheric dynamo. Descending phases of the cycles 17, 18, and 20 exhibit multiple peaks and step-like complex patterns. From cycle 22 to 24, the peak height of the sunspot group area reduced continuously.

\subsection{Periodicities in KO sunspot number}
\label{subsec:psn}
\begin{figure}[!h]
\centering
\includegraphics[width=0.42\textwidth]{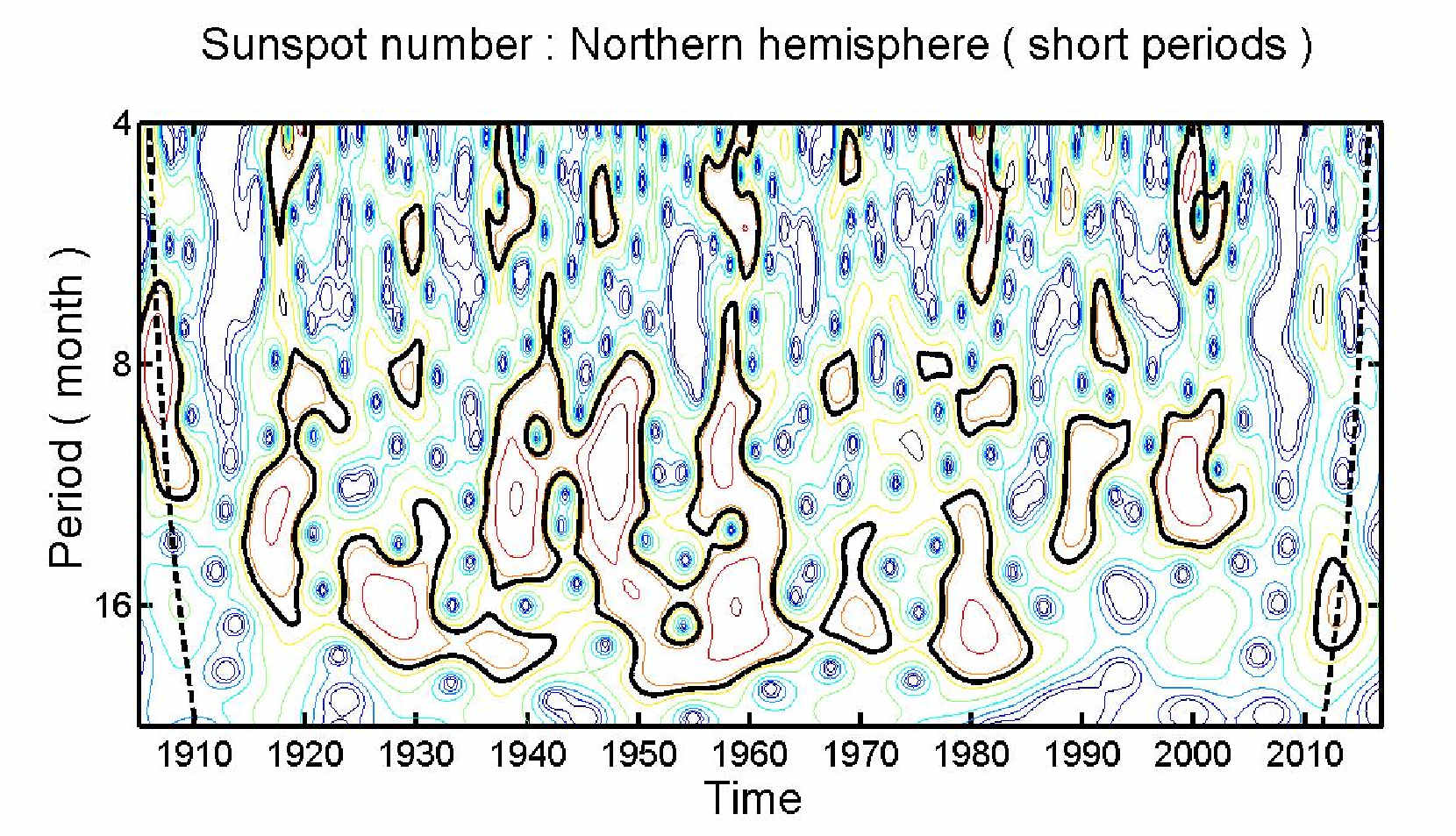}
\includegraphics[width=0.45\textwidth]{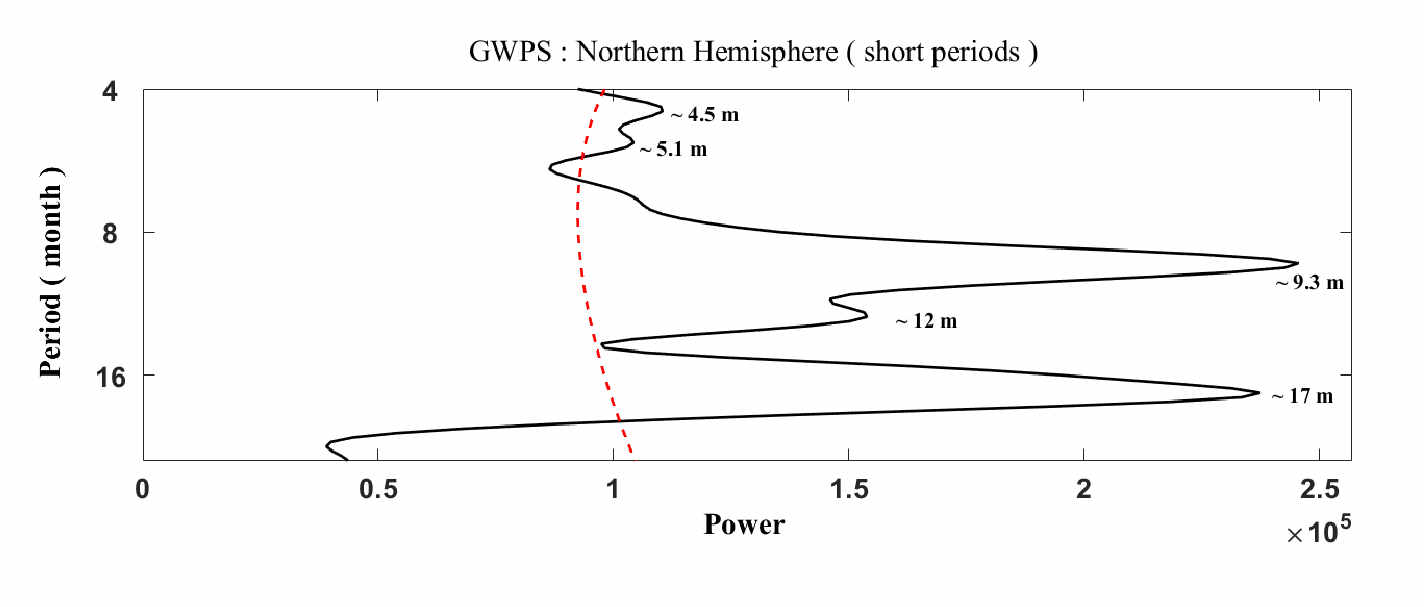} \\
\includegraphics[width=0.42\textwidth]{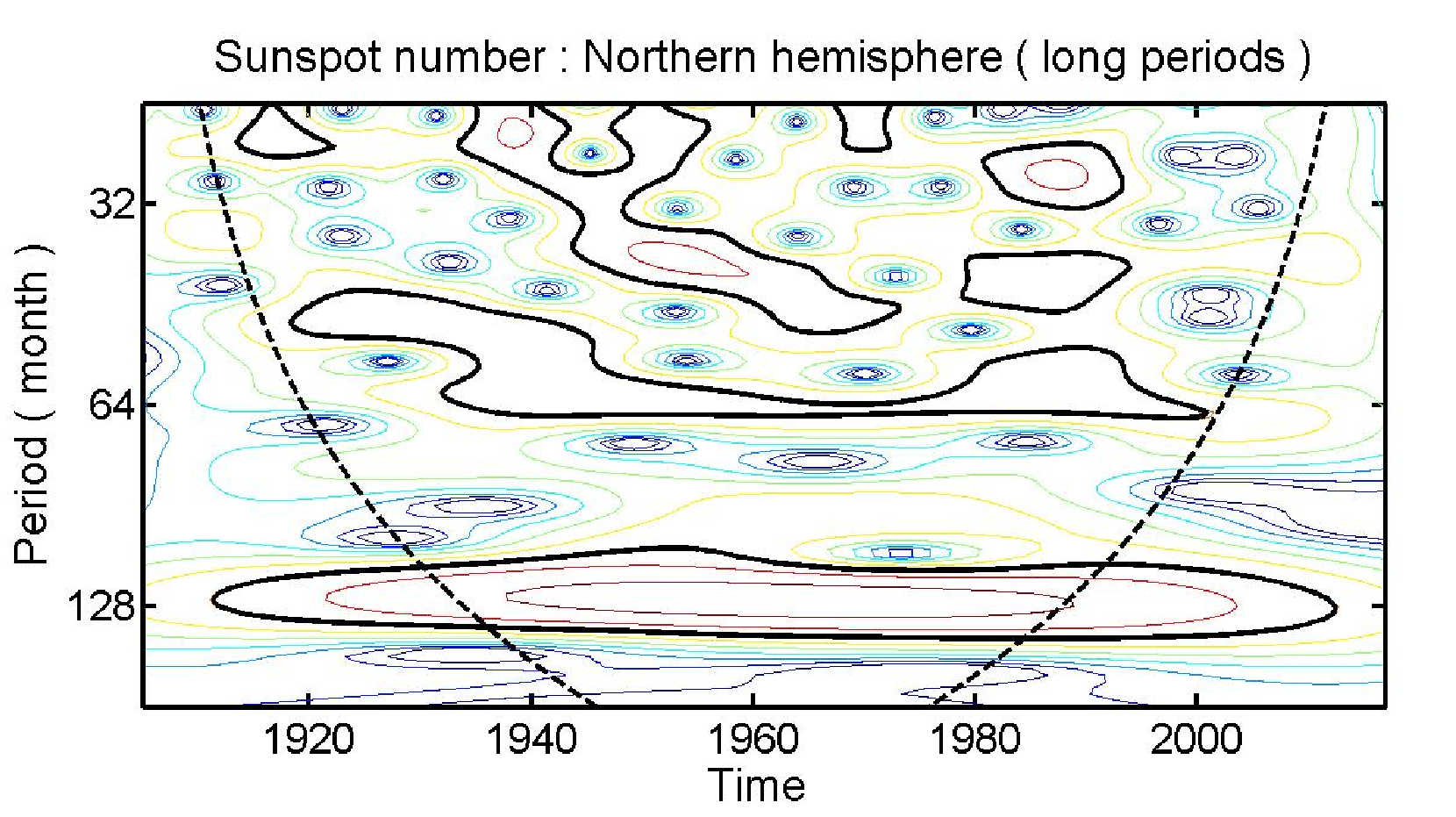}
\includegraphics[width=0.45\textwidth]{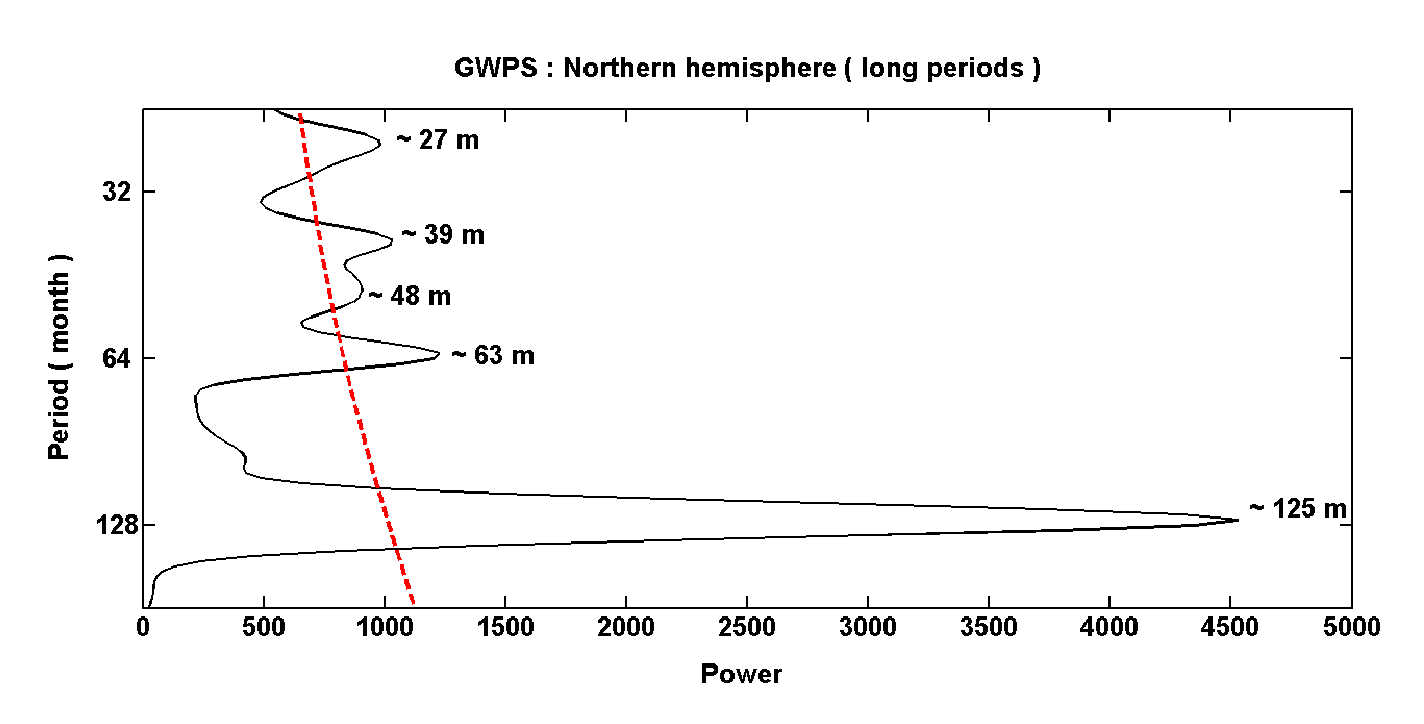} \\
\caption{(a) Morlet wavelet spectra of the monthly KO sunspot number time series in the northern hemisphere. (b) Global power spectra for short term periods mainly to study Rieger group of periodicities. (c) Local morlet wavelet power spectra to investigate the dynamical behaviors of QBOs (1.4 to 4 years) and other long term periods including solar cycle period. (d) Similar to panel (b) for other long term periods including Schwabe cycle. Dotted lines in all global wavelet power spectra represent a 95\% confidence level.}
\label{fig:4}
\end{figure}

\begin{figure}[!h]
\centering
\includegraphics[width=0.42\textwidth]{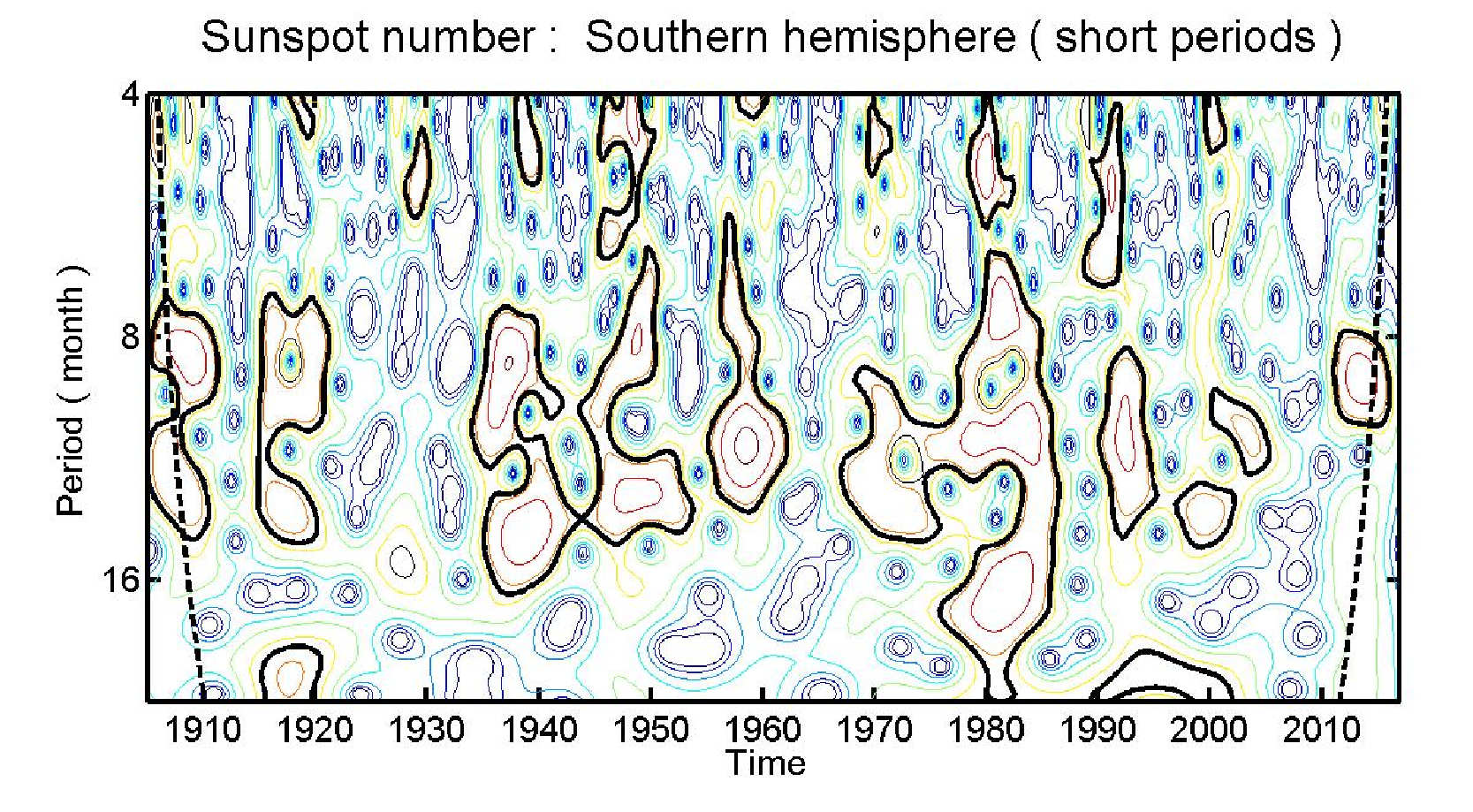}
\includegraphics[width=0.45\textwidth]{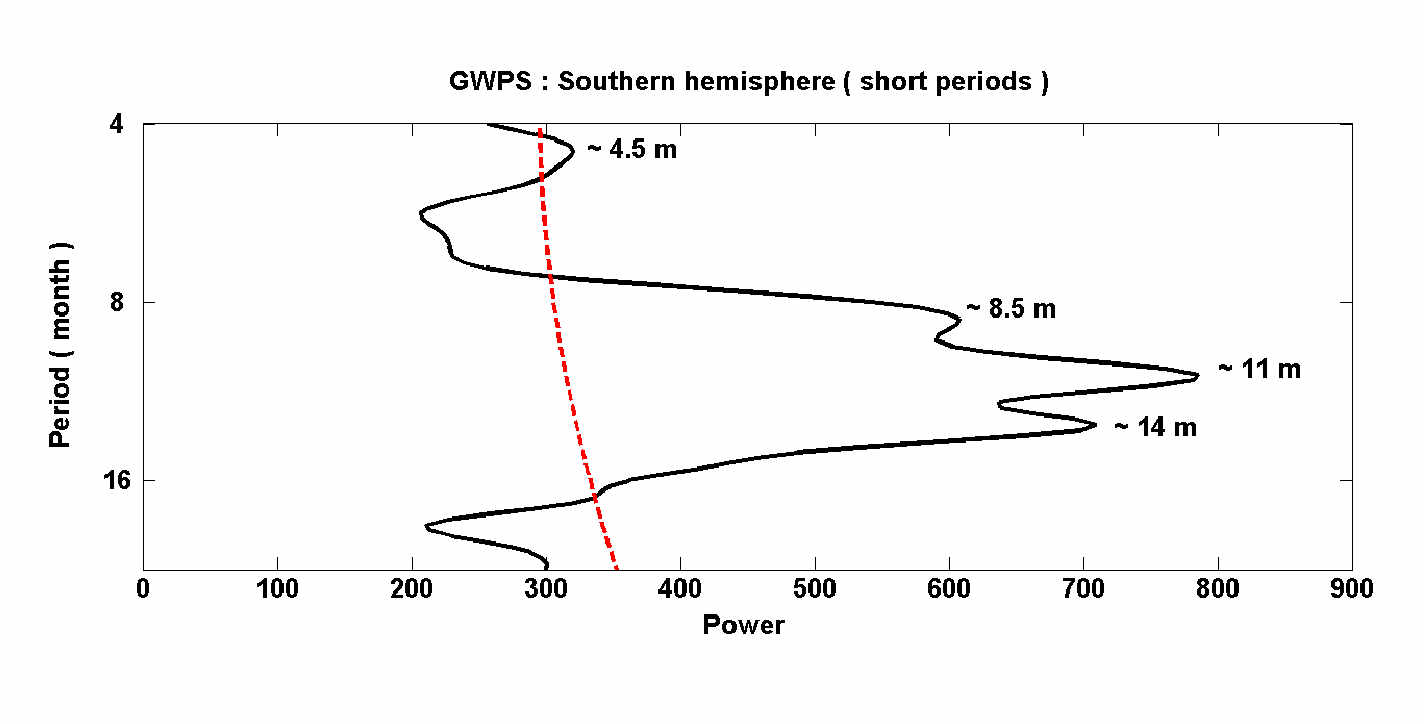} \\
\includegraphics[width=0.42\textwidth]{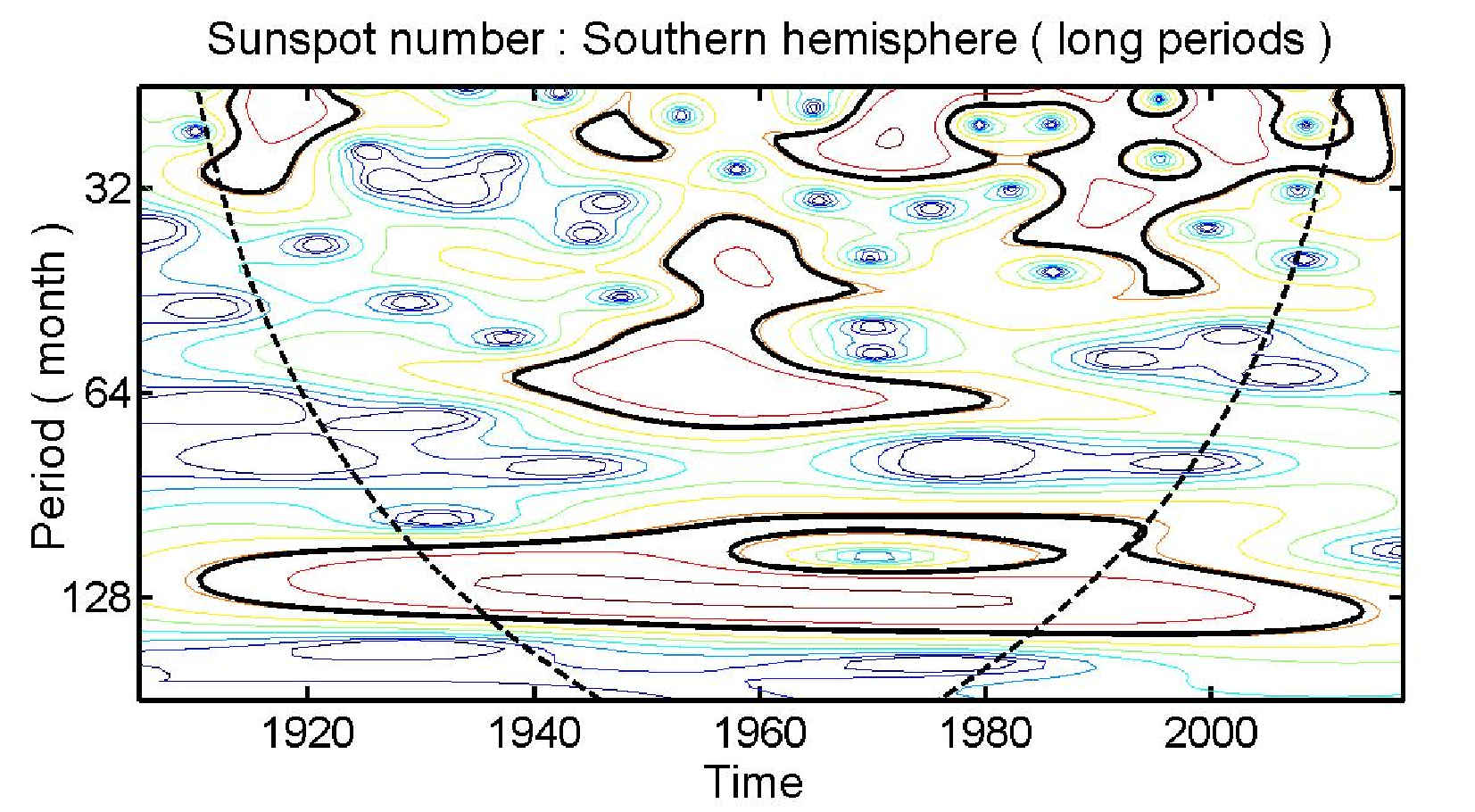}
\includegraphics[width=0.45\textwidth]{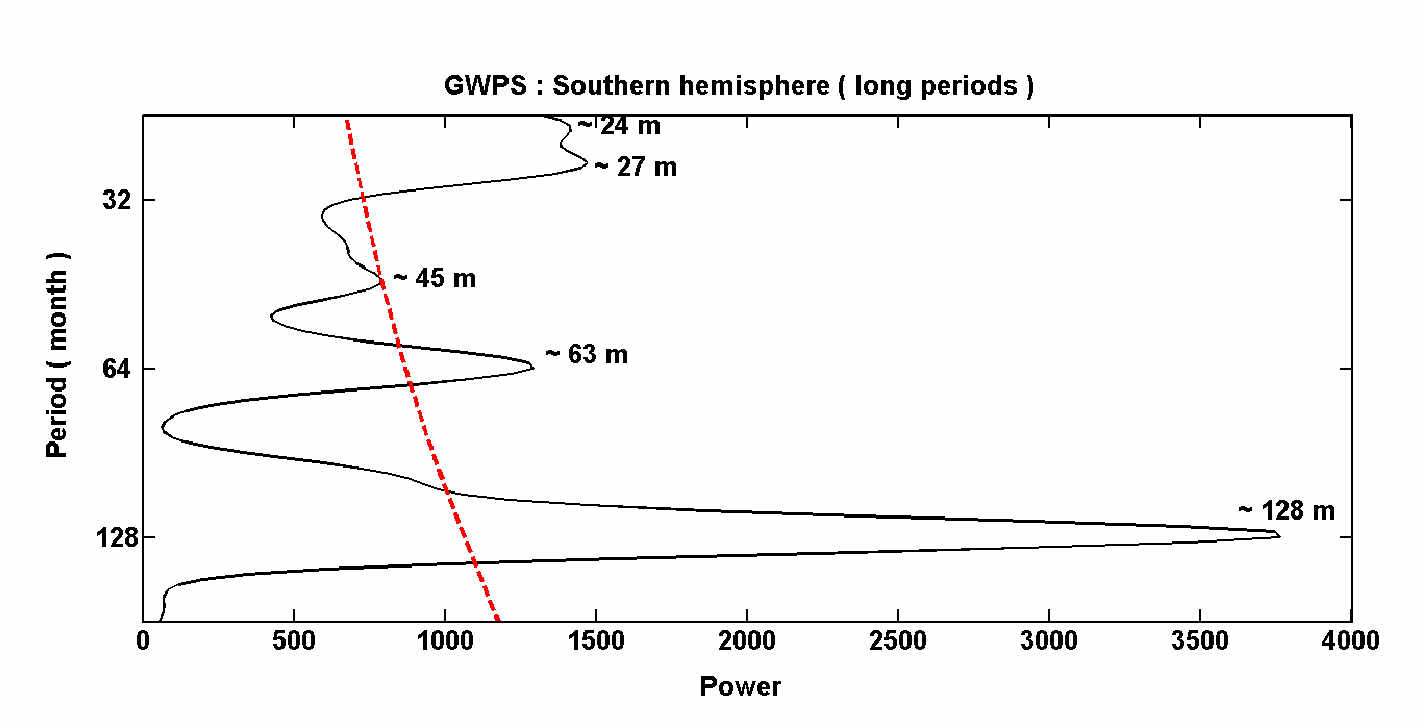} \\
\caption{Same as Figure~\ref{fig:4}, but for the southern hemisphere.} 
\label{fig:5}
\end{figure}

\begin{figure}[!h]
\centering
\includegraphics[width=0.42\textwidth]{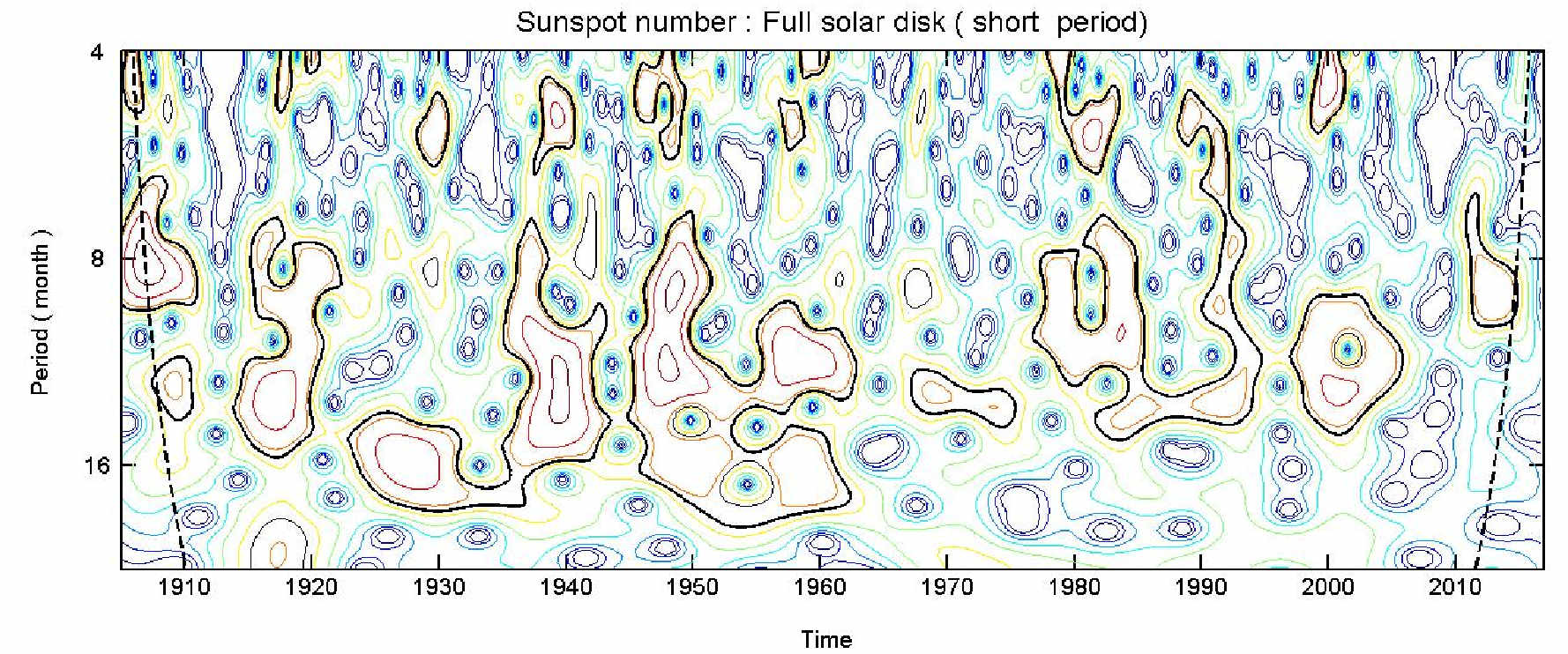}
\includegraphics[width=0.45\textwidth]{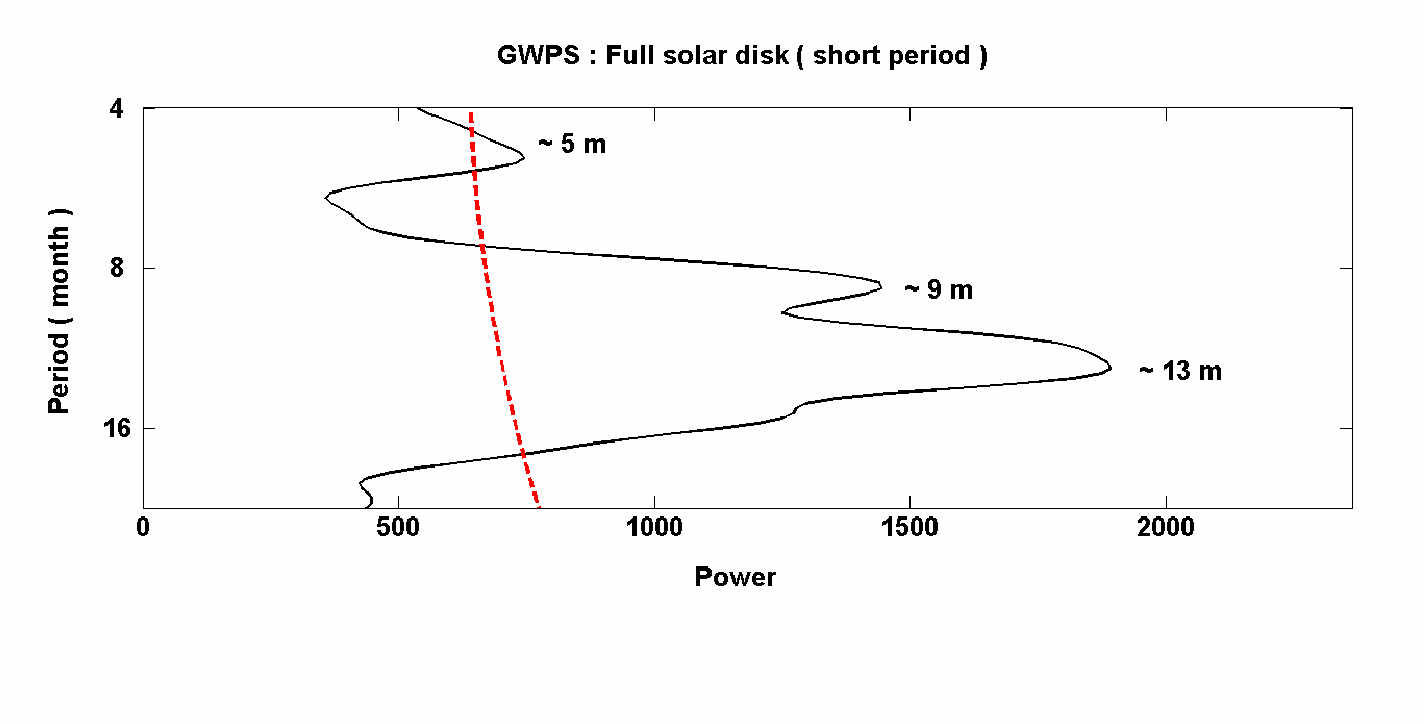} \\
\includegraphics[width=0.42\textwidth]{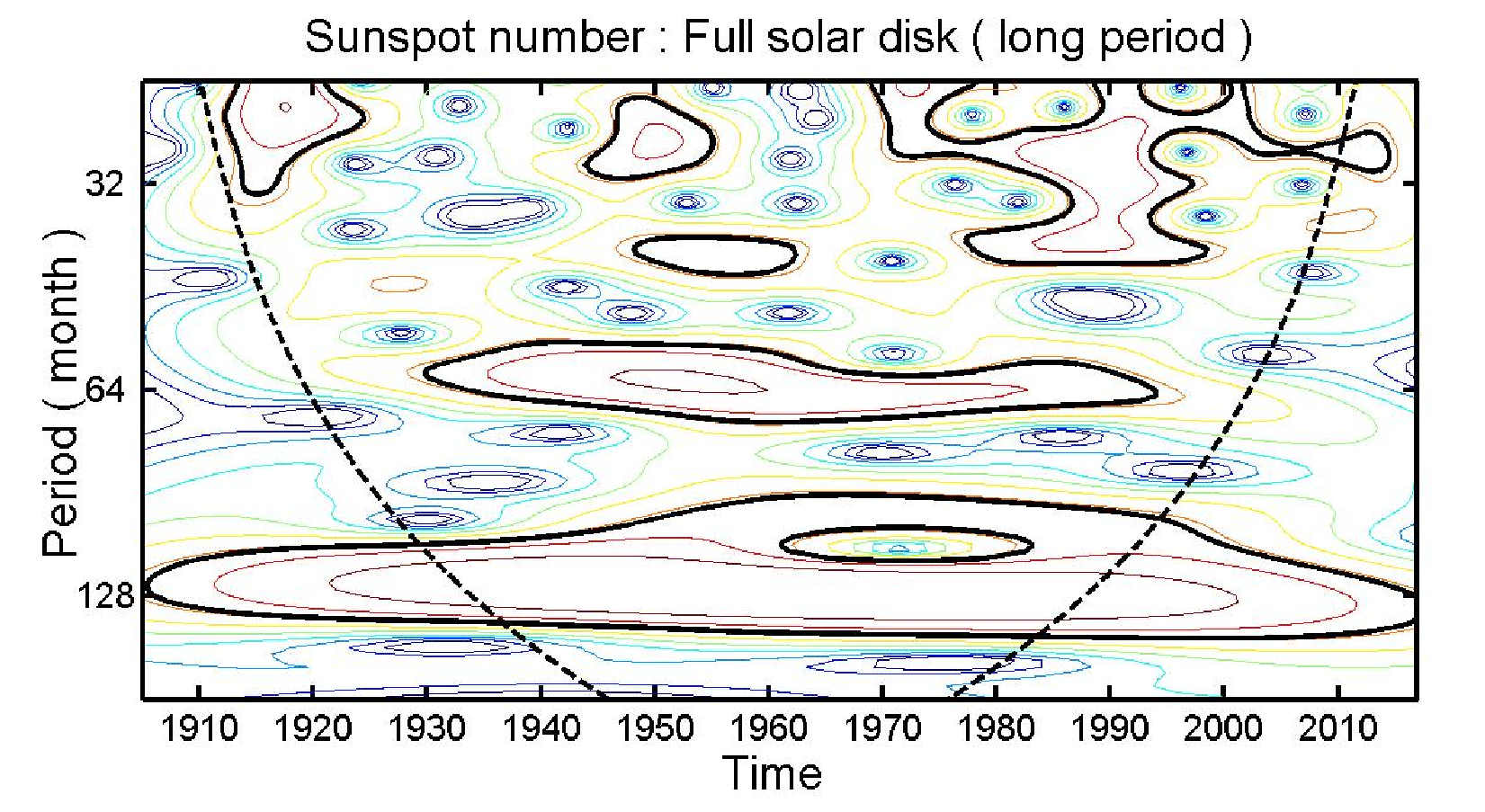}
\includegraphics[width=0.45\textwidth]{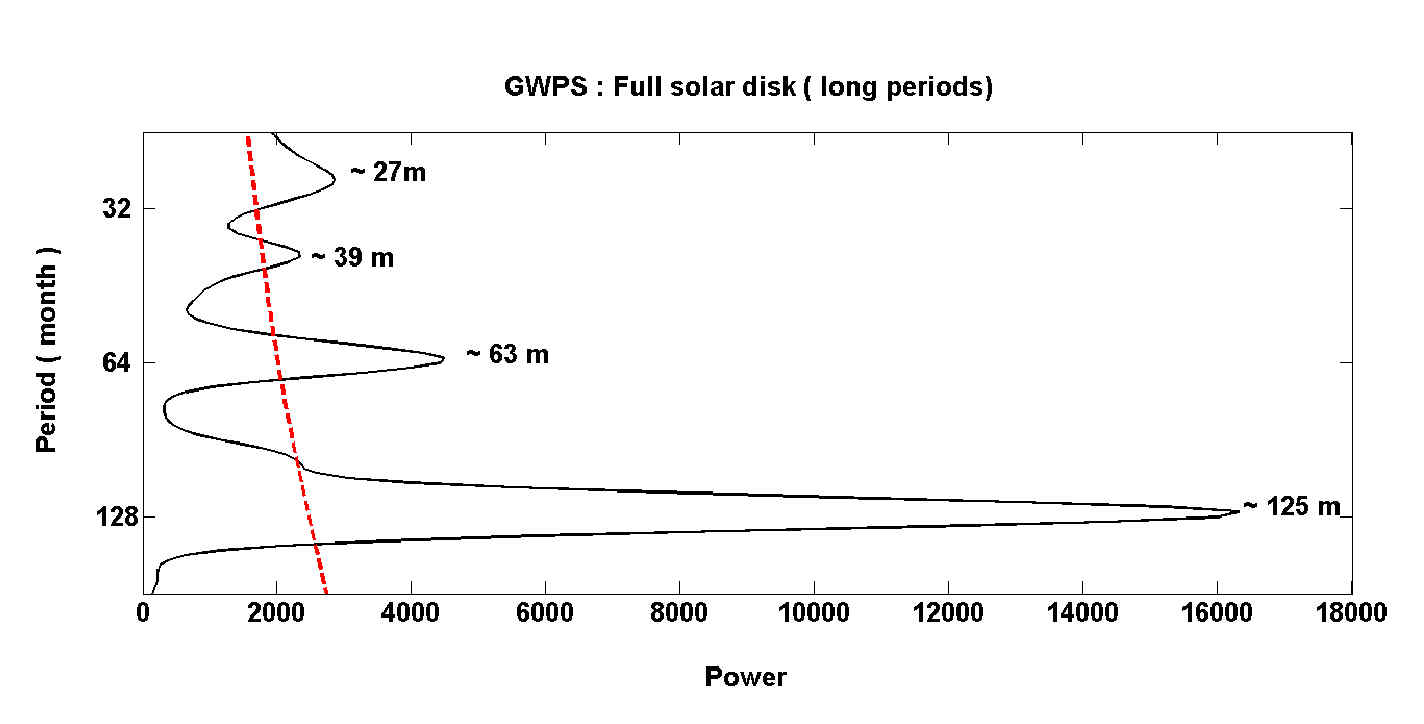} \\
\caption{Same as Figure~\ref{fig:4}, but for the whole disk data.}
\label{fig:6}
\end{figure}

The results of employing the Morlet Wavelet Analysis to the monthly sunspot number of the northern and southern hemisphere as well as the whole solar disk are displayed in Figures~\ref{fig:4}, \ref{fig:5}, and \ref{fig:6} respectively. Here, we focused on the mid-term periodic and quasi-periodic variations in the monthly sunspot number for solar cycles 14 to 24.

Figure~\ref{fig:4}(top-left) represents the short periods present in the northern hemispheric sunspot number in the local wavelet spectrum. This figure indicates the presence of Rieger and Rieger type periods in different solar cycles like 15, 17, 19, 21, and 23. The length of these groups of periods changes from 4 to 6 months. Quasi-periodicities of varying length between 300 and 365 days were clustered mainly in cycles 14, 15, 17, and 20, 22, and 23. The QBO's in the range of 1.2 -- 1.4 yrs were found during cycles 16 to 18 and also from the maximum phase of cycle 18 to the end of cycle 20 and partly in cycle 21.

The Rieger group of oscillations were detected in different parts of cycles 16 to 23 in the southern hemisphere [Figure~\ref{fig:5}(top-left)]. Intermediate-term periods in the range of 6 to 14 months were seen in cycle 15, cycles 17 to 20, descending phase of cycle 20 to the end of cycle 22 and 23.  

Rieger type of periods was present in the data of whole solar disk in scattered form during parts of the cycles 16, 17, 18, 21, 22, and 23 [Figure~\ref{fig:6}(top-left) ]. The intermediate-term periods in the range of 7 months to 1 year were detected in solar cycles, 14, 17, 18, 19, 21, 22, 23, and 24. It is found that QBOs in the range of 1.2 -- 1.4 yrs were concentrated in cycles 16 to 19 and cycle 23. 

Figure~\ref{fig:4}(bottom-left) indicates that long--term QBOs in the range of 2--3.5 years were present during cycles 17 to 20, cycles 22 and 23 in the northern hemisphere. On the other hand, these periods were significant, mainly during cycles 19 to 24 in the southern hemisphere [Figure 5(bottom-left)]. These QBOs were detected in cycles 14, 15, 17, and 20 to 24 in the sunspot number of the whole disk data [Figure~\ref{fig:6}(bottom-left)]. A long contour of period $\sim$~5~yrs were appeared during cycles 15 to 23 in the northern hemisphere but persisted during the descending phase of cycle 17 to ascending phase of cycle 21 in the southern hemisphere. However, this period is significant during cycles 16 to 22 for the full solar disk. These three data sets exhibit a strong appearance of $\sim$~11 yrs sunspot cycle period. 

Figures~\ref{fig:4}(top-right) \& (bottom-right), \ref{fig:5}(top-right) \& (bottom-right), and \ref{fig:6}(top-right) \& (bottom-right) represents the GWPS of the corresponding local wavelet plots of north, south, and total disk data where the peaks of the significant periods are marked. 


\begin{table}[h]
\centering
\caption{Periods determined using the GWPS applied to KO sunspot number data for Cycles 14 to 24}
\begin{tabular}{|c| c|}
\hline
Sunspot number Data &  Major Periods in months ($>$ 95\% confidence level)  \\
\hline 
Northern Hemisphere & $\sim$~5, $\sim$~11, $\sim$~17, $\sim$~27, $\sim$~39, $\sim$~48, $\sim$~63, $\sim$~125 \\
Southern Hemisphere & $\sim$~4.5, $\sim$~8.5, $\sim$~11, $\sim$~14, $\sim$~24, $\sim$~27, $\sim$~45, $\sim$~63, $\sim$~128  \\
Whole Solar Disk & $\sim$~5, $\sim$~9, $\sim$~13, $\sim$~27, $\sim$~39, $\sim$~63, $\sim$~125 \\
\hline
\end{tabular}
\label{tab:1}
\end{table}

\begin{figure}[!h]
\centering
\includegraphics[width=0.42\textwidth]{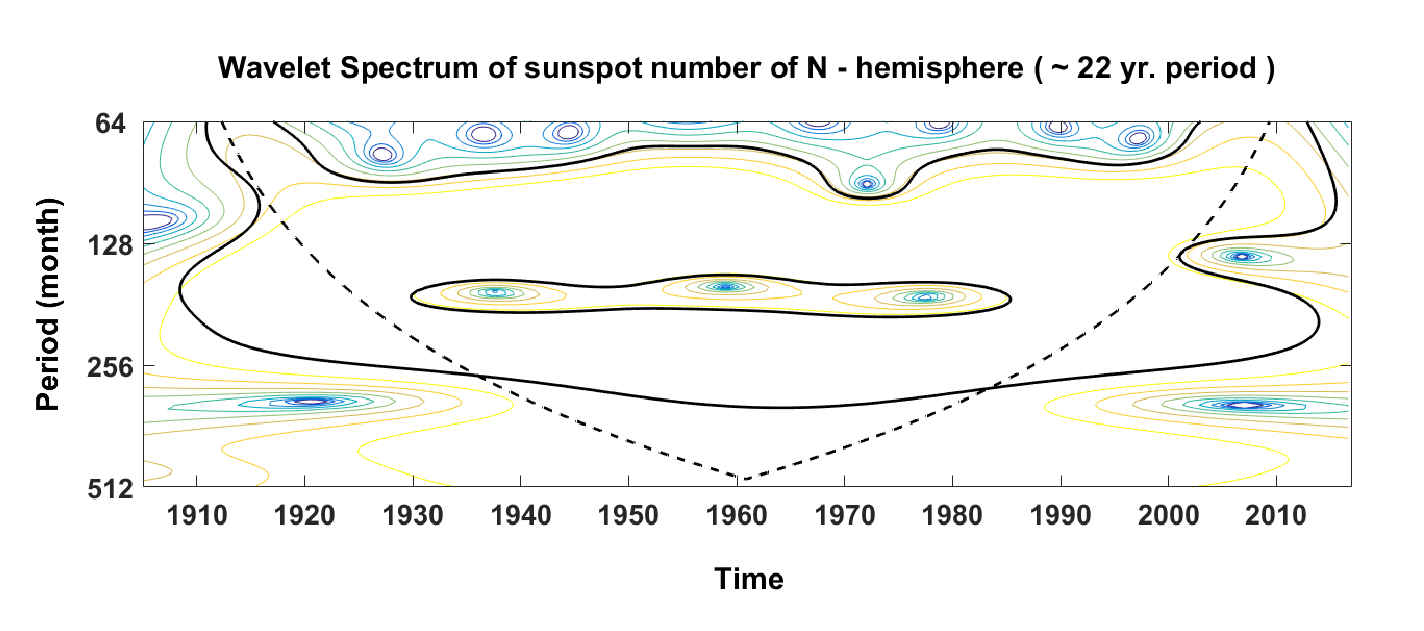}
\includegraphics[width=0.45\textwidth]{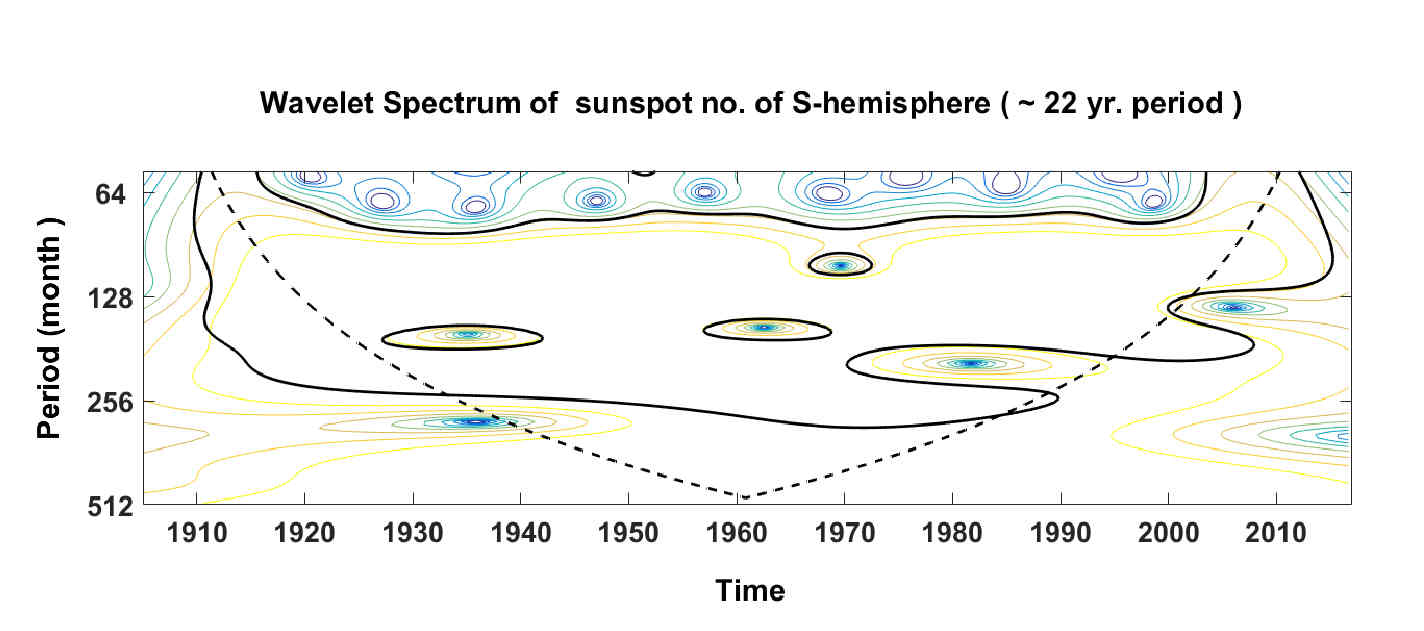} \\
\includegraphics[width=0.45\textwidth]{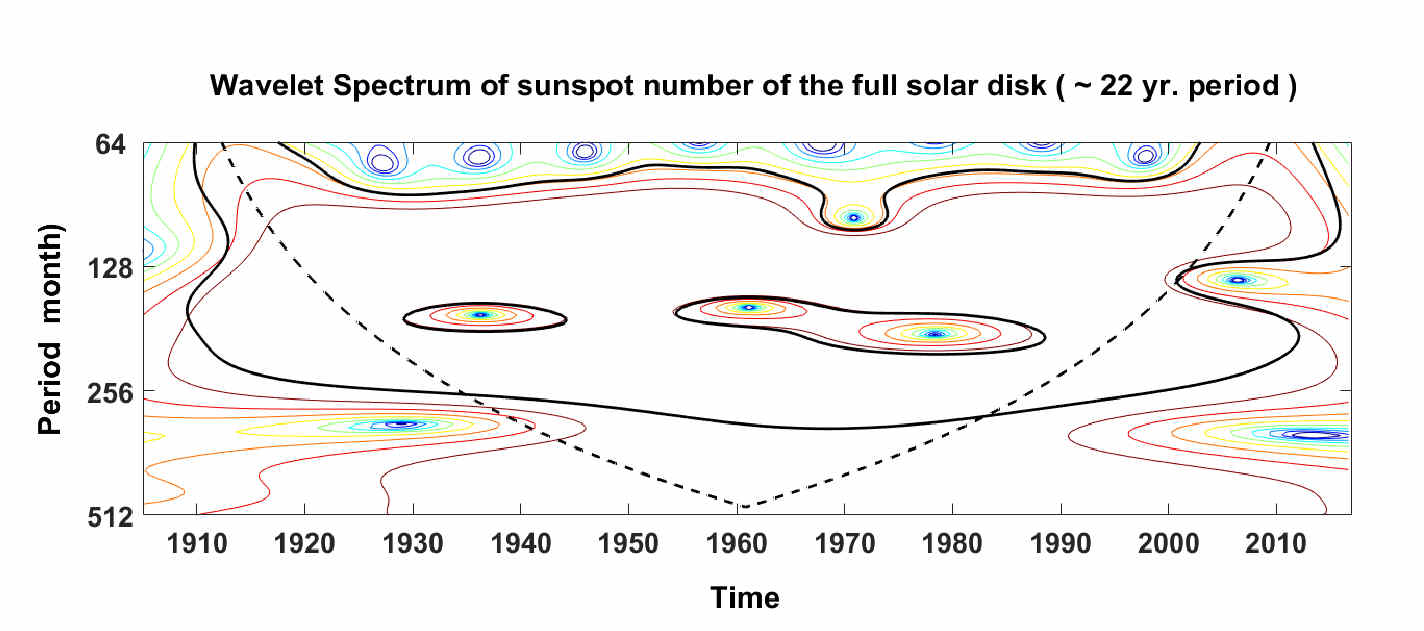}
\includegraphics[width=0.42\textwidth]{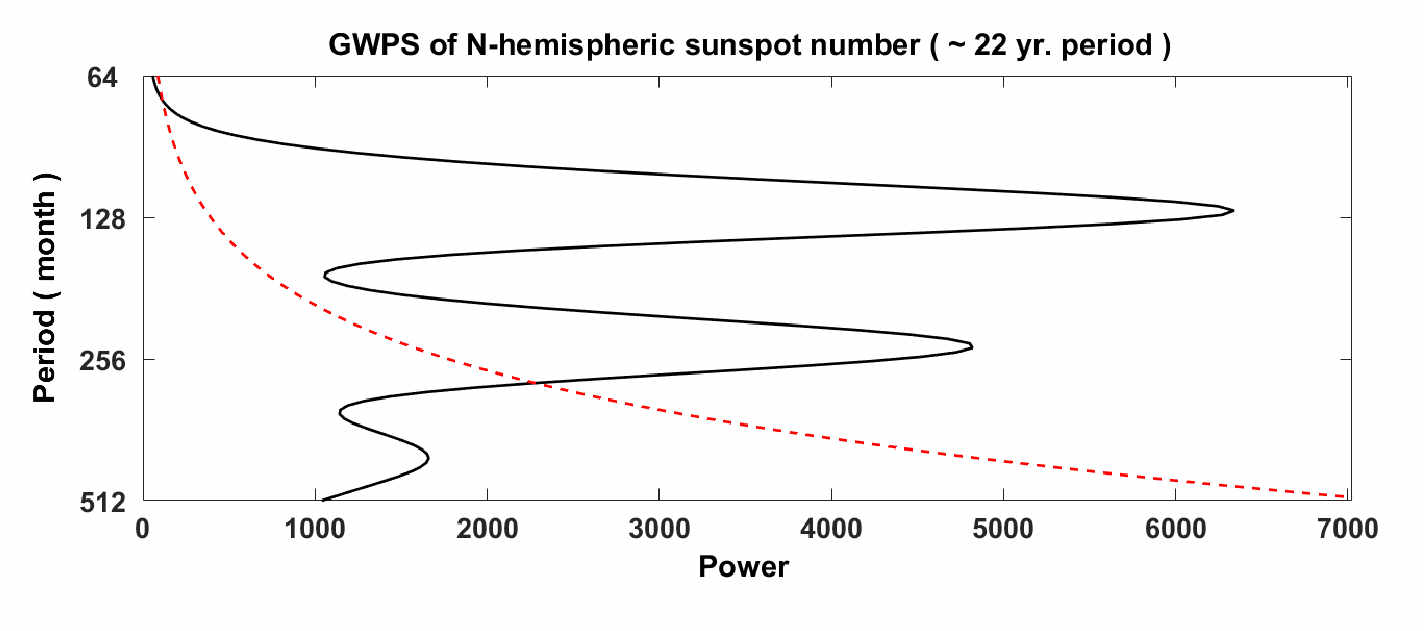} \\
\includegraphics[width=0.42\textwidth]{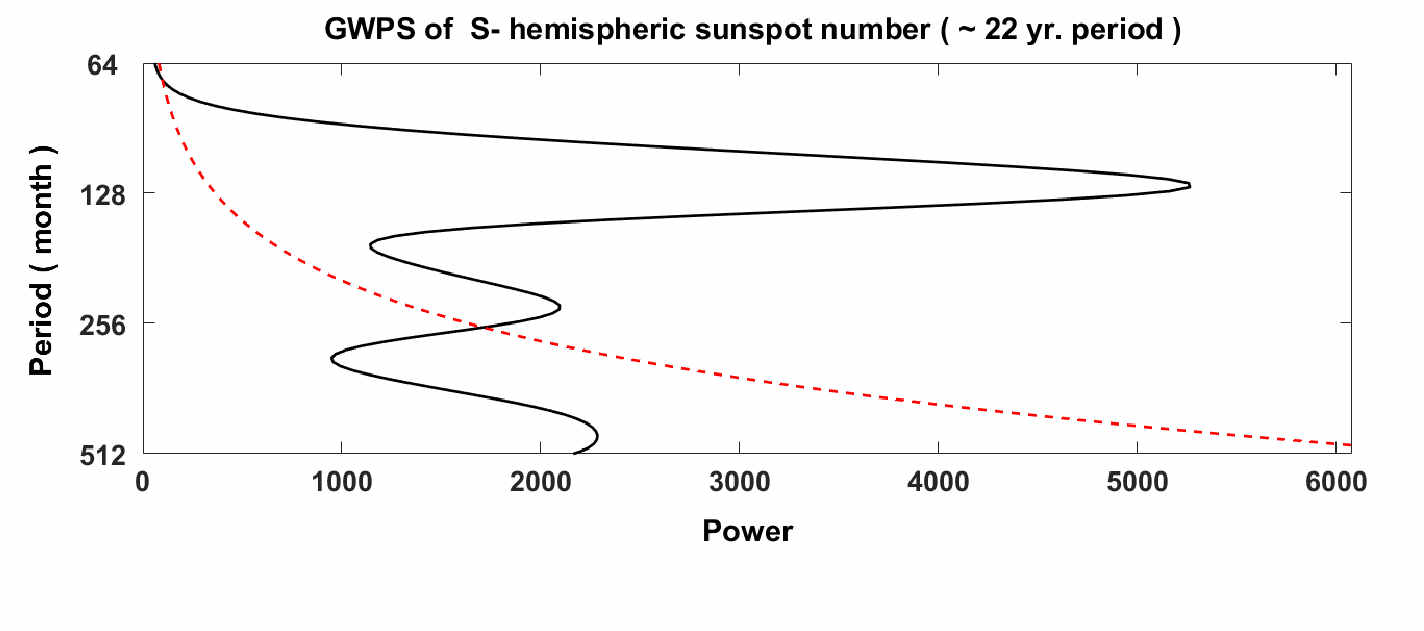}
\includegraphics[width=0.42\textwidth]{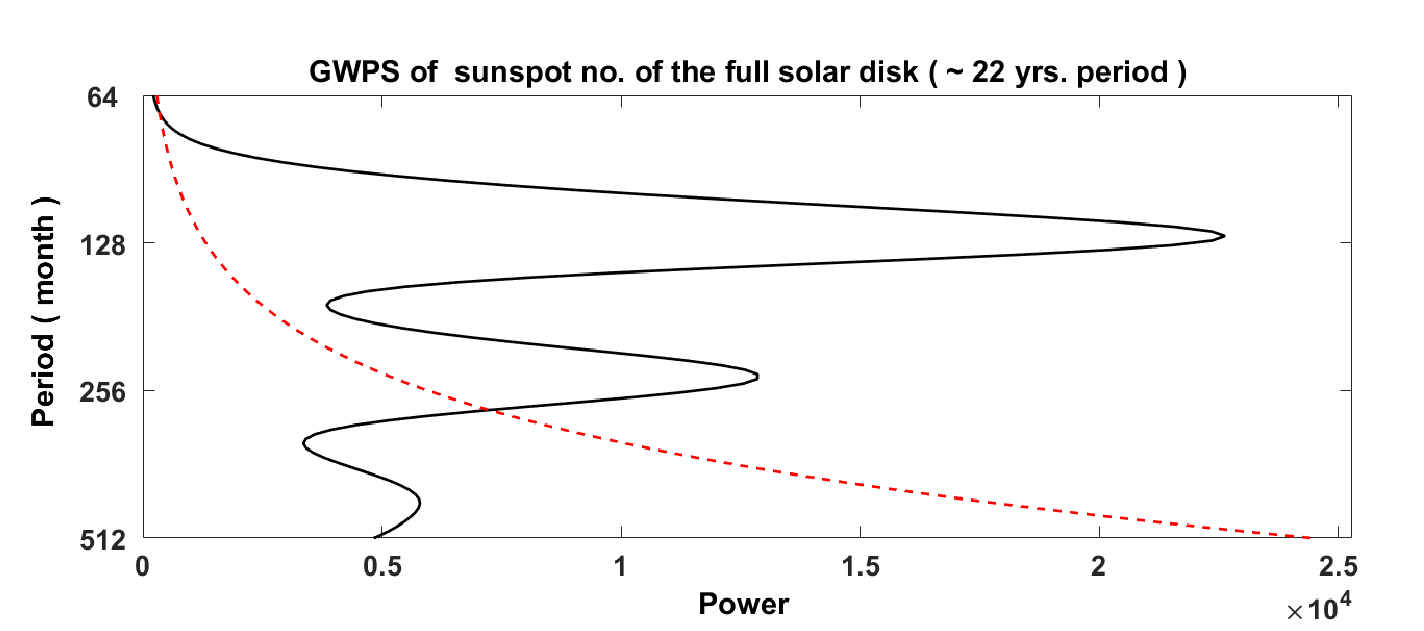} \\
\caption{The wavelet plots showing the presence of $\sim$~22 years Hale cycle in the hemispheric and full-disk  sunspot number data sets.}
\label{fig:7}
\end{figure}

Table~\ref{tab:1} shows that GWPS plots have statistically significant periods, which are also significant in the local wavelet plots. Especially the Rieger type of periods, QBOs, $\sim$5 year, and solar cycle periodicity are prominent in the GWPS plots. Along with the most prominent $\sim$~11 years sunspot cycle, sunspot also exhibit $\sim$~22 years cycle which is known as Hale cycle. Near the maximum epoch of the sunspot cycle, sun's polar magnetic field flips and returns to its original state after completing two solar cycles and hence the length of this Hale cycle / Hale polarity law is $\sim$22 years. It is commonly assumed that this Hale cycle is governed by the 22 year magnetic dynamo cycle in the presence of Sun's remnant magnetic field \citep{Mursula2002, 2019MNRAS.489.4329H}. Previously few studies indicated that $\sim$~22 years Hale cycle is related with some atmospheric phenomena like air circumfluence oscillation, variations of the air pressure, air temperature etc. \citep[][and references therein]{2012AJ....144....6Q}. However, in the power spectrum analysis this cyclic nature is suppressed due to presence of very prominent $\sim$~11 year solar cycle period. We have studied the nature and variations of this magnetic cycle in KO sunspot number data sets by wavelet technique with $\omega_{0}$ = 6 (after reducing the power of Schwabe cycle by applying moving average method.) Figure~\ref{fig:7}, shows the required plots in this context. From GWPS plots it is found that the length of the Hale cycle is slightly higher in the northern hemisphere than the southern one.

\subsection{Periodicities in KO sunspot group area}
\label{subsec:psga}

\begin{figure}[!h]
\centering
\includegraphics[width=0.42\textwidth]{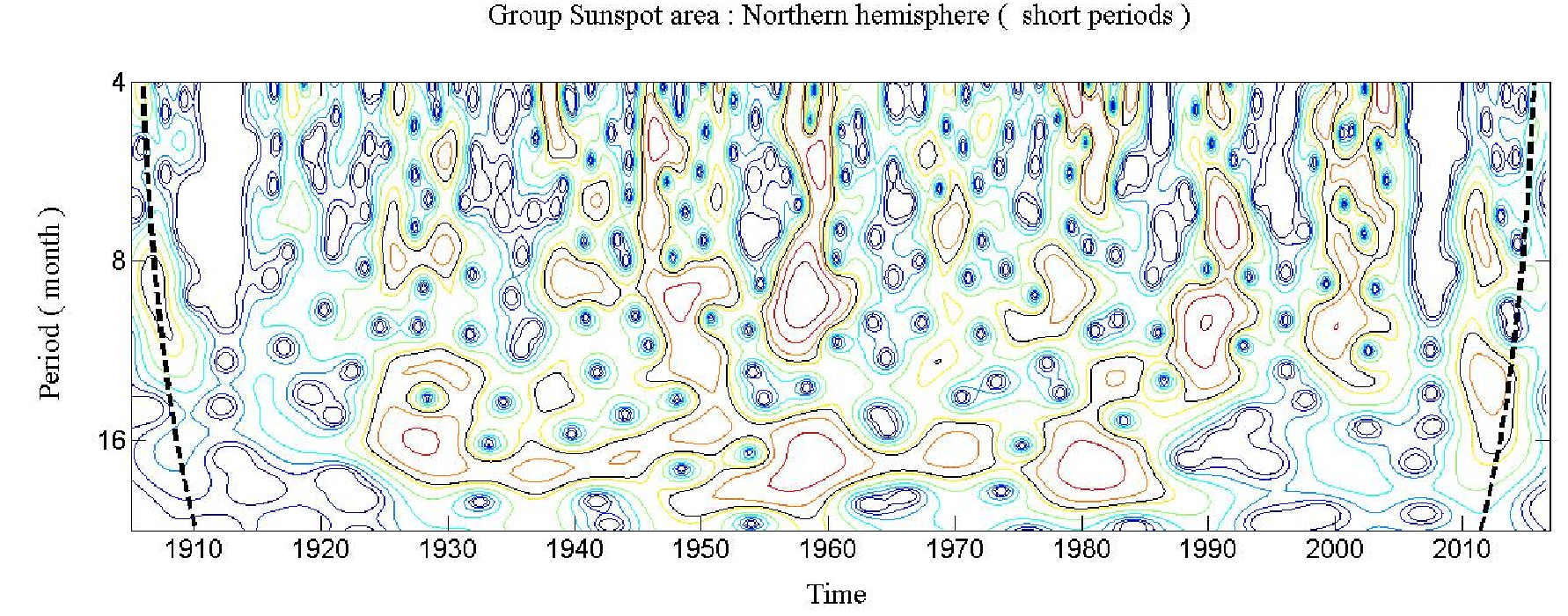}
\includegraphics[width=0.42\textwidth]{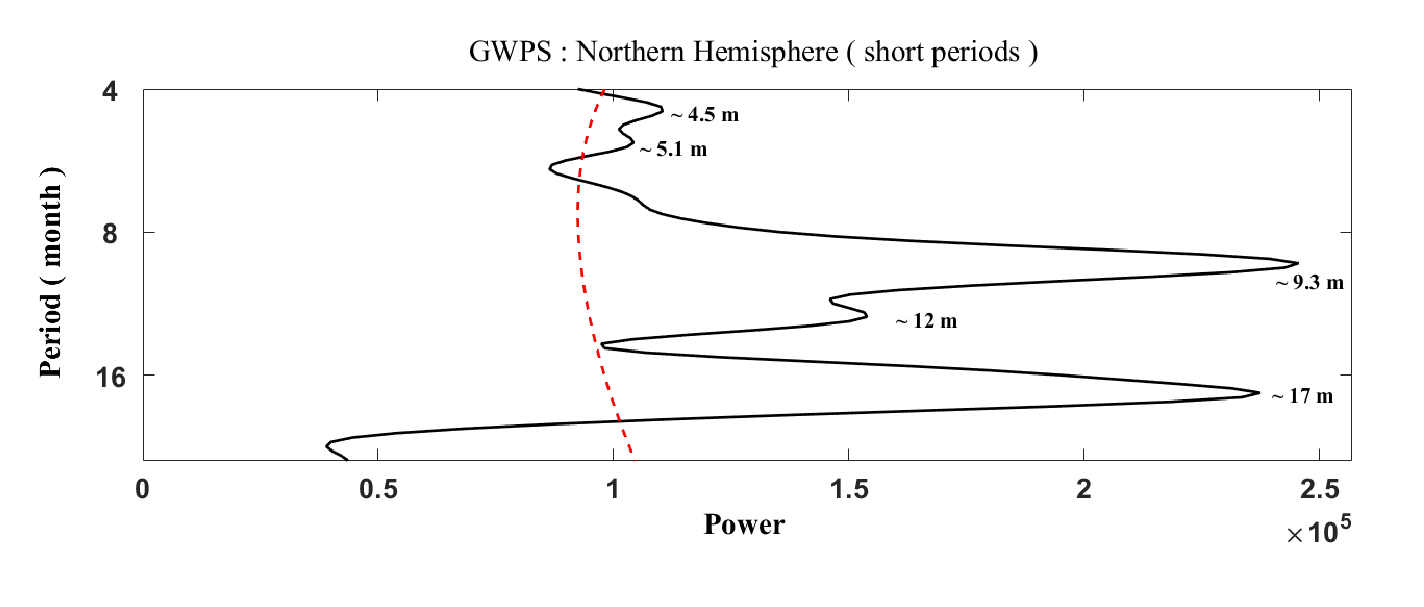} \\
\includegraphics[width=0.42\textwidth]{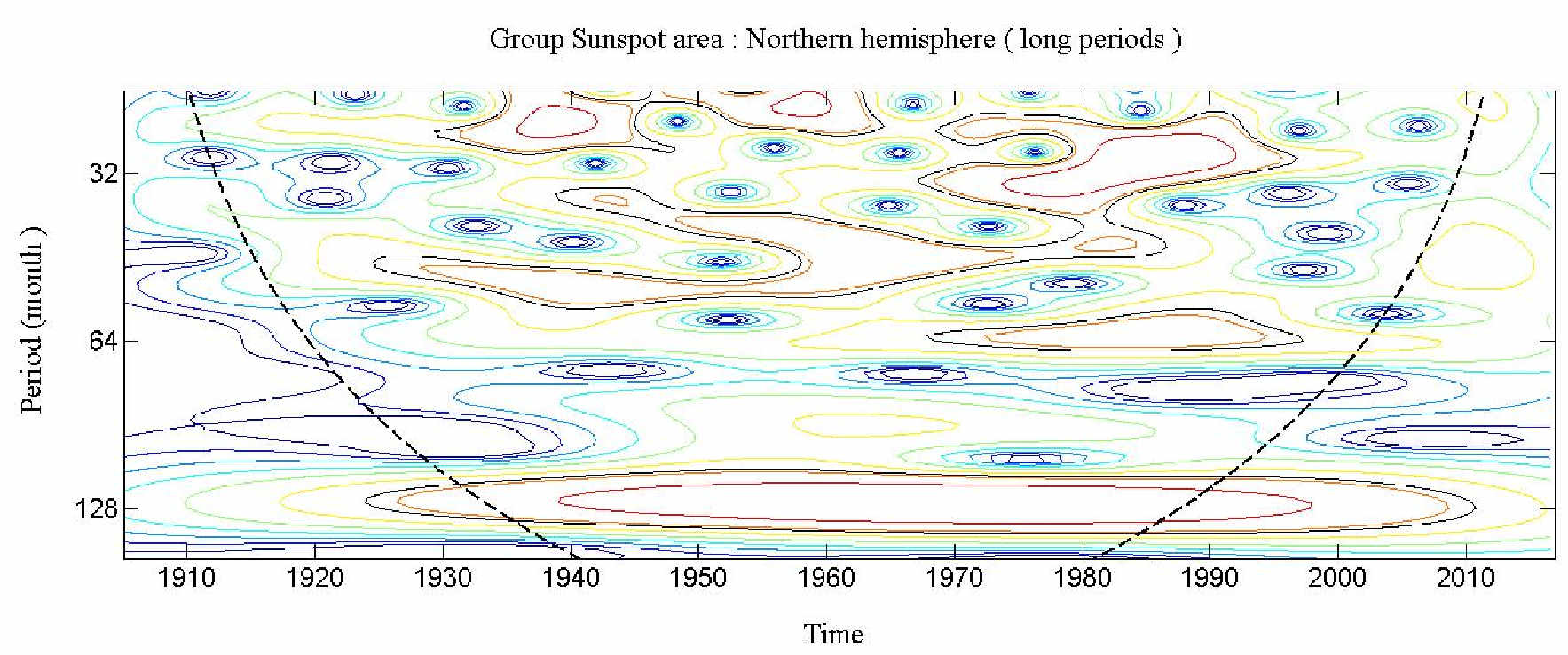}
\includegraphics[width=0.42\textwidth]{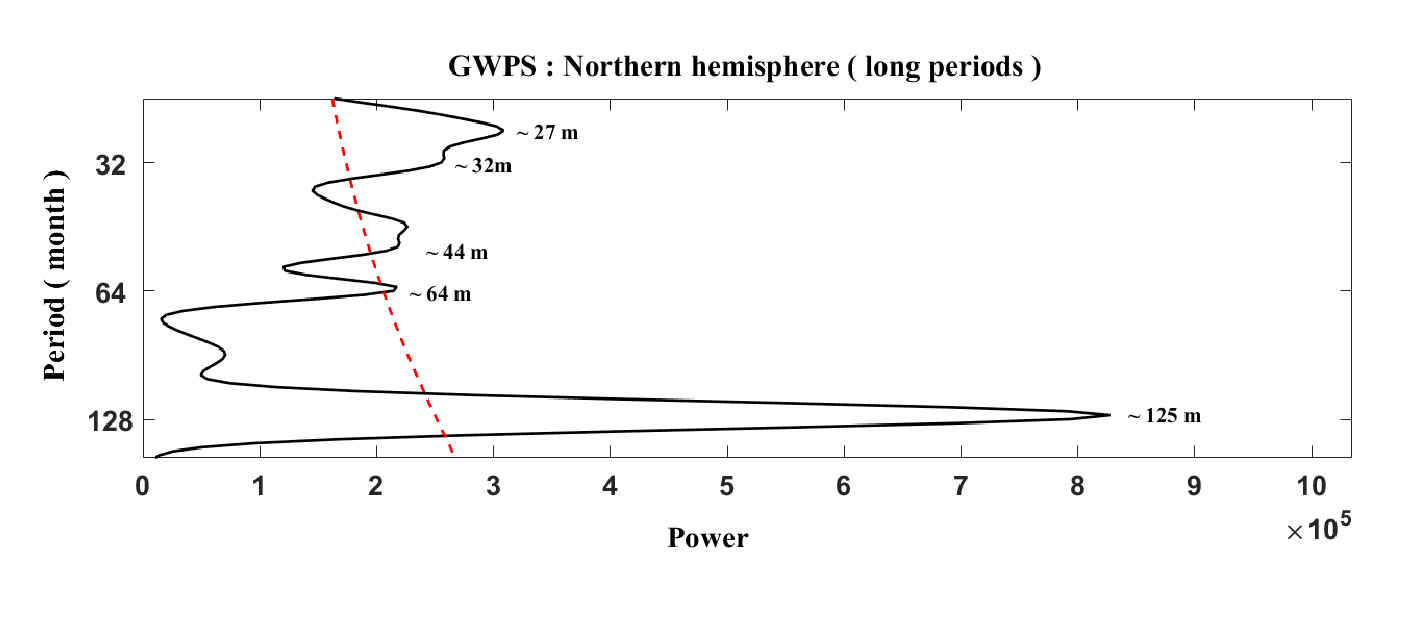} \\
\caption{(a) Morlet wavelet spectra of the monthly KO sunspot group area time series in the northern hemisphere. (b) Global power spectra for short term periods mainly to study Rieger group of periodicities. (c) Local morlet wavelet power spectra to investigate the dynamical behaviors of QBOs (1.4 to 4 years) and other long term periods including solar cycle period. (d) Similar to panel (b) for other long term periods including Schwabe cycle. Dotted lines in all global wavelet power spectra represent a 95\% confidence level.}
\label{fig:8}
\end{figure}

\begin{figure}[!h]
\centering
\includegraphics[width=0.42\textwidth]{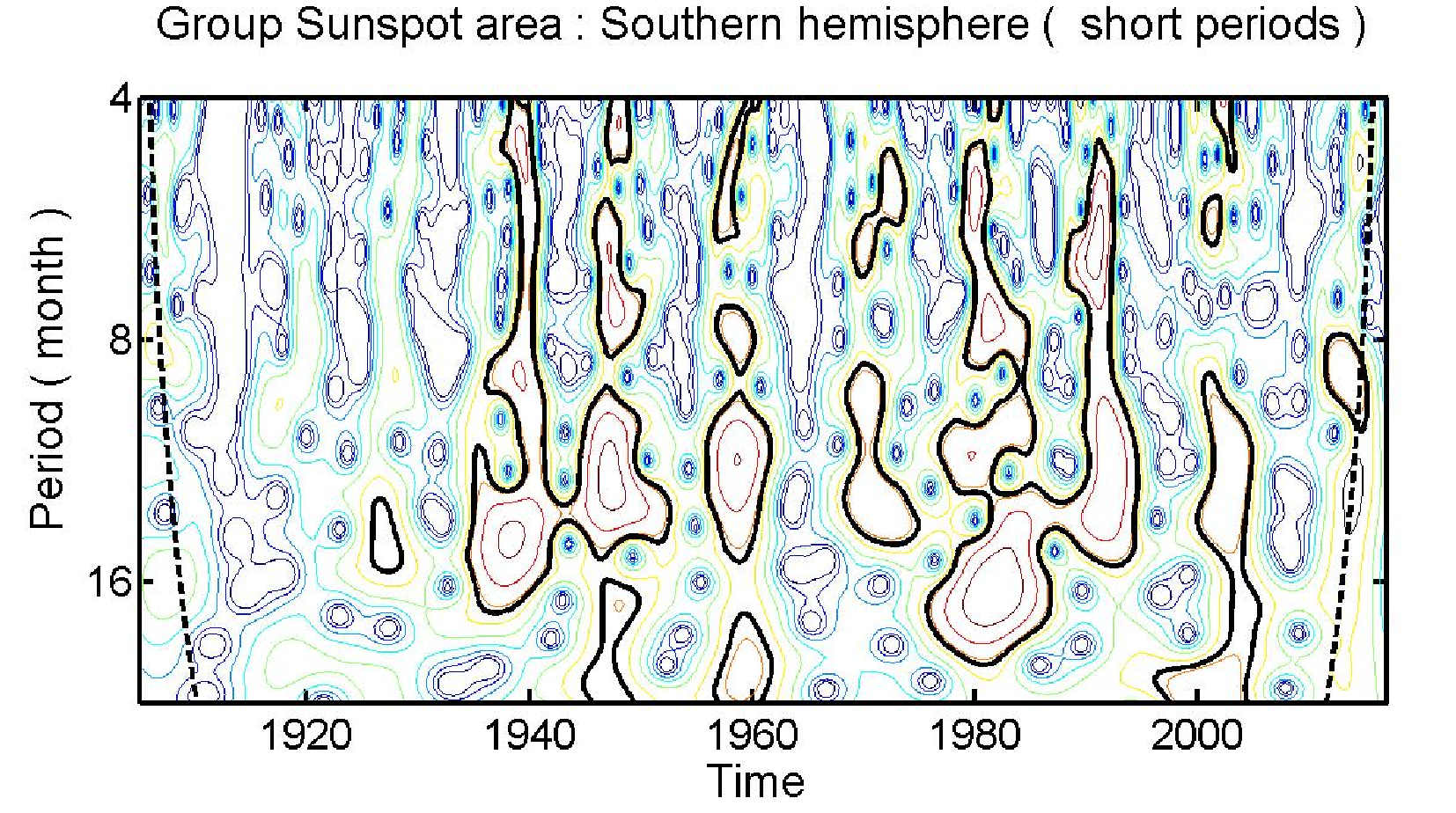}
\includegraphics[width=0.45\textwidth]{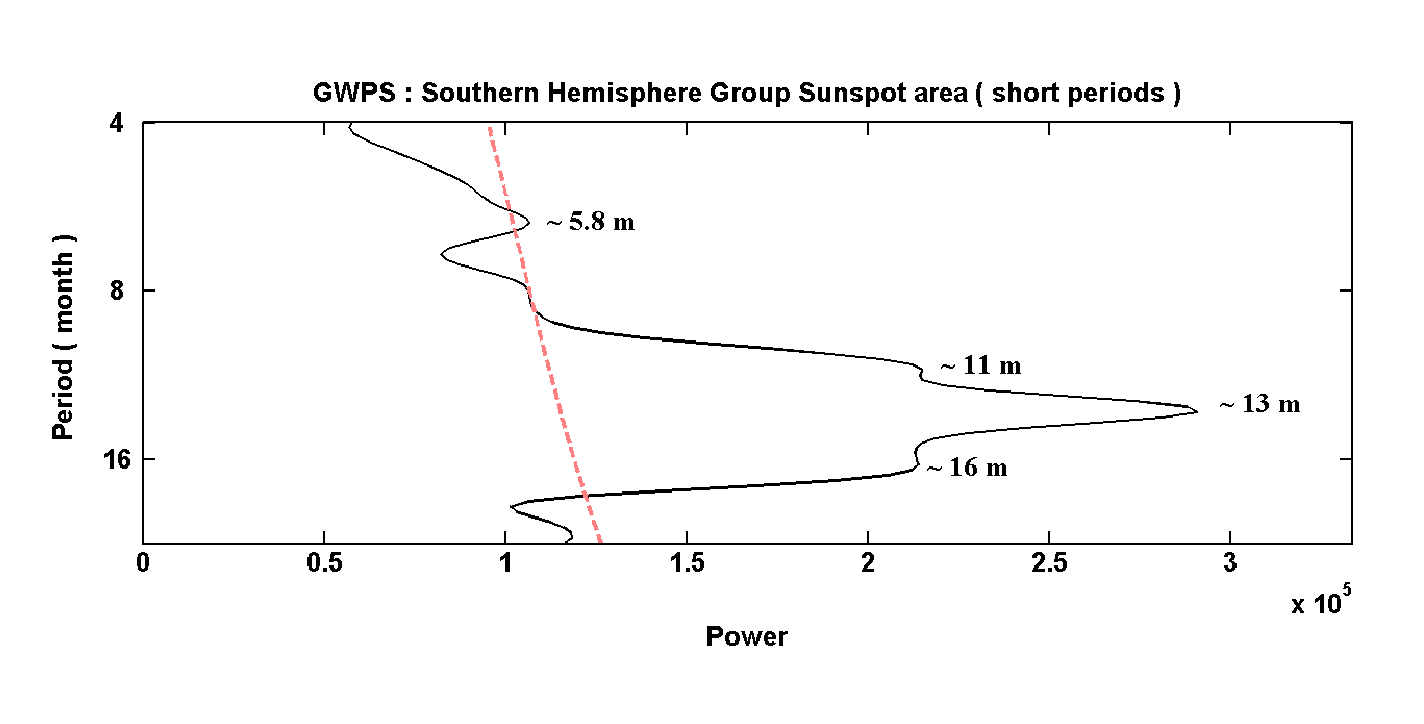} \\
\includegraphics[width=0.42\textwidth]{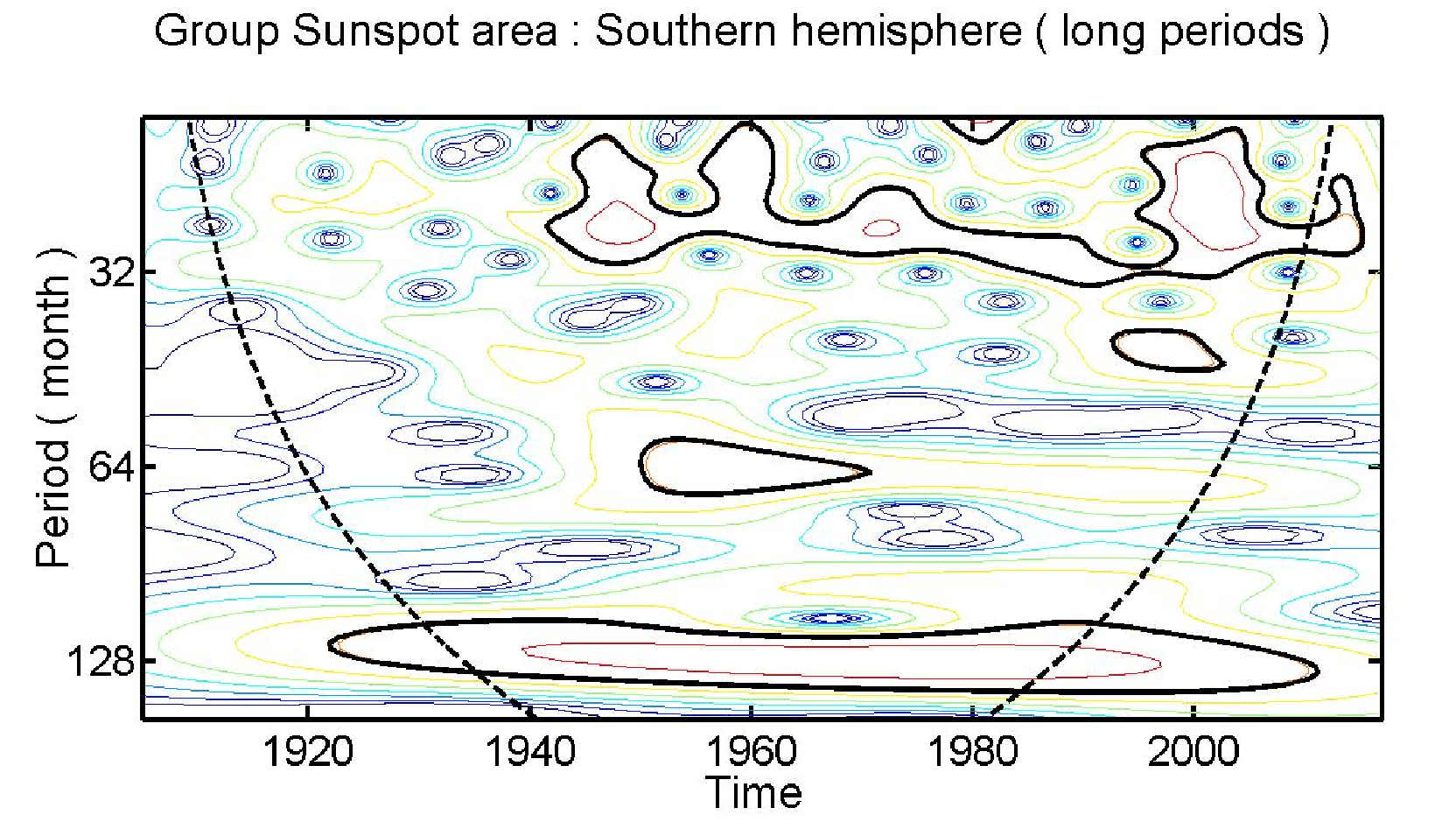}
\includegraphics[width=0.45\textwidth]{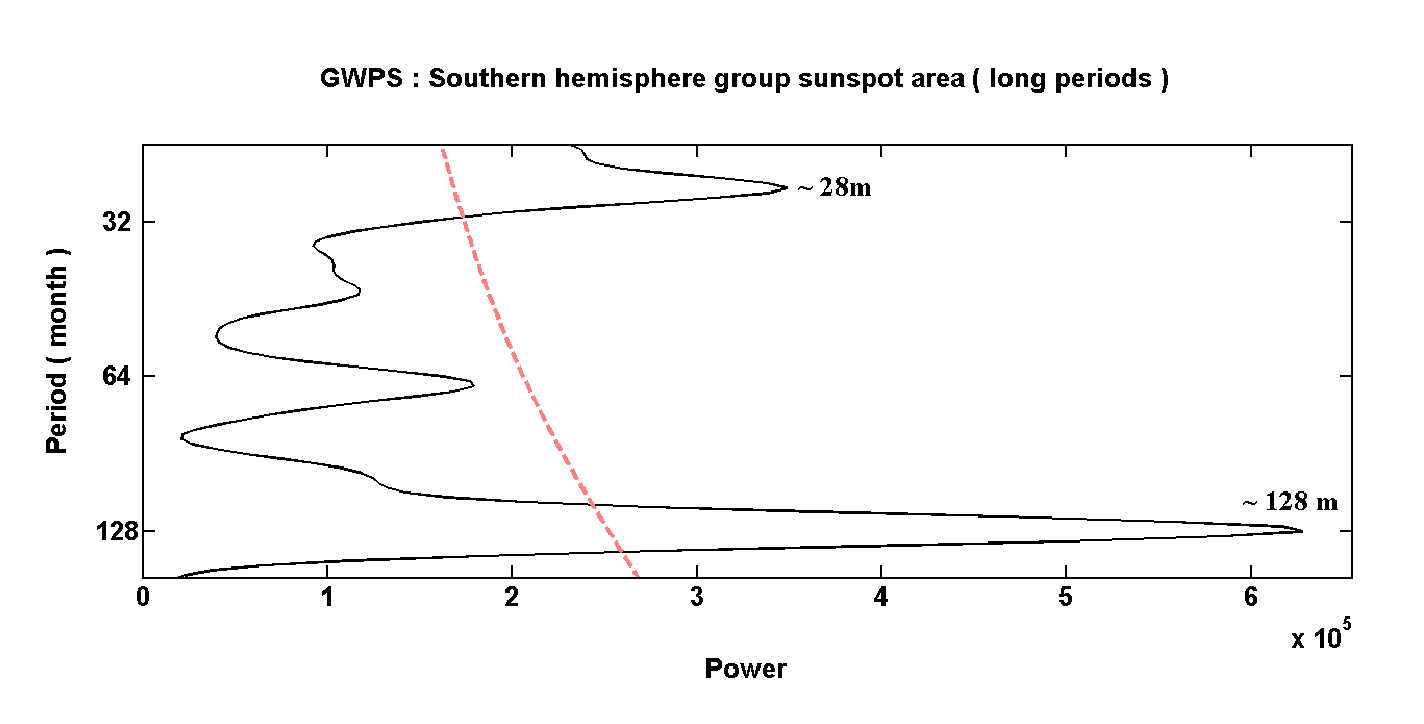} \\
\caption{Same as Figure~\ref{fig:8}, but for the southern hemisphere.}
\label{fig:9}
\end{figure}

\begin{figure}[!h]
\centering
\includegraphics[width=0.42\textwidth]{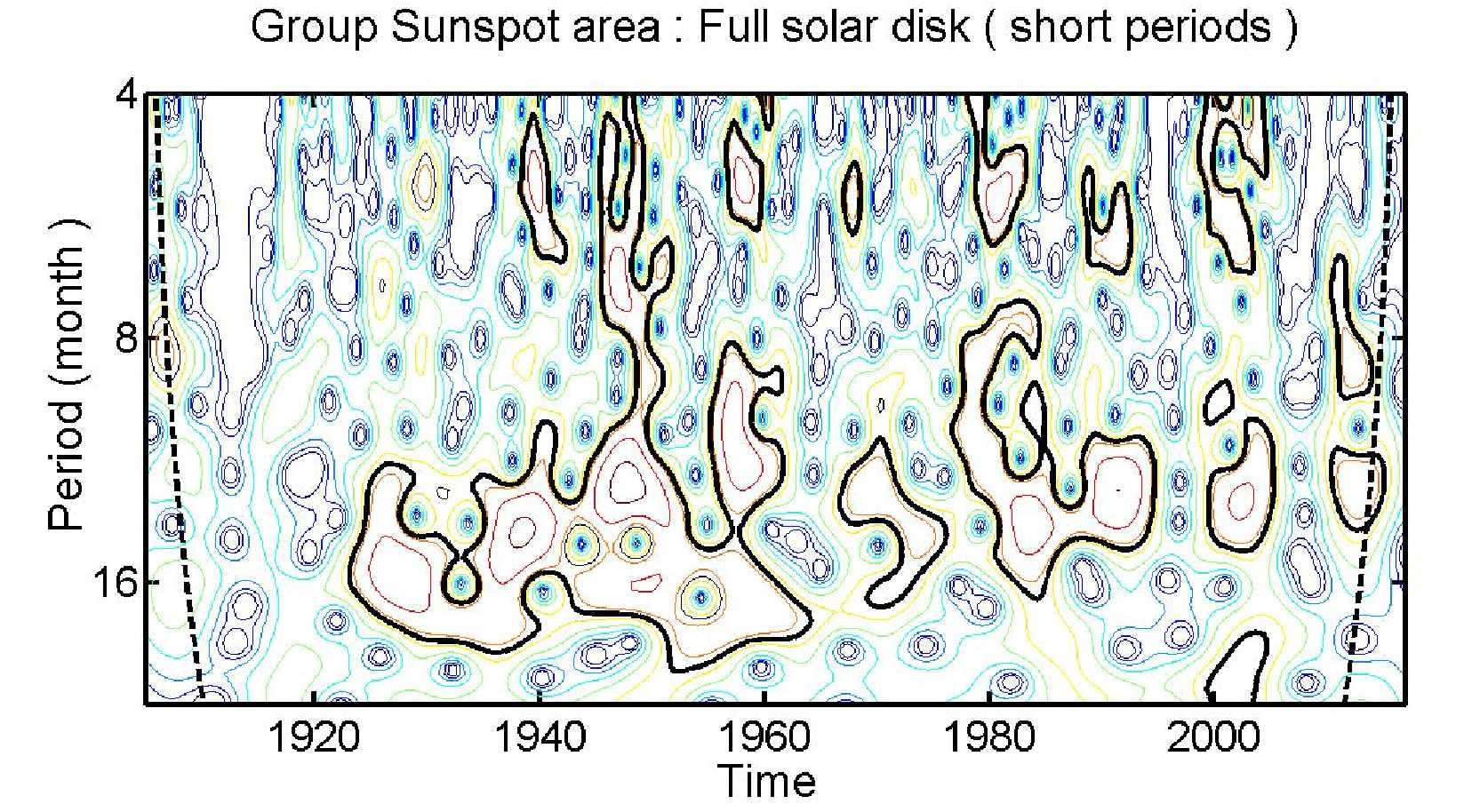}
\includegraphics[width=0.45\textwidth]{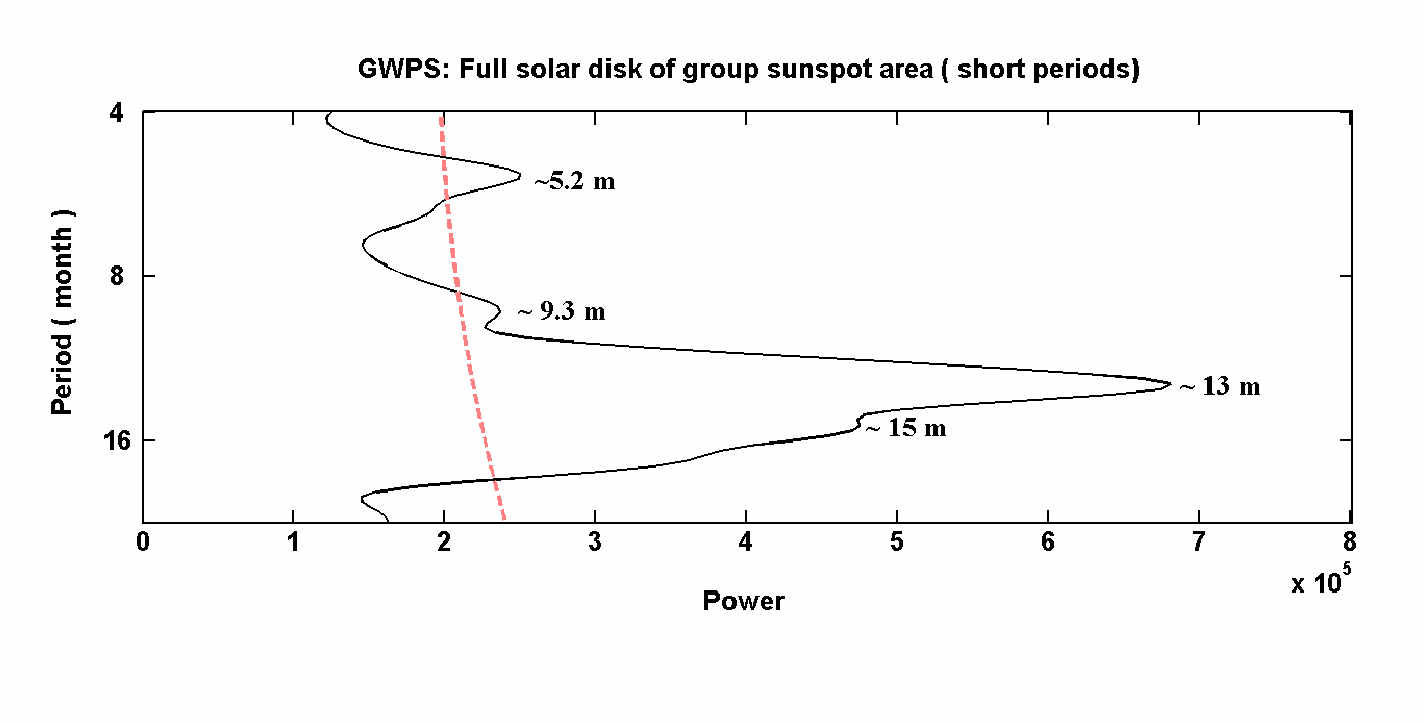} \\
\includegraphics[width=0.42\textwidth]{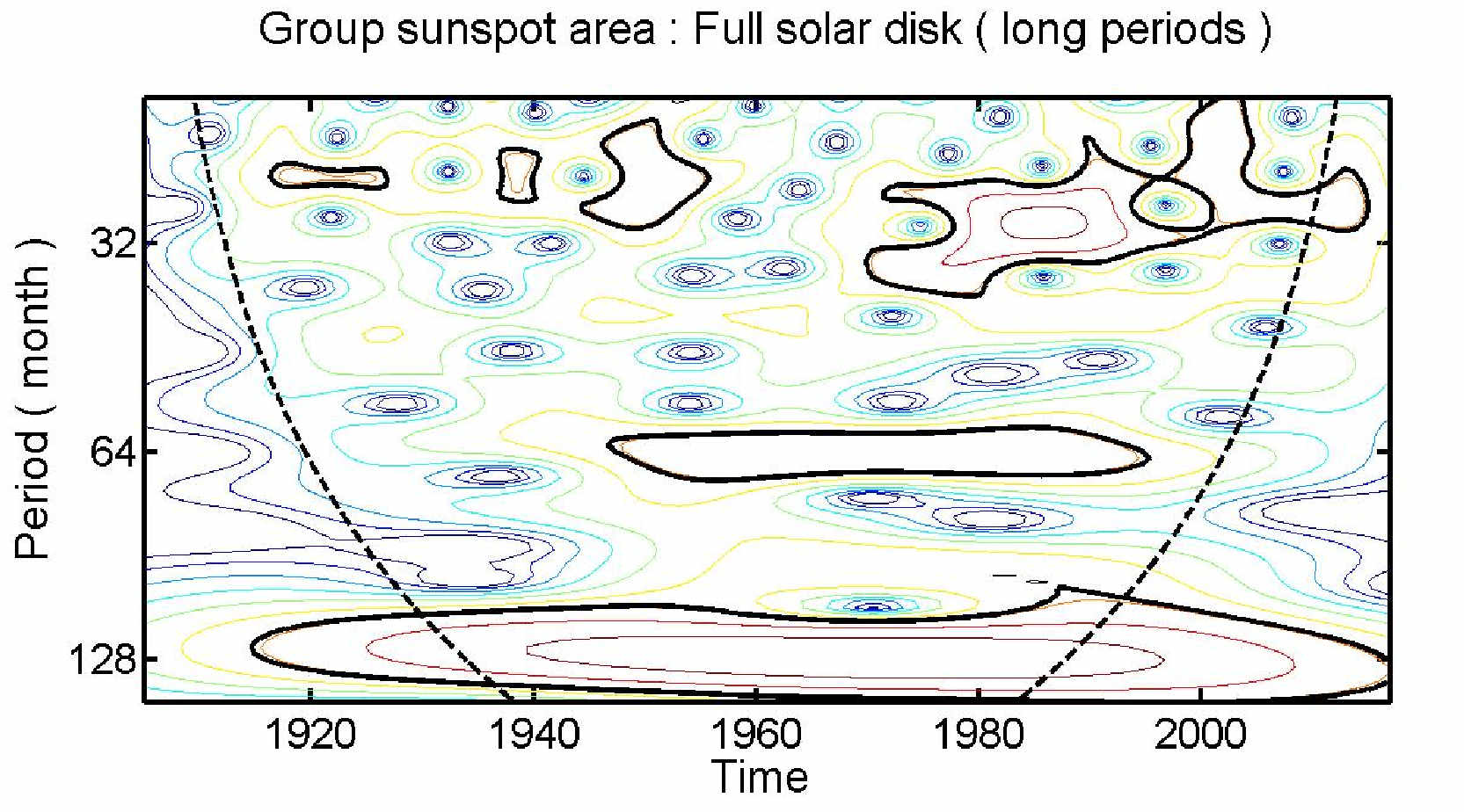}
\includegraphics[width=0.45\textwidth]{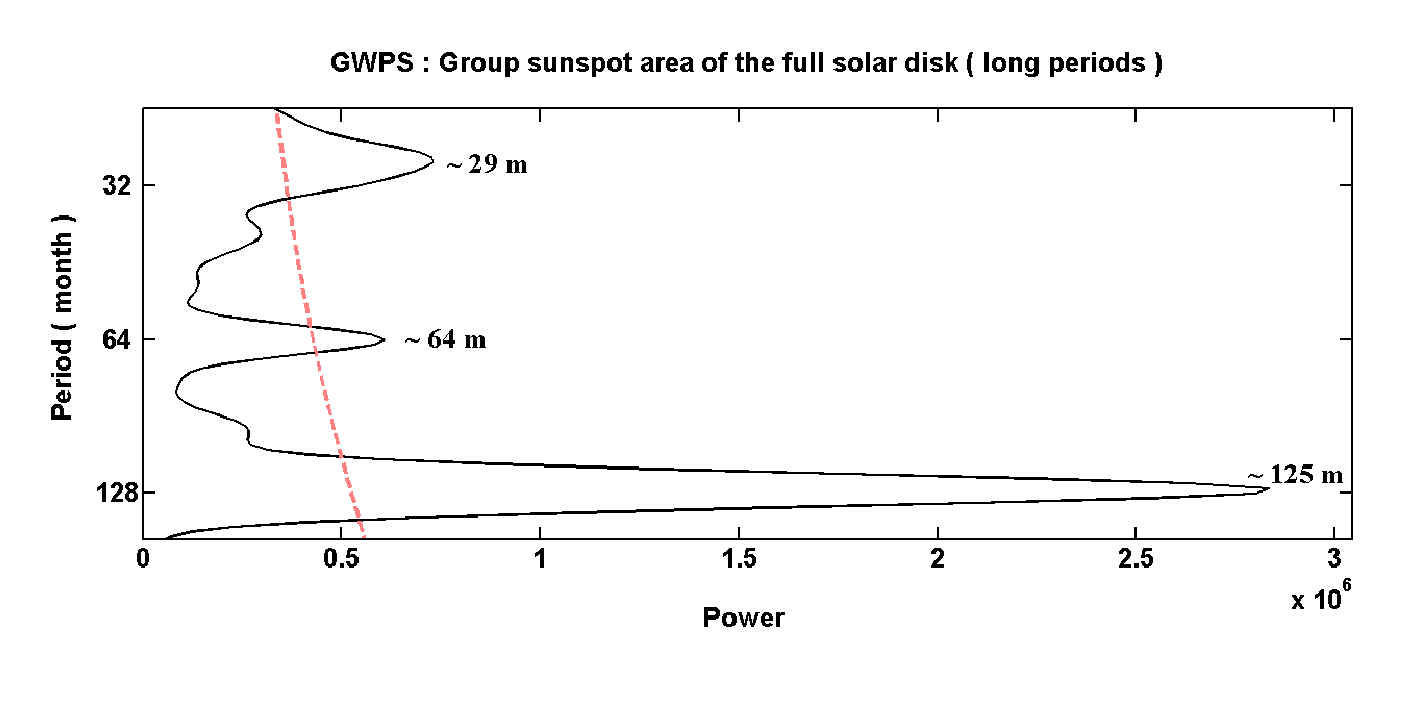} \\
\caption{Same as Figure~\ref{fig:8}, but for the whole hemisphere.}
\label{fig:10}
\end{figure}

Figures~\ref{fig:8}, \ref{fig:9}, and \ref{fig:10} show the results of the periodic variations of the KO monthly sunspot group area data after Morlet wavelet analysis.

In the northern hemisphere, the Rieger and Rieger type of periods are present mainly during different phases of the cycles 17, 18, 19, 21, 22, and 23 [Figure~\ref{fig:8}(top-left)]. The QBO's in the range of  1.2 -- 1.5 years were strongly pronounced continually from cycles 16 to 21 and during cycle 24. The Rieger-type periods were found in the southern hemisphere during cycles 17, 18, 19, 21, and 22 in the scattered form [Figure~\ref{fig:9}(top-left)]. We have detected such types of QBO's in the southern hemisphere, in scattered form during 1935 -- 1950; $\sim$1958 -- 1962; with varying lengths from $\sim$1976 to 1990 and from $\sim$1998 to 2004. Figure \ref{fig:10}(top-left) and \ref{fig:10}(bottom-left) represents the local wavelet spectrum of the sunspot group area of the whole solar disk. These plots show Rieger and Rieger type of periods in different parts of the cycles 17, 19, 20, 21, 22, and 23. We found that QBOs in the range of 1.2 -- 1.5 years were significant continuously from cycles 16 to 20, 21, 22, and 23 in this time series.

The QBO's in the range of 2 -- 2.5 years were significant from cycles 16 to 19 and cycles 20 to 22 (Figure~\ref{fig:8}(bottom-left)) in the northern hemisphere. This plot indicates the existence of other QBOs (3 -- 4 years) between cycles 16 and 22. On the other hand, the southern hemisphere shows the presence of only 2 -- 2.5 years period from cycles 17 to 24 [Figure~\ref{fig:9}(bottom-left)]. Solar cycle 16, 18, and from cycles 21--24 provided the presence of a 2--2.5 year period for the whole solar disk group sunspot area data set [Figure~\ref{fig:10}(bottom-left)]. A quasi-periodicity of around 5-years is present in all sunspot group area data sets. Wavelet power spectrum showed the robust existence of solar cycle periodicity in all sunspot area data sets under study.

Figures~\ref{fig:8}(top-right) and (bottom-right), \ref{fig:9}(top-right) and (bottom-right) and \ref{fig:10}(top-right) and 9(bottom-right) represent the GWPS of the corresponding local wavelet plots of the sunspot group area where the peaks of the significant periods are marked. Table~\ref{tab:2} shows the results of the GWPS plots. 

\begin{table}[h]
\centering
\caption{Periods determined using the GWPS applied to KO sunspot group area data for Cycles 14 to 24}
\begin{tabular}{|c|c|}
\hline
Sunspot number Data &  Major Periods in months ($>$ 95\% confidence level)  \\
\hline 
Northern Hemisphere & $\sim$~4.5, $\sim$~5, $\sim$~9.3, $\sim$~12, $\sim$~17, $\sim$~27, $\sim$~32, $\sim$~44, $\sim$~64, $\sim$~125 \\
Southern Hemisphere & $\sim$~5.8, $\sim$~11, $\sim$~13, $\sim$~16, $\sim$~28, $\sim$~128  \\
Whole Solar Disk & $\sim$~5.2, $\sim$~9.3, $\sim$~13, $\sim$~15, $\sim$~29, $\sim$~64, $\sim$~125 \\
\hline
\end{tabular}
\label{tab:2}
\end{table}

\begin{figure}[!h]
\centering
\includegraphics[width=0.42\textwidth]{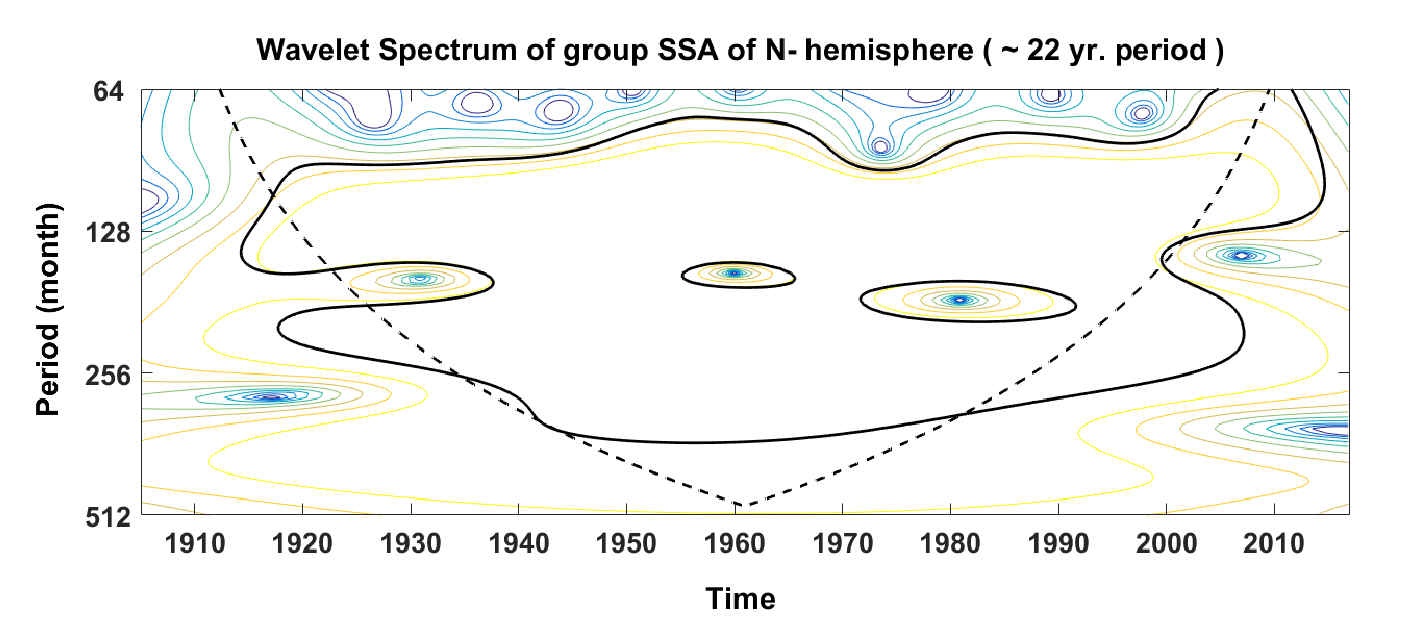}
\includegraphics[width=0.42\textwidth]{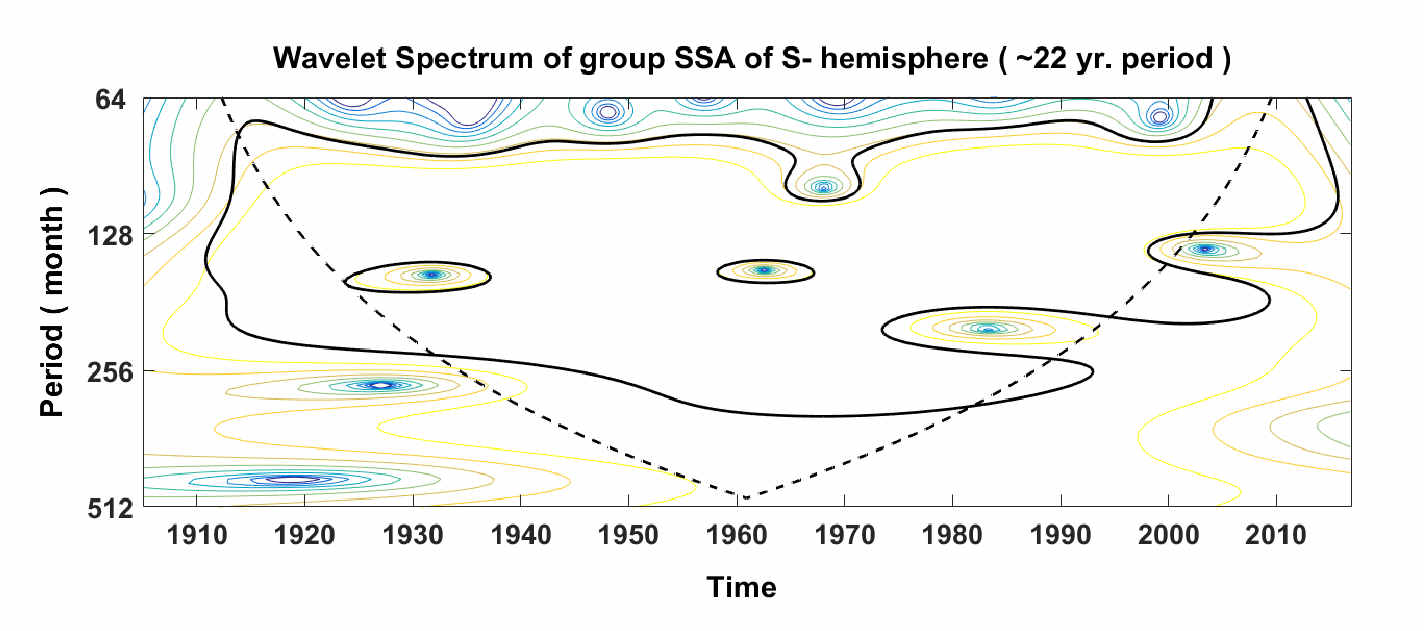} \\
\includegraphics[width=0.42\textwidth]{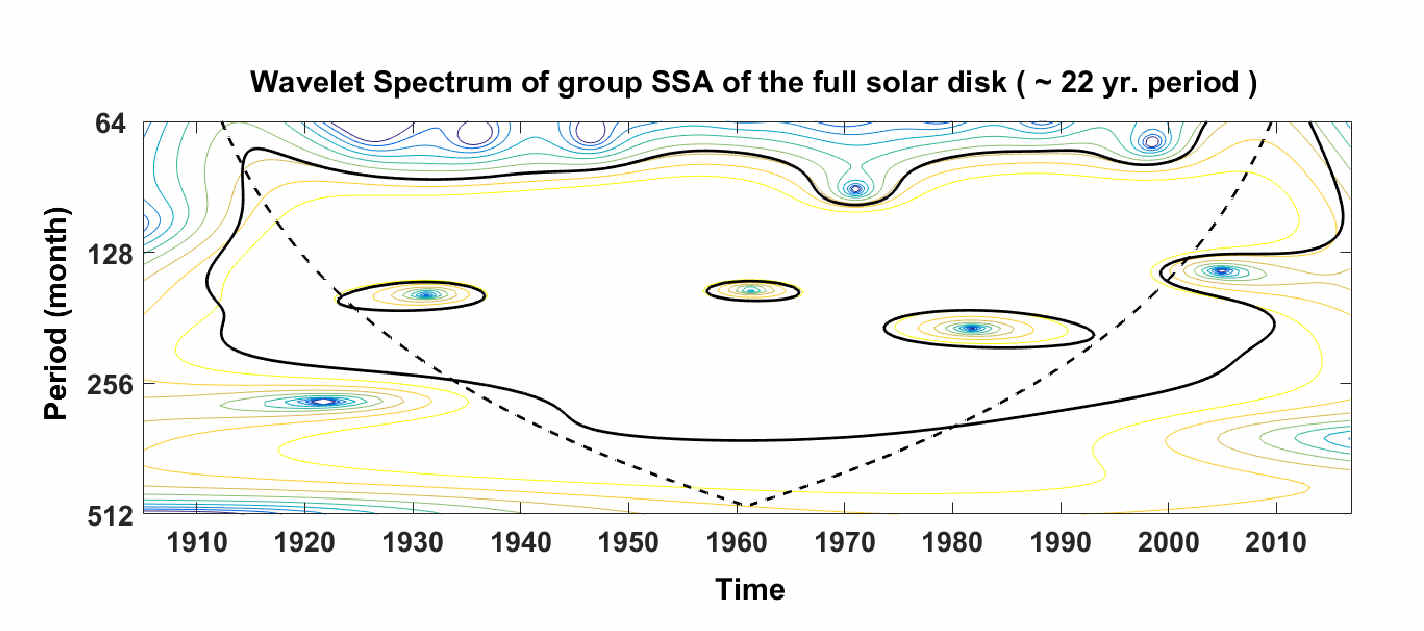}
\includegraphics[width=0.42\textwidth]{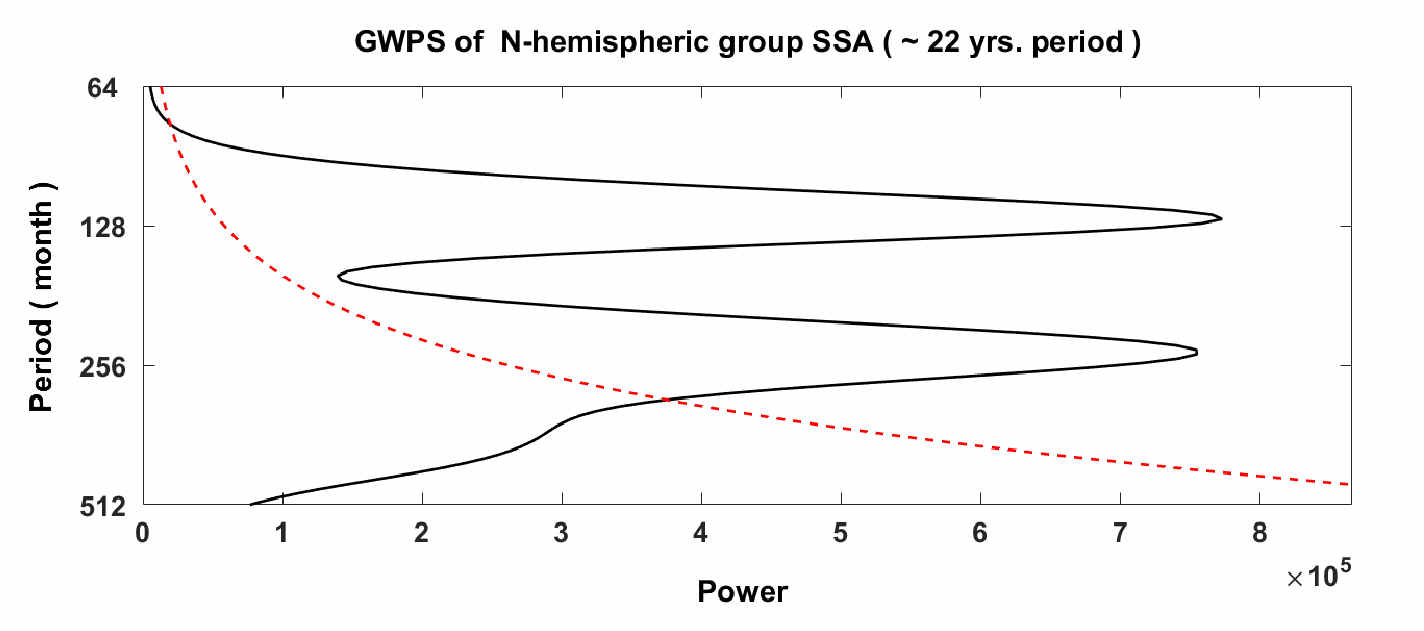} \\
\includegraphics[width=0.42\textwidth]{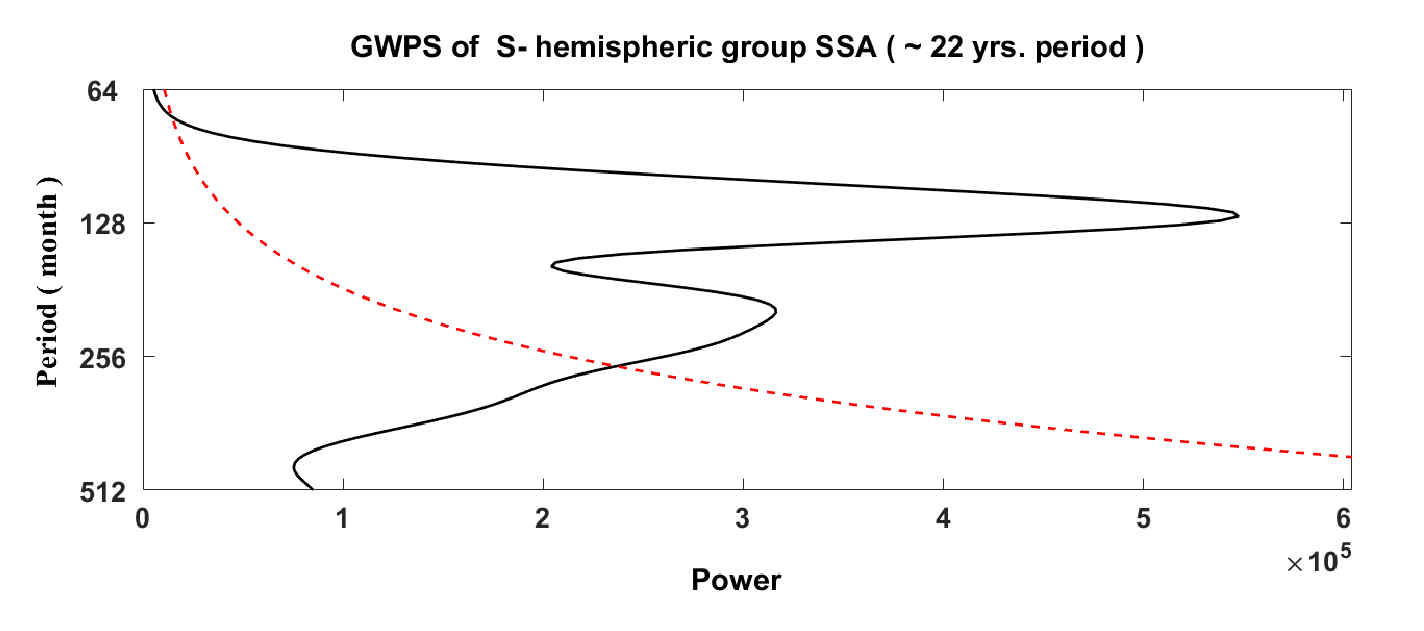}
\includegraphics[width=0.42\textwidth]{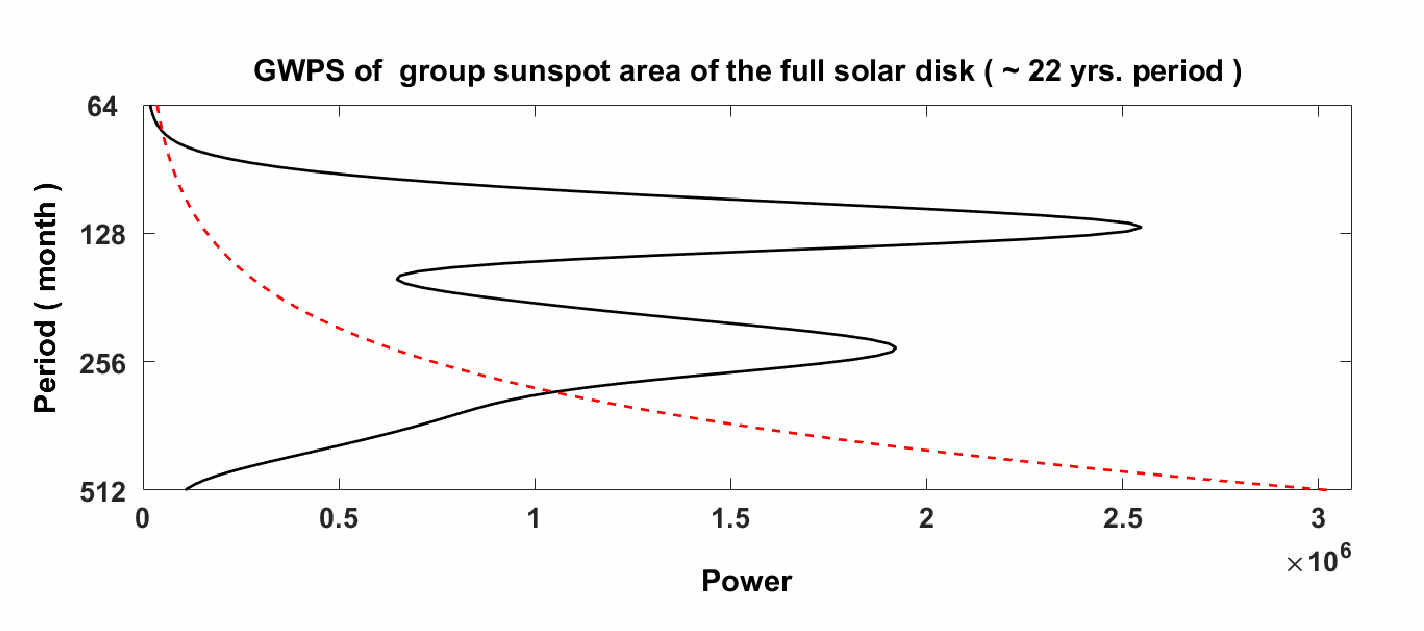} \\
\caption{The wavelet plots showing the presence of $\sim$~22 years Hale cycle in the hemispheric and full-disk group sunspot area data sets.}
\label{fig:11}
\end{figure}

Table~\ref{tab:2} indicates that peaks in the GWPS plots have similarities with the periods of local wavelet spectra of the sunspot group area time series. Rieger type of periods, QBOs in the range of 1.2 -- 3 years, $\sim$~5 year period, and Schwabe cycle period (10--11 years) were prominent in GWPS. To detect the presence and evolution of the $\sim$~22 year magnetic cycle in the group sunspot area data, we have adopted the same technique as applied in the sunspot number data. Figure~\ref{fig:11}, display the wavelet plots which indicate the presence of Hale cycle in the hemispheric and full-disk group sunspot area data sets. From the GWPS plots it is clear that length of this cycle is slightly lower in southern hemisphere.

\subsection{Spatio--temporal evolution of QBOs (1 -- 2.5~yrs.)} 

QBO's in the range of 1.2 to 2.5 years are considered as most prevalent quasi periodicity shorter than about 11-years sunspot cycle and were related with the double peak nature of the solar cycle \citep{2015LRSP...12....4H}. These groups of periods were detected in different solar and heliospheric activity indices, including helioseismic proxies, asymmetry time series as well as in the galactic cosmic ray data sets \citep[e.g.,][etc.]{2012ApJ...749...27V, 2014SSRv..186..359B, 2015SoPh..290.3095B, 2019MNRAS.488..111D, 2021SoPh..296....2R}. It has been reported that QBOs are intermittent in nature without any stable period. The amplitude of the QBOs is modulated in the course of about an 11-year cycle, being highest nearly at the maximum solar epoch and becoming weaker during the descending/ minimum phase of the solar cycle \citep{2014SSRv..186..359B}.  Some studies indicated that QBOs around two years are correlated with the second solar dynamo mechanism \citep[][etc.]{1998ApJ...509L..49B, 2012MNRAS.420.1405B, 2014ARep...58..936O}.

\begin{figure}[!h]
\centering
\includegraphics[width=0.42\textwidth]{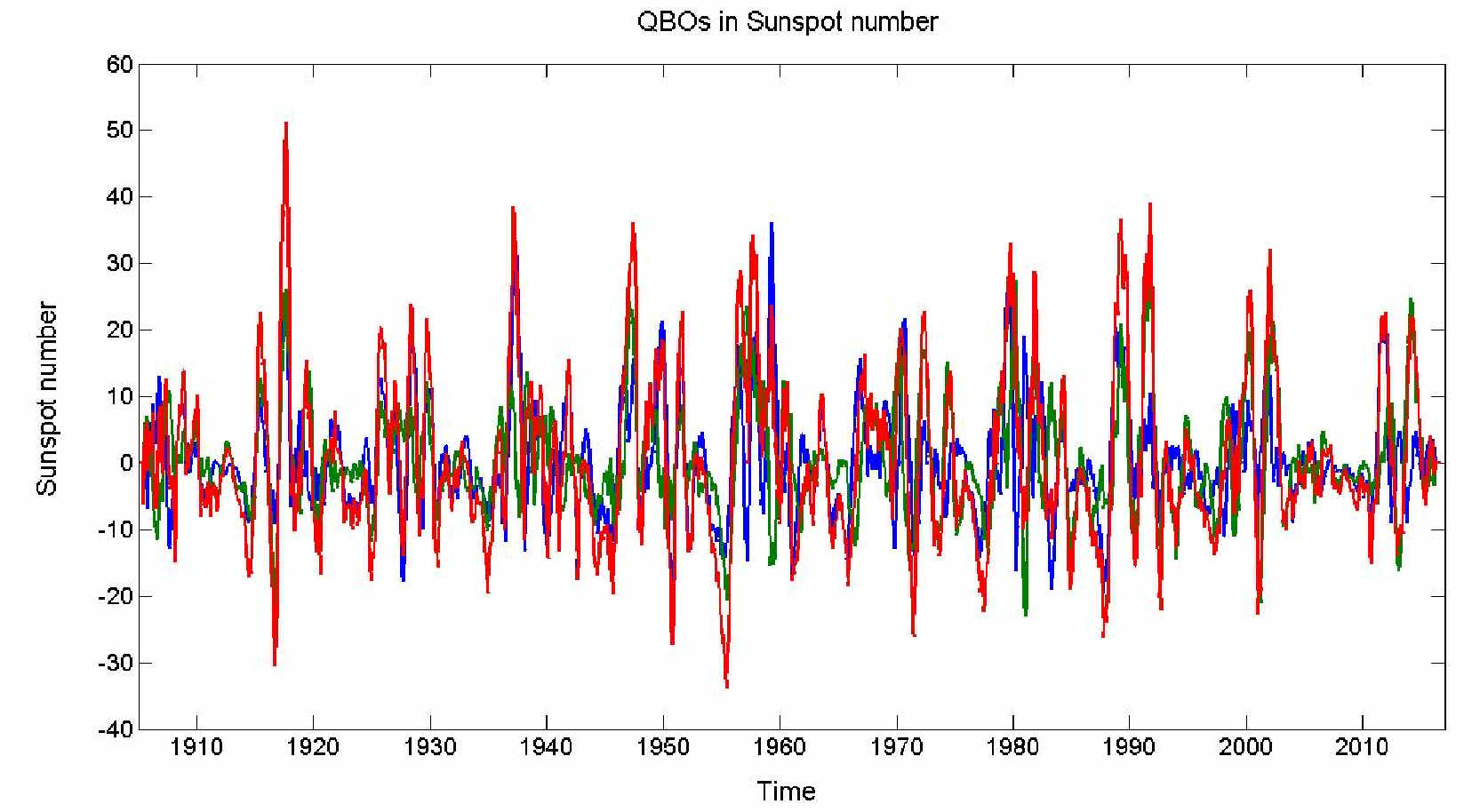}
\includegraphics[width=0.47\textwidth]{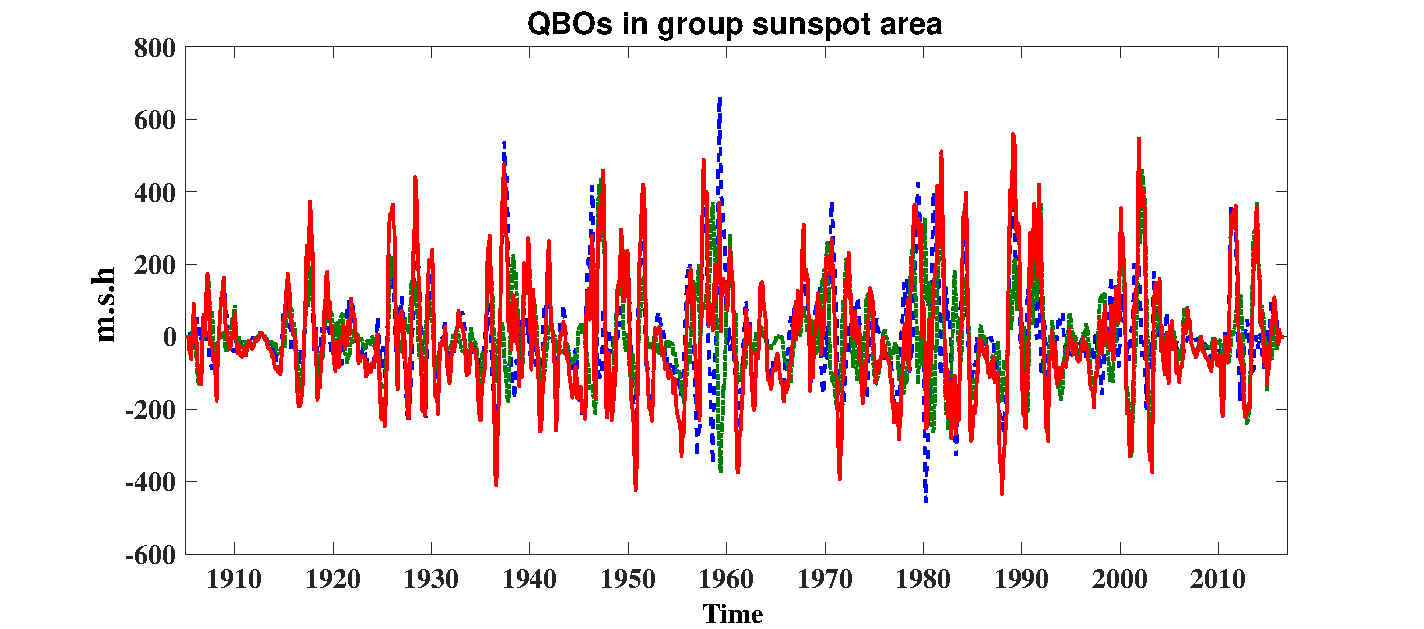} \\
\caption{Left: The nature of the QBOs (1.2 -- 2.5 years), isolated from the KO sunspot number data sets. Right: Same as left side plot but for KO sunspot group area time series. In both cases ``blue'' curve for Northern hemisphere; ``green'' curve for Southern hemisphere and ``red'' indicates full solar disk.}
\label{fig:12}
\end{figure}

Here, we have made an effort to study the nature and variations of the QBO's in the range of about two years in both KO sunspot number and sunspot group area time series. To retrieve the QBOs from all data sets under study, we passed the data through a simple pass-band filter which consists of 10-months and 30-months smoothing with a consequent subtraction of the latter data from the former one following the recipe of \citet{2014SSRv..186..359B}. Figures~\ref{fig:12}(left) and \ref{fig:12}(right) shows the evolution/ nature of the data sets of sunspot number and sunspot group area after using the above-mentioned filters, respectively. Next, we have applied the Morlet wavelet tool on these extracted data sets considering $\omega_{0}$ = 6 under ``red-noise background''  to explore the spatio-temporal evolution of the QBOs (1.2 -- 2.5 yrs.). In all cases, we have drawn the plots of GWPS considering the ``red-noise background''. 

\subsection{QBOs in sunspot number}

\begin{figure}[!h]
\centering
\includegraphics[width=0.42\textwidth]{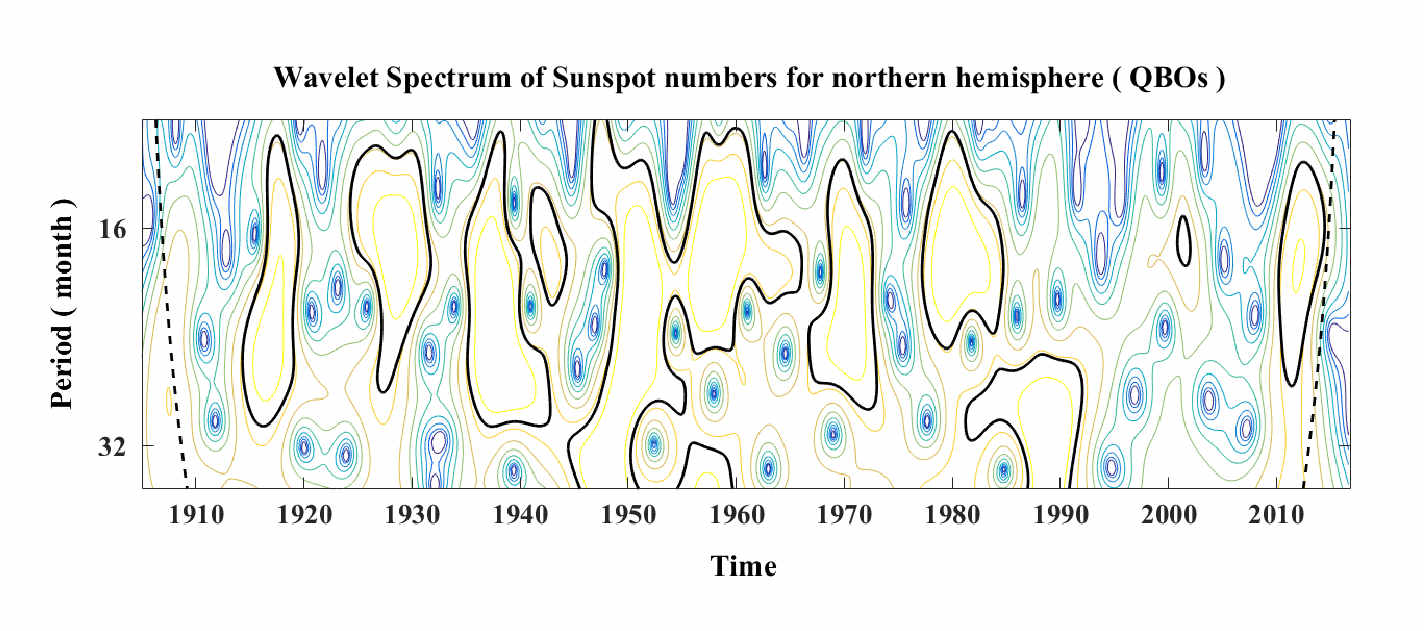}
\includegraphics[width=0.42\textwidth]{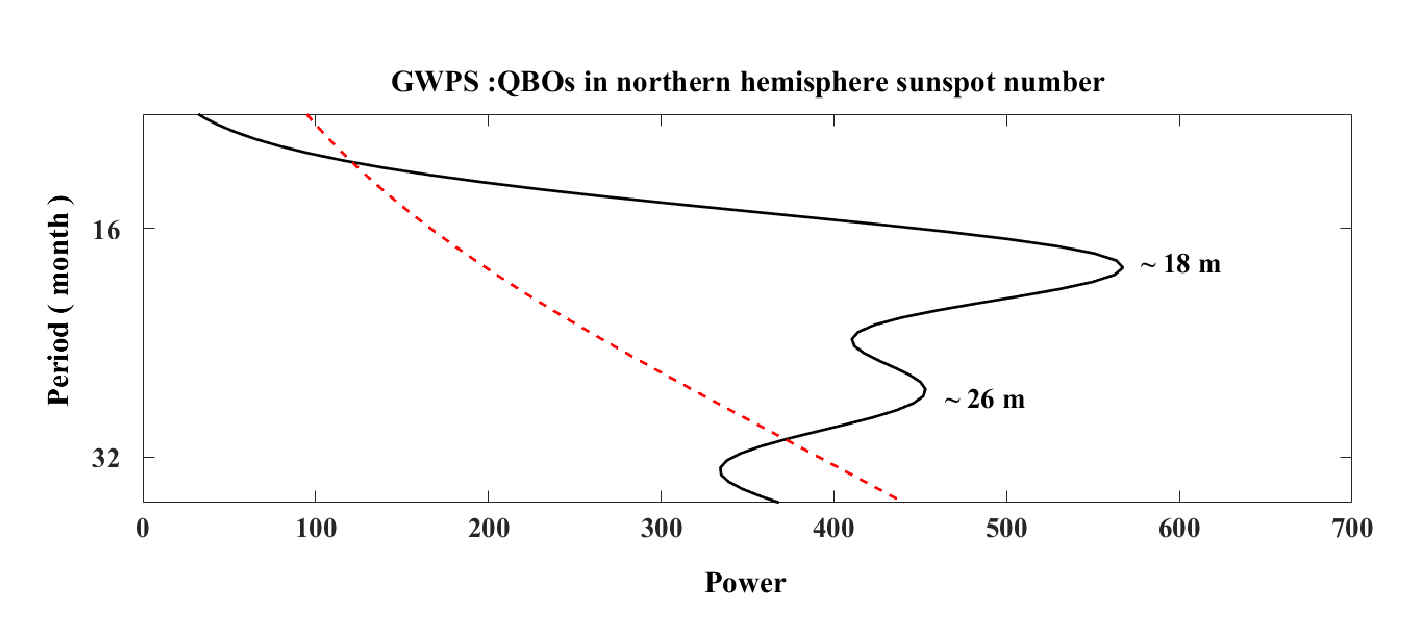} \\
\includegraphics[width=0.42\textwidth]{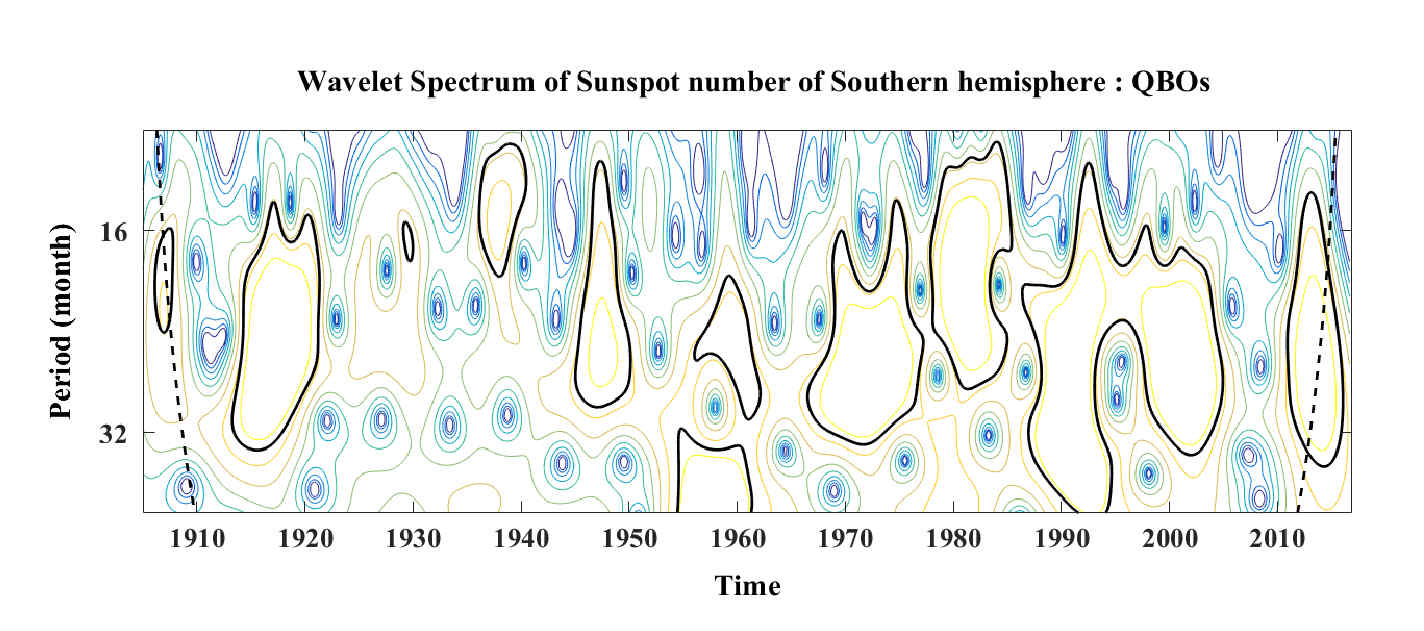}
\includegraphics[width=0.42\textwidth]{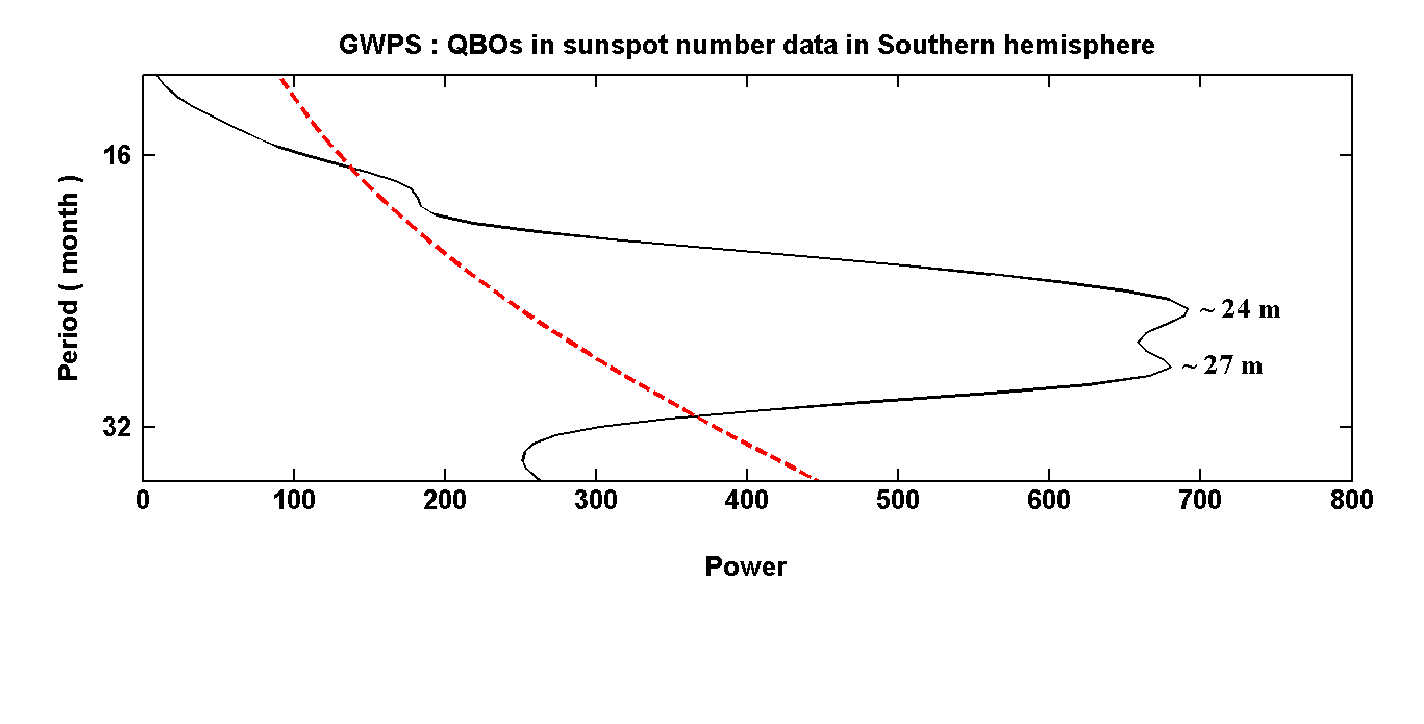} \\
\includegraphics[width=0.42\textwidth]{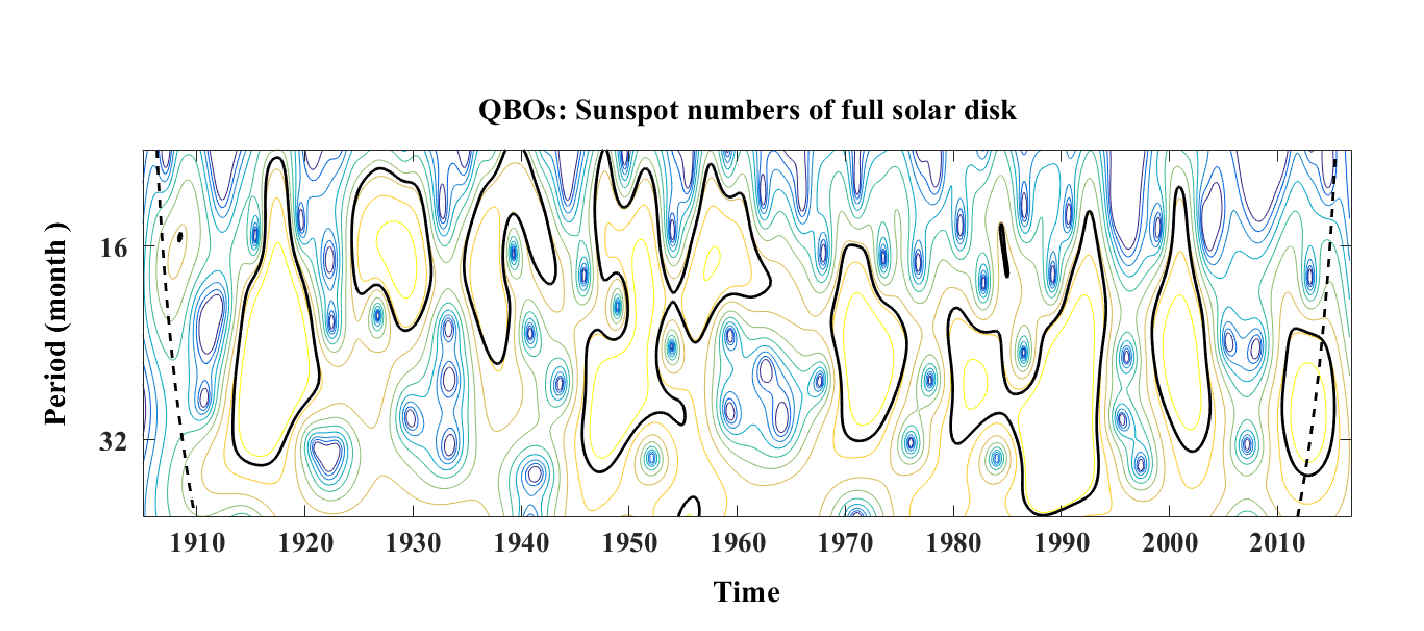}
\includegraphics[width=0.42\textwidth]{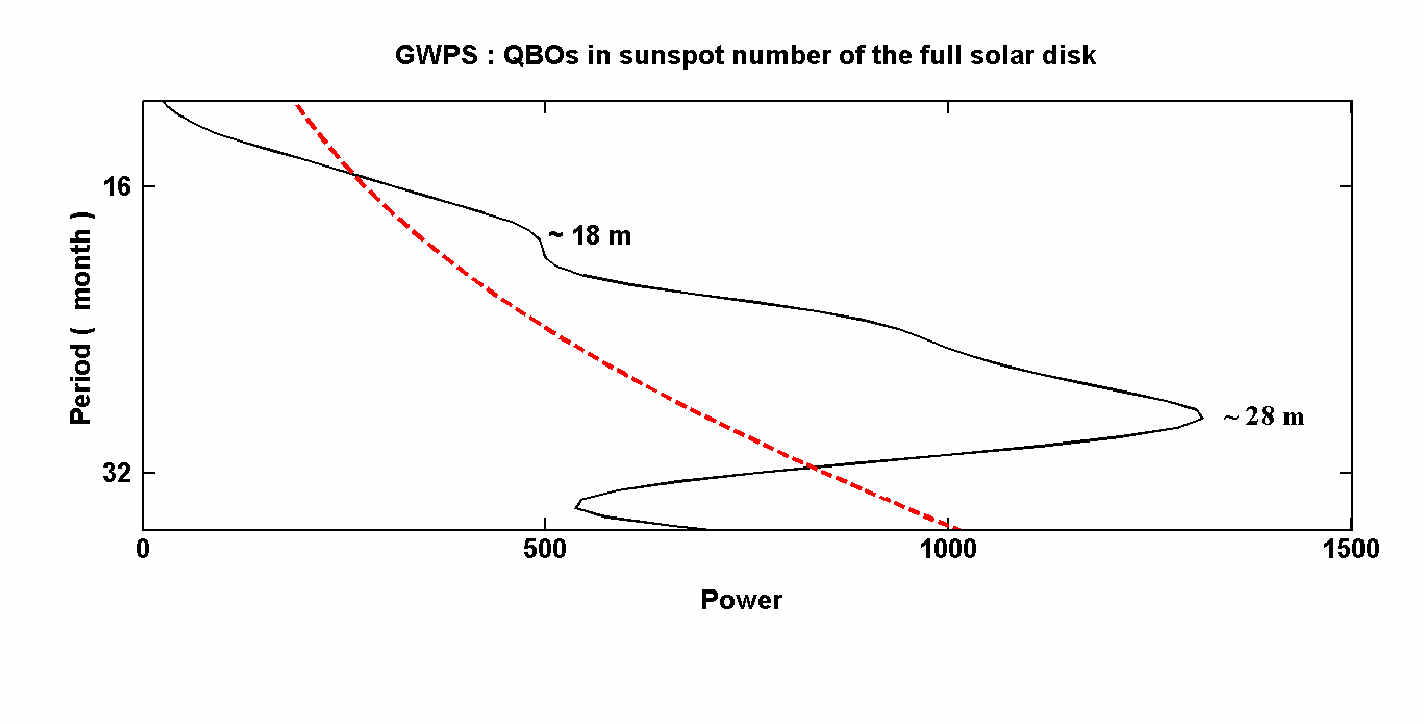} \\
\caption{Top-left: Morlet wavelet spectra of the northern hemispheric sunspot number data as derived from Figure~\ref{fig:12}(top-left) to study the nature and evolution of the QBOs (1.2 -- 2.5 years). Top-right: GWPS of the Figure~\ref{fig:13}(top-left). Middle-left: Same as (Top-left), but for the southern hemisphere data. Middle-right: GWPS of middle-left. Bottom-left: Same as top-left, but for full-disk data. Bottom-right: GWPS of bottom-left plot.}
\label{fig:13}
\end{figure}

Figure~\ref{fig:13}(top-left) represents the evolution of QBOs in sunspot number data in the northern hemisphere. Periods of length from 1.5 years to 2 were present from $\sim$~1915 to $\sim$~1925. On the other hand, a big contour of varying length between 1.3 to 2.5 years was prominent from 1935 to 1980 (during cycle 17--21); and around 2010. In southern hemisphere these periods were significant during 1910--1925; during 1945--1950; and during 1960--2015 (covering cycle 19 to 24) (Figure~\ref{fig:13}(middle-left)). For the sunspot number of the full solar disk, these periods were appeared during 1915--1925; 1930-1940; and 1950--1965 and continually from 1970 -- 2010 (during cycle 20--24) (Figure~\ref{fig:13}(bottom-left)). These results indicate that the temporal evolution of this group of periods was different in the opposite hemispheres. Table~\ref{tab:3} shows the results of the GWPS plots of the wavelet analysis.

\begin{table}[!h]
\centering
\caption{QBOs determined using the GWPS applied to KO sunspot number data for Cycles 14 to 24}
\begin{tabular}{|c|c|}
\hline
Sunspot number Data &  Major Periods in months ($>$ 95\% confidence level)  \\
\hline 
Northern Hemisphere & $\sim$~18, $\sim$~26 \\
Southern Hemisphere & $\sim$~24, $\sim$~27  \\
Whole Solar Disk & $\sim$~18, $\sim$~28 \\
\hline
\end{tabular}
\label{tab:3}
\end{table}

\subsection{QBOs in sunspot group area}

\begin{figure}[!h]
\centering
\includegraphics[width=0.42\textwidth]{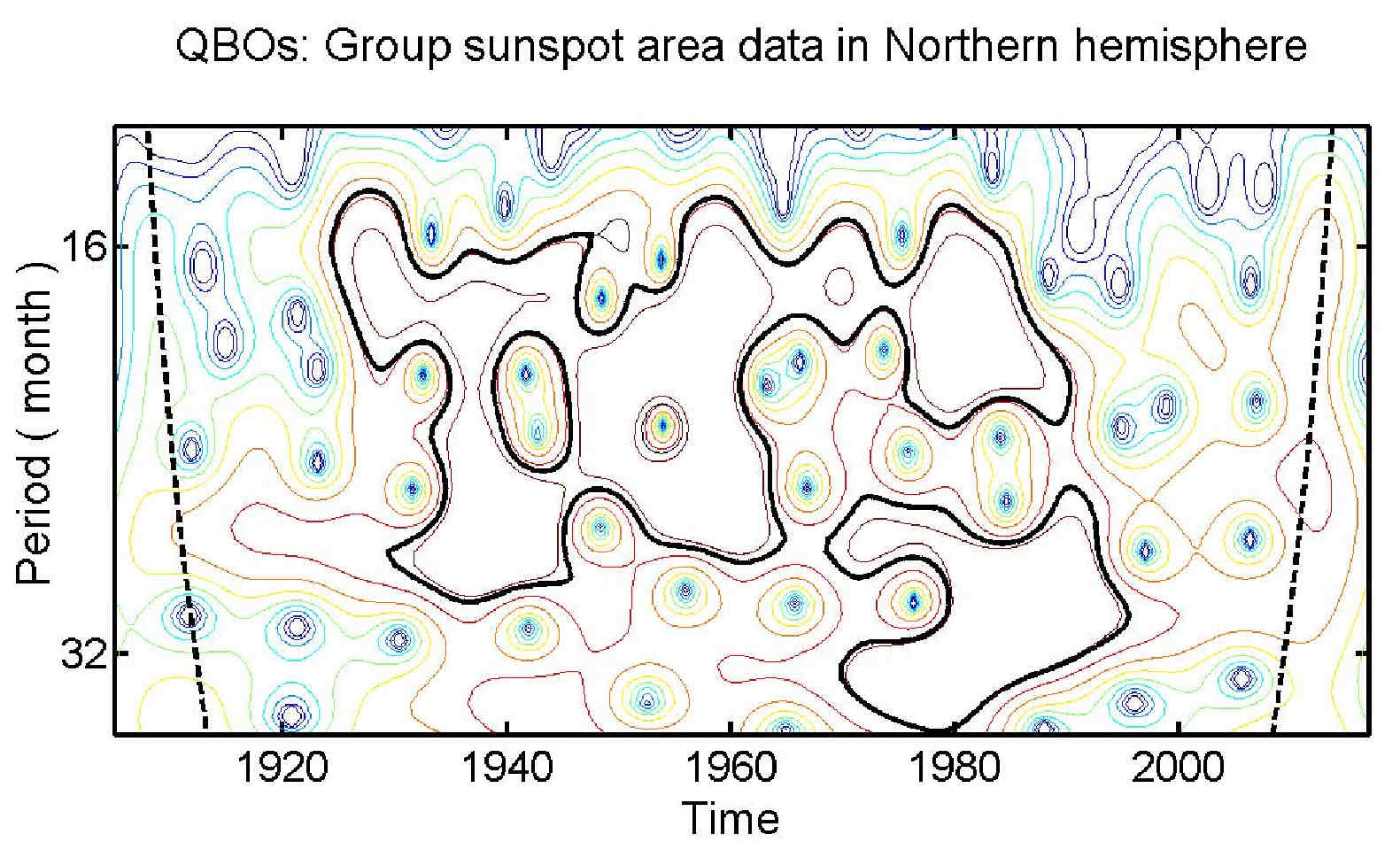}
\includegraphics[width=0.45\textwidth]{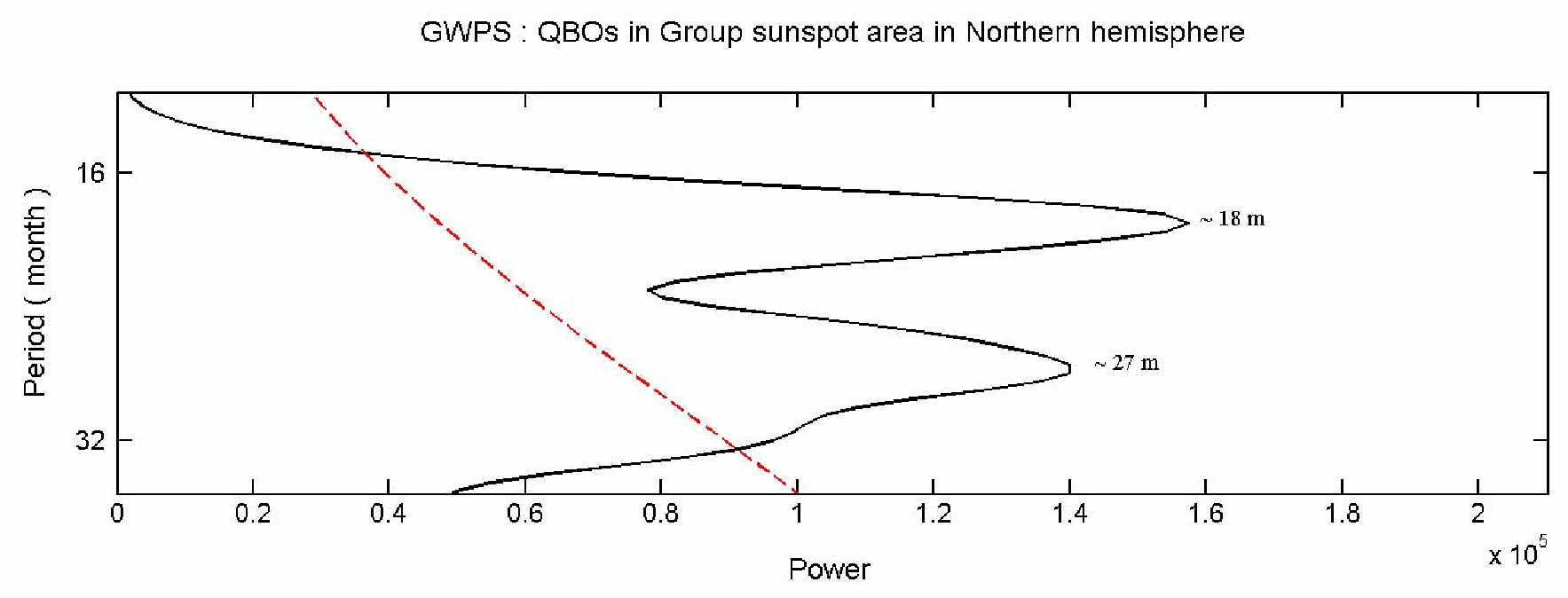} \\
\includegraphics[width=0.42\textwidth]{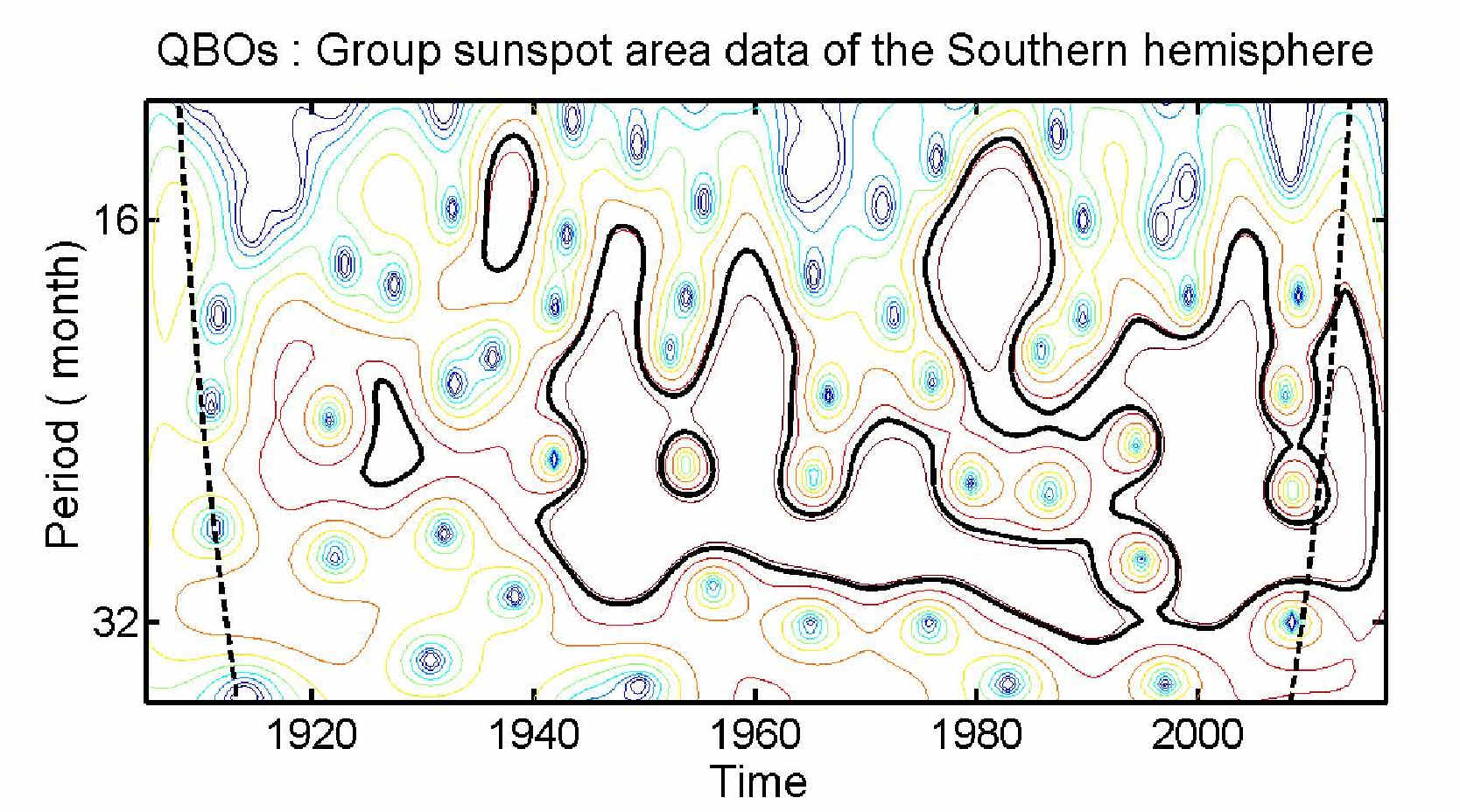}
\includegraphics[width=0.45\textwidth]{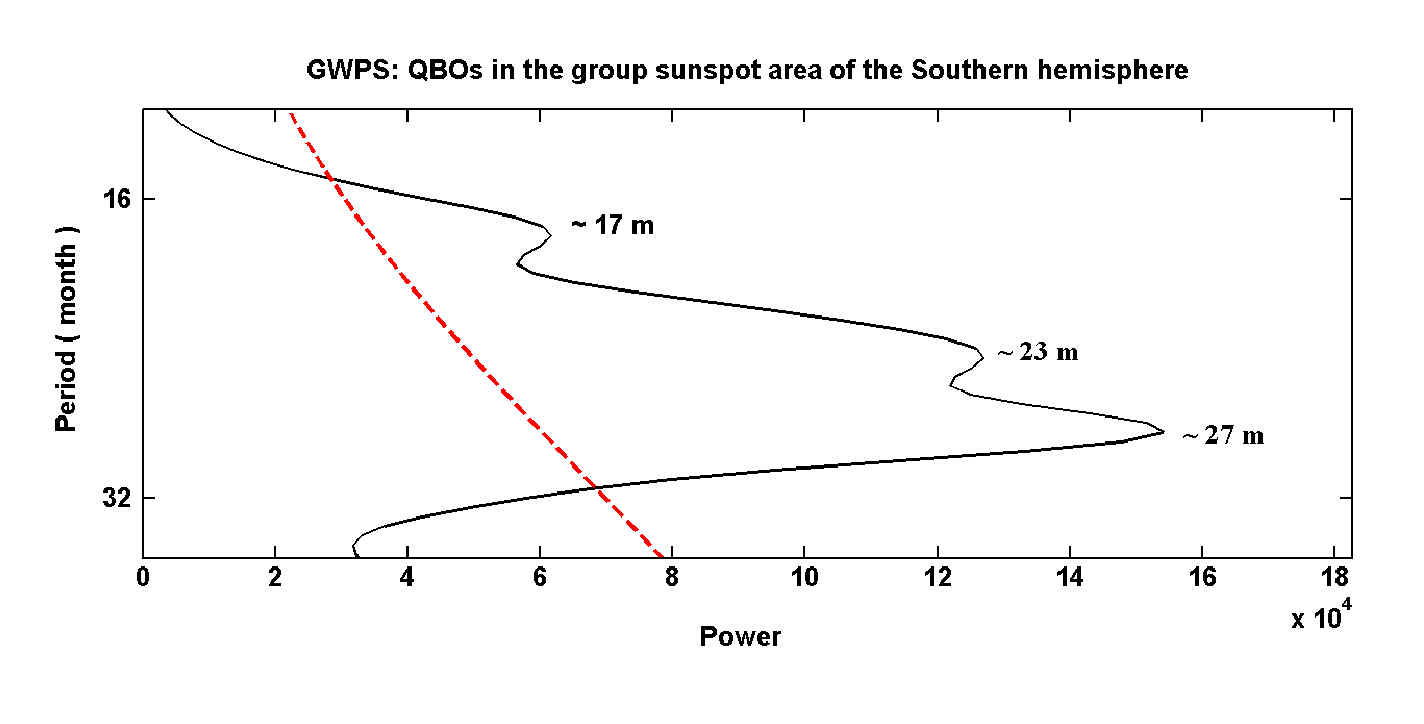} \\
\includegraphics[width=0.42\textwidth]{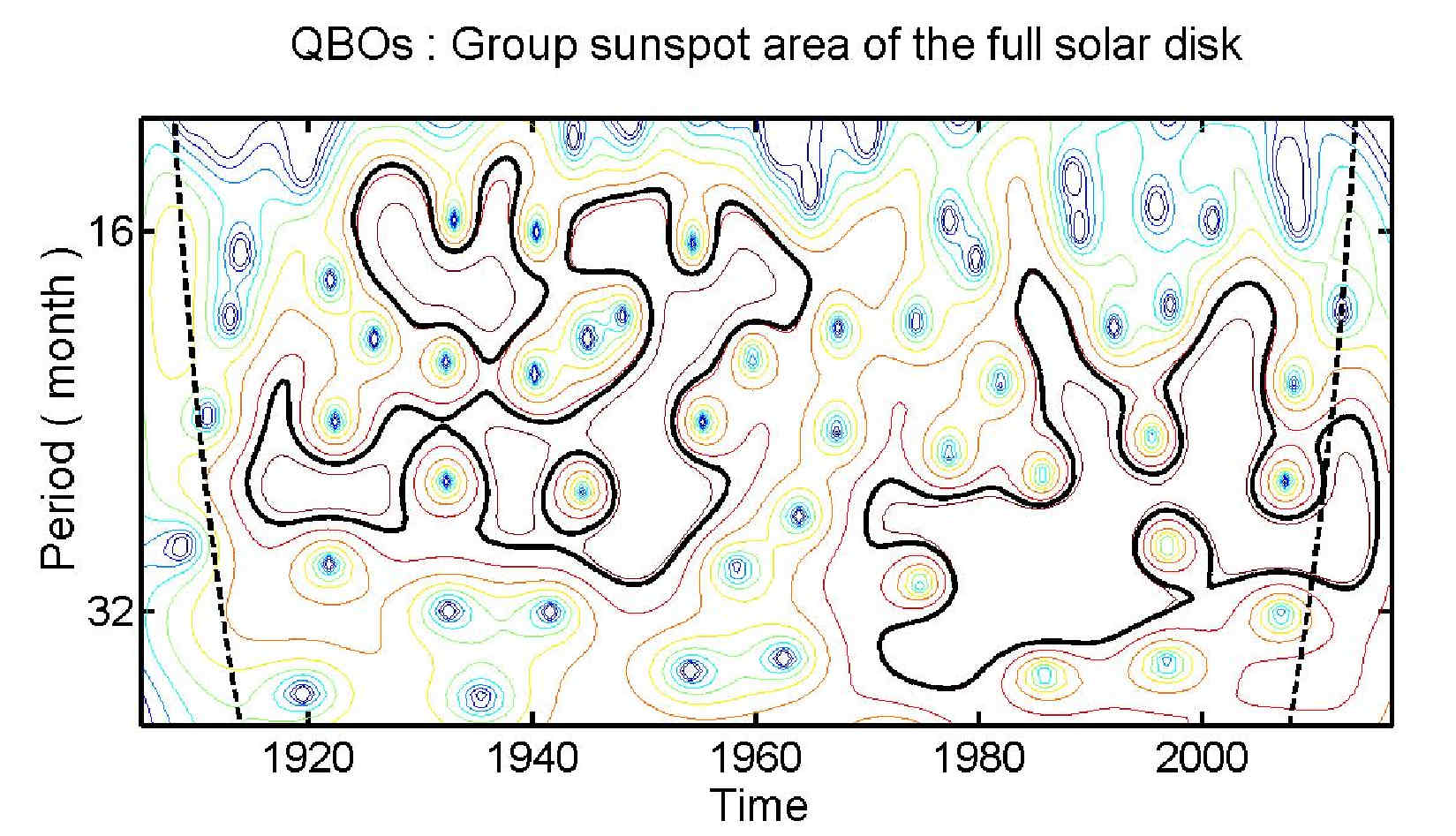}
\includegraphics[width=0.45\textwidth]{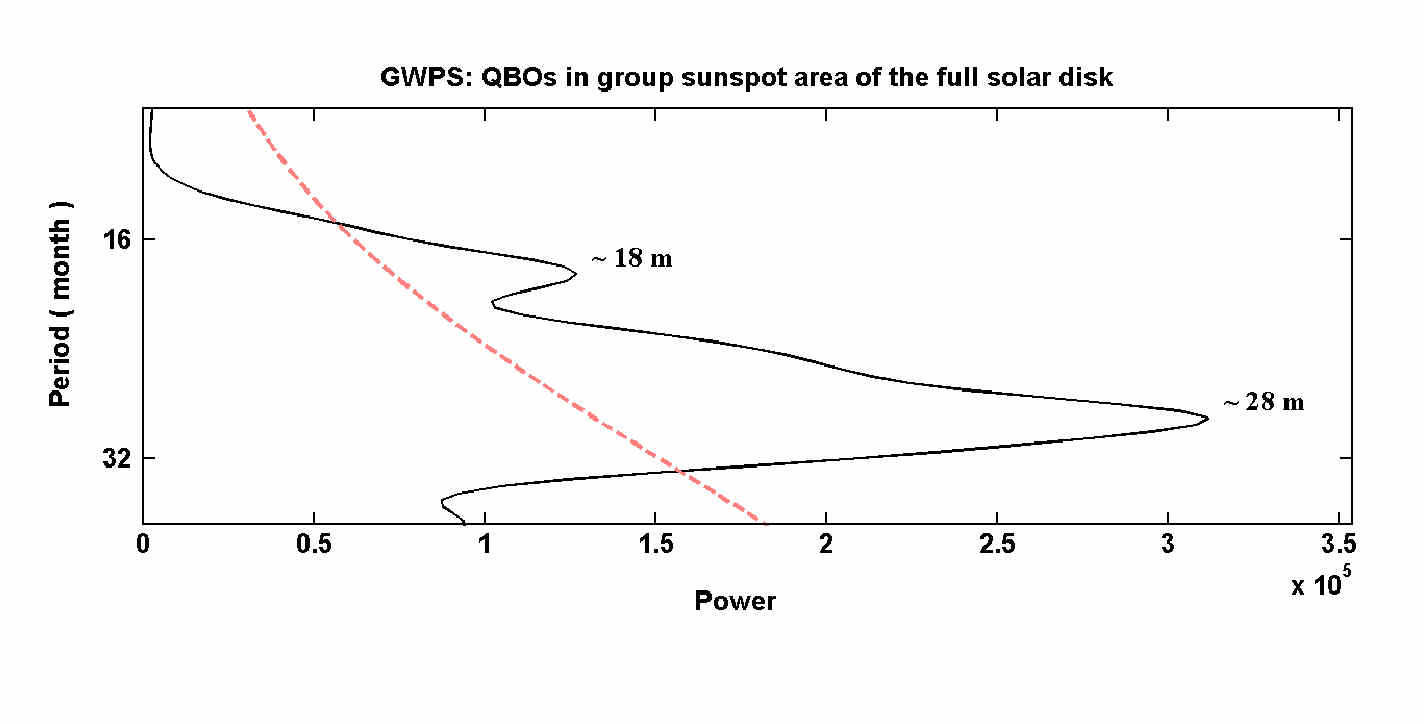} \\
\caption{Same as Figure~\ref{fig:13}, but for the sunspot group area data.}
\label{fig:14}
\end{figure}

Figure~\ref{fig:14}(Top-left) shows the evolution of QBOs in sunspot group area in the northern hemisphere. The long contours of varying lengths between 1.3 -- 2.2 years were significant during cycles 16 to 22 (1930 -- 1990). Another period of length between 2 -- 2.5 years was present from 1970 to 1995 (cycle 20 to 23). On the other hand, in the case of the southern hemisphere, these groups of periods were prominent from 1940 to 1995 (cycle 17 to 23) as well as 1976 -- 2015 (cycle 21 to 24) in a different form (Figure~\ref{fig:14}(Middle-left)). When the full solar disk is concerned (Figure~\ref{fig:14}(Bottom-left)), these groups of periods were present from 1910 to 1960 (cycle 14 to 19) with varying lengths from 1.3 to 2 years. Again appeared from 1975 to 2015 (cycle 21 to 24), and the length changed from 1.4 to 2.5 years. The outcome of the GWPS analysis is displayed in Table~\ref{tab:4}.

\begin{table}[!h]
\centering
\caption{QBOs determined using the GWPS applied to KO sunspot group area data for Cycles 14 to 24.}
\begin{tabular}{|c|c|}
\hline
Sunspot number Data &  Major Periods in months ($>$ 95\% confidence level)  \\
\hline 
Northern Hemisphere & $\sim$~18, $\sim$~27 \\
Southern Hemisphere & $\sim$~17, $\sim$~23, $\sim$~27  \\
Whole Solar Disk & $\sim$~18, $\sim$~28 \\
\hline
\end{tabular}
\label{tab:4}
\end{table}

\begin{figure}[!h]
\centering
\includegraphics[width=0.42\textwidth]{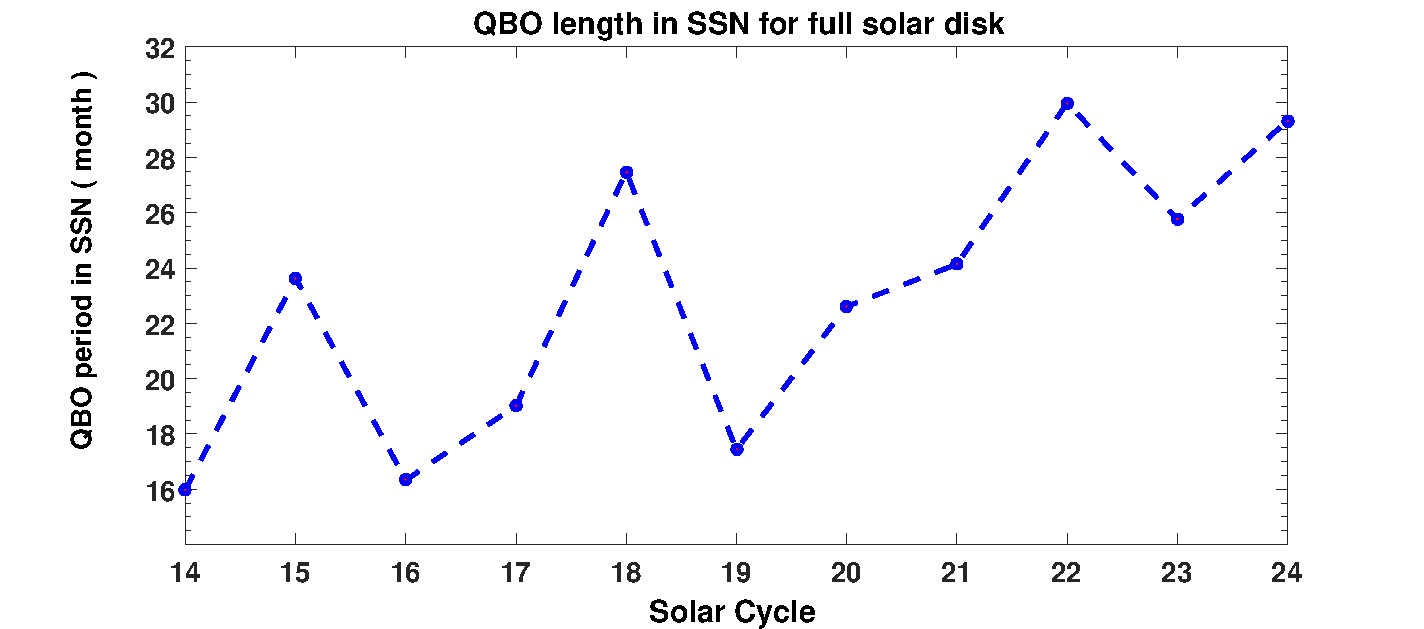}
\includegraphics[width=0.42\textwidth]{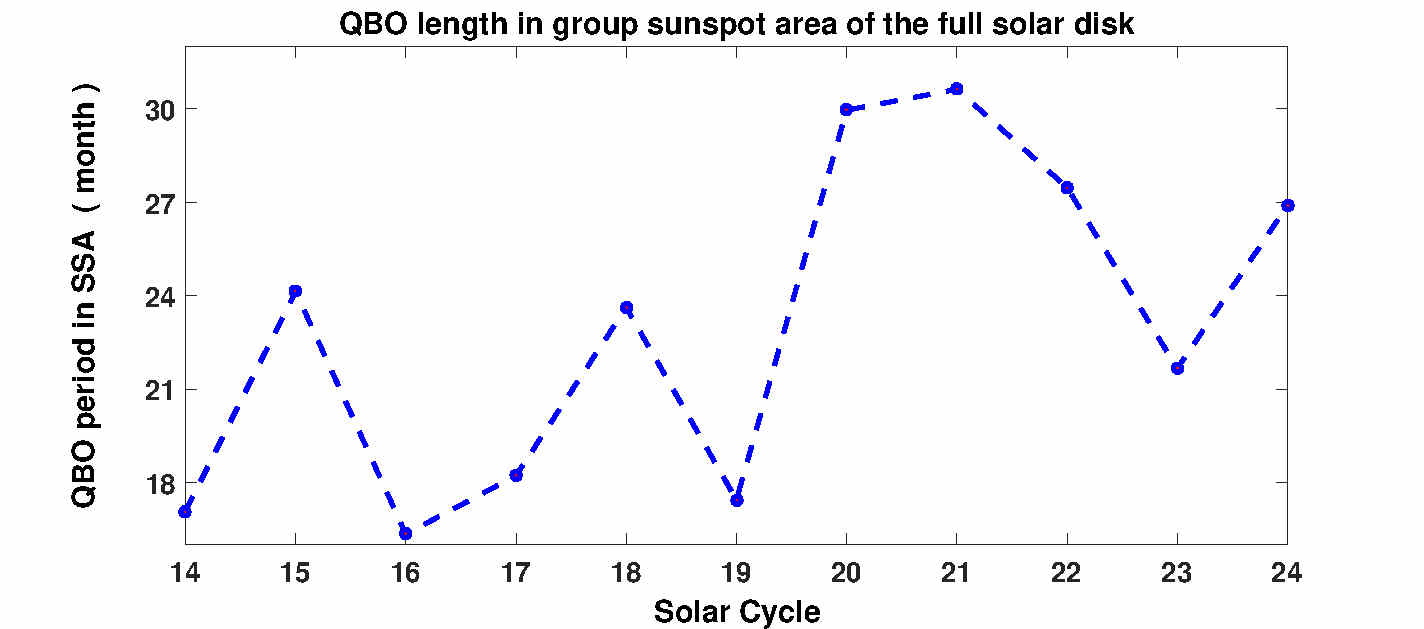} \\
\caption{The display of QBO length and cycle strength.}
\label{fig:15}
\end{figure}

\begin{figure}[!h]
\centering
\includegraphics[width=0.7\textwidth]{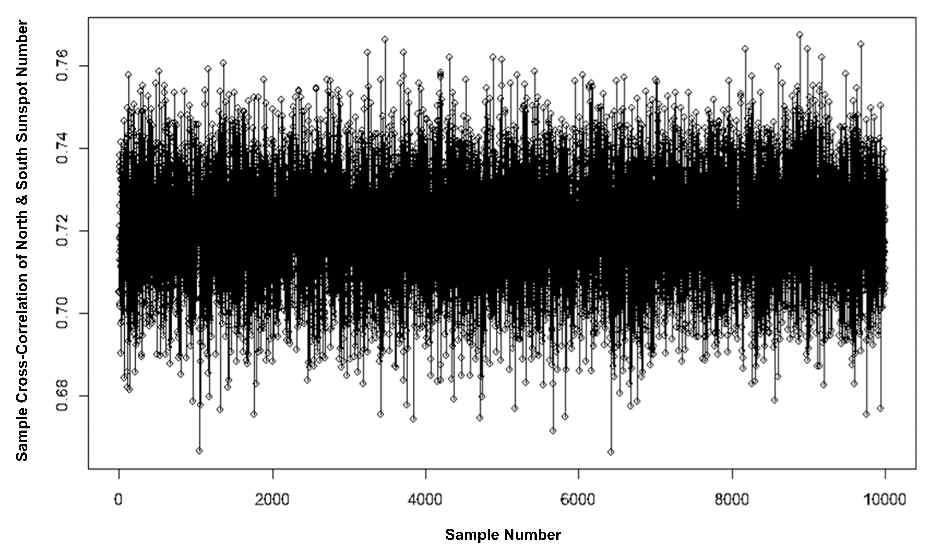} \\
\includegraphics[width=0.7\textwidth]{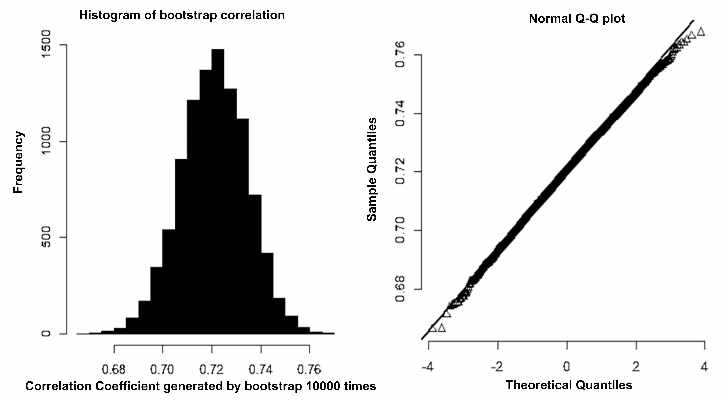} \\
\caption{Top: The cross correlation between the sample North and South hemispheric sunspot number time series generated by bootstrap method. The boot strap for sunspot number. Bottom-left: The histogram of bootstrap generated cross correlations. Bottom-right: The Q-Q plot of generated cross correlations data.}
\label{fig:16}
\end{figure}

\begin{figure}[!h]
\centering
\includegraphics[width=0.8\textwidth]{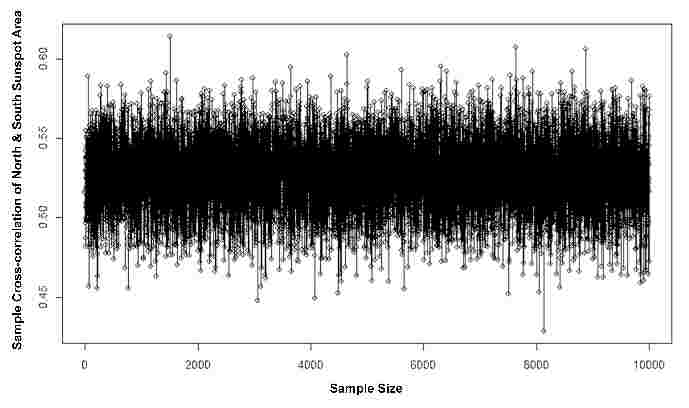} \\
\includegraphics[width=0.8\textwidth]{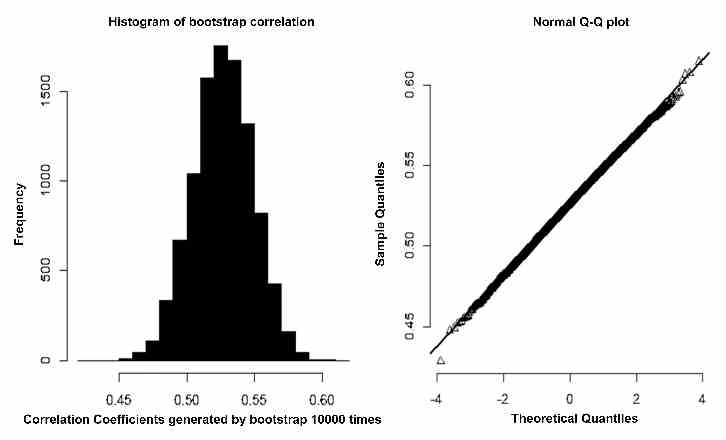} \\
\caption{Same as Figure~\ref{fig:16}, but for group sunspot area data.}
\label{fig:17}
\end{figure}

\citet{2016ApJ...826...55G} have shown that Rieger periodicity appeared in all solar cycles from 14 to 24 in sunspot data sets and is strongly correlated with the solar cycle strength. These authors indicated the presence of a periodicity from 185 to 195 days during the relatively weak solar cycles 14--15 and 24 and a periodicity in the range of 155--165 days during the stronger solar cycles 16--23 in the daily sunspot area / number time series measured at the Greenwich Royal Observatory and the Royal Observatory of Belgium(ROB). In this article, we have made an attempt to study the correlation between QBO length and solar cycle strength considering sunspot number and group sunspot area of the full solar disk measured at KO. Although, the length of QBOs vary from $\sim$~1 year to $\sim$~4 years, but, we have restricted our analysis for those QBOs whose length varies between 1.2 years to 2.5 years. To evaluate the length of the QBO periods during each cycle, we utilized again wavelet analysis in every solar cycle separately of both the sunspot number and group sunspot area full disk data-sets. The QBO periods which resulted from the GWPS covering an interval of 1.2--2.5 years and above 95 \% confidence level, are shown in Table~\ref{tab:5} and displayed in Figure~\ref{fig:15}.

\begin{table}[!h]
\centering
\caption{The QBO's seen in sunspot number and area of each cycle is listed here.}
\begin{tabular}{|c|c|c|c|c|}
\hline
Cycle No. &  T-SSN QBO (months) & Average SSN & T-SSA QBO (months) & Average group SSA (m.s.h)  \\
\hline 
cycle 14(decay phase) &  15.9 & 58.9 & 17.1 & 435.04 \\
Cycle 15 & 23.6 & 73.3 & 24.13 & 497.27 \\
Cycle 16 & 16.7 & 68.24 & 16.32 & 689.10 \\
Cycle 17 & 19.01 & 96.07 & 18.21 & 880.54 \\ 
Cycle 18 & 27.47 & 108.89 & 23.62 & 914.23 \\
Cycle 19 & 17.43 & 129.10 & 17.78 & 1398.36 \\
Cycle 20 & 22.61 & 86.88 & 29.96 & 968.30 \\
Cycle 21 & 24.134 & 111.40 & 30.63 & 1245.77 \\
Cycle 22 & 29.96 & 106.73 & 27.47 & 1147.8 \\
Cycle 23 & 25.758 & 82.23 & 21.66 & 955.40 \\
Cycle 24 & 29.32 & 64.35 & 26.89 & 694.48 \\
\hline

\end{tabular}
\label{tab:5}
\end{table}

Table~\ref{tab:5} and Figure~\ref{fig:15}, shows that the QBO type periodicities in the range of 1.2--2.5 years are not directly correlated with the long-term variations in the solar cycle amplitude. However, we have noted that in some cycles both the sunspot number and group sunspot area data display nearly similar QBO type of periods. Cycle 19 had the highest amplitude and the QBO length is $\sim$~1.4 years in both data sets. In some relatively strong solar cycles like cycle 18, 21, and 22, the length of the QBOs vary between 2 and 2.5 years. On the other hand, some weak cycles (15, 23, and 24) also possessed similar type of QBOs ($\sim$~2 to 2.5 years). The significant QBO for cycle 16 is $\sim$~1.3 years and this cycle is a weaker one which contradicts with the results of the strongest cycle 19. Hence, with the KO sunspot data sets we did not detect any direct correlation between cycle strength and QBO's length.

\section{Model fitting and Statistical Significance tests}
\label{sec:mod}
 
In the previous sections, we have determined different properties as well as periodic and quasi-periodic variations present in the KO sunspot number / group sunspot area data sets. To find the reliability of the significance levels of these periodicities, we have performed a model fitting to the data sets in Autoregressive model (AR model) \citep{Box1970, Chatfield2001}. For this purpose we have applied ``Auto-regressive Integrated Moving Average Model (ARIMA)'' \citep{Montgomery2015}, which was previously used to build different models for time series analysis and cycle strength forecasting \citep{Abdel-Rahman2018}. Goodness-of-fit of the model has been checked with the Kolmogorov-Smirnov (K--S) and Anderson-Darling (AD) test.  AD test \citep{Stephens1974} is a robust statistical analysis which is used to test if a sample of data (here sunspot number and area) occurred from a population following any specific distribution. It is a modified version of the K--S test and provides more weight to the tails of the distribution than the K-S test. We have utilized generalized ARIMA (p, d, and q) model which is a mixed integrated model where p is the number of lag observations in the model (lag order); q indicates the size of the moving average window (order of the moving average) and d represents the number of times that the raw observations are differenced to make the time series stationary (degree of differences).

In the ARIMA model, d=1 means taking the differences in the series only once (differences is first order) and the difference continue up to the series stationary. i. e., d=1, 2, 3, …etc. The main aim of this model is to investigate carefully and rigorously the past observations of any time series to develop an appropriate model which can forecast future values for the series. We have determined the Akaike information criterion (AIC) which is an estimator of prediction error; the Akaike's Information Corrected Criterion (AICc) and the Bayesian information criterion (BIC) for each of the data sets under study. The AIC and the BIC measure the performance of the fitted model i.e., goodness of fit and its complexity \citep{Kass1995}. Detailed description about the above mentioned statistical tests/ models are available in \citet{Hyndman2021}. Below, we have provided the results of different statistical tests.

Before applying the ARIMA model, we have used BOX-COX transformation of  all the  data sets (time series) under study and  have padded all missing data by one as it needs all data should be positive. 

\subsection{ARIMA model fitting and residue analysis in sunspot number}

After performing the Box-Cox transformation (with optimal lambda = 0.384) for the northern and southern hemispheric sunspot number data, we have modelled northern hemisphere using ARIMA(3, 0, 2); southern hemisphere using ARIMA (1, 0, 2) and obtained the minimum values of AIC, AICc, and BIC. After this, we have performed the Anderson-Darling normality test on the residue of ARIMA model as a input data and obtained the values of A (test Statistic) and p-value. The two-sample Kolmogorov-Smirnov test is also performed on the data of Residue of ARIMA model and Random Normal data (two sample Test) and obtained values D (test statistic) p-values from the test. All these values are listed in Table~\ref{tab:6} for the northern and southern hemispheric sunspot number data sets. The obtained values show that the null hypothesis is true, that means standardized residues may be considered which is nothing but white noise is drawn from N(0,1).

\begin{table}[!h]
\centering
\caption{The values of AIC, AICc, BIC obtained after using the ARIMA(3, 0, 2) (ARIMA(1, 0, 2)) northern (southern) model, Anderson-Darling normality (AD) test and Kolmogorov-Smirnov test for the northern and southern hemispheric sunspot number data are listed.}
\begin{tabular}{|c|c|c|c|c|c|c|c|}
\hline

\multicolumn{1}{|c|} - & \multicolumn{3}{|c|} {ARIMA(p, d, q)} & \multicolumn{2}{|c|} {AD} & \multicolumn{2}{|c|} {KS} \\
\hline
SSN & AIC & AICc & BIC & A & p & D & p \\
\hline
NH & 5333.38 & 5333.44 & 5364.59 & 0.364 & 0.439 & 0.018 & 0.982 \\
SH & 5447.83 & 5447.87 & 5473.83 & 0.573 & 0.136 & 0.022 & 0.891 \\

\hline

\end{tabular}
\label{tab:6}
\end{table}

\subsection{ARIMA model fitting and residue analysis in group sunspot  area data}

A similar analysis is done on the sunspot group area data as well. After performing the Box-Cox transformation (with optimal $\lambda$ = 0.210 for the northern and 0.303 for the southern) of northern (southern) sunspots group area have been modeled by ARIMA(1,1,3) (ARIMA(2,1,2)) for the northern (southern) hemispheric data. The values obtained for this model, the AD model and Kolmogorov-Smirnov test data are shown in Table~\ref{tab:7}. 

\begin{table}[!h]
\centering
\caption{The values of AIC, AICc, BIC obtained after using the ARIMA(1, 1, 3) (ARIMA(2, 1, 2)) northern (southern) model, Anderson-Darling normality (AD) test and Kolmogorov-Smirnov test for the northern and southern hemispheric group sunspot area data are listed.}
\begin{tabular}{|c|c|c|c|c|c|c|c|}
\hline

 \multicolumn{1}{|c|} - & \multicolumn{3}{|c|} {ARIMA(p, d, q)} & \multicolumn{2}{|c|} {AD} & \multicolumn{2}{|c|} {KS} \\
 \hline
SSN & AIC & AICc & BIC & A & p & D & p \\
\hline
NH & 6407.5 & 6407.55 & 6433.51 & 1.659 & 0.00029 & 0.0334 & 0.442 \\
SH & 7715.67 & 7715.71 & 7741.68 & 0.955 & 0.015 & 0.0307 &0.551  \\

\hline

\end{tabular}
\label{tab:7}
\end{table}

AD test indicates that null hypothesis is rejected which means that standardized residues cannot be thought as a white noise drawn from N(0,1). However, K -- S test states that null hypothesis may be considered true and this means standardized residues may be considered nothing but white noise drawn from N(0,1).

\section{Wave strapping analysis of sunspot activities}
From Figure~\ref{fig:2} and \ref{fig:3}, we have found several spatio-temporal properties of sunspot number and group sunspot area of both the hemispheres as well as the full solar disk. Our wavelet analysis also detected a number of mid-term quasi-periods including well known  $\sim$~11 and $\sim$~22 years. To assess the significance of the sample cross-correlation between two time series i.e., northern and southern hemispheric sunspot number / group sunspot area we have used Wave strapping method \citep{Percival2001}. Wave strapping is an important sampling method in the wavelet domain which joins two major concepts in statistical signal processing: wavelet analysis and bootstrapping. This statistical tool is mainly appropriate for those time series which exhibit long term memory properties, i.e., that exhibit correlation over long periods. Classical Fourier analysis method distinguishes the frequency components of a signal/ time series, whereas wavelet analysis considers the signals whose frequency components change over time. Wavelets are basically some `little waves' that are finitely extended and this technique decomposes a signal/ time series in an ortho normal basis of these wavelets \citep[][etc.]{Ogden1997, 1998BAMS..79..61T}. On the other hand, bootstrapping is a re-sampling technique with replacement from the observed data, and makes use of the re-sampled sequences to evaluate the properties of a given estimator through its empirical distribution, without any pre assumption of any kind of theoretical distribution (e.g., Gaussian/ exponential distribution etc.) \citep{Robert2004}. In this technique, the simulation is done by the random selections of the events from the original experimental data set with returning them back to the initial experimental set. Here, the selection of any particular event from the original data set can be done many times or even not for a single time. This method is widely applied for the calculation of the statistical errors of any data set when its direct evaluation is difficult or not realizable at all. The bootstrap methodology provides different types of confidence intervals like Normal Representation; Basic; Percentile, Bca (bias-corrected, accelerated). \citet{Barbe1995}  have provided a detailed description of the bootstrap method.
Here, we have applied bootstrapping method between North and South hemispheric sunspot number data.
Cross-correlations between North and South data have been calculated after taking 10,000 samples generated from North and South hemispheric sunspot number time series by bootstrap method. We have made plots of the Histogram of bootstrap generated sample cross-correlation and the Q-Q plot of generated cross-correlations data (Figure~\ref{fig:16}).

\begin{figure}[!h]
\centering
\includegraphics[width=0.8\textwidth]{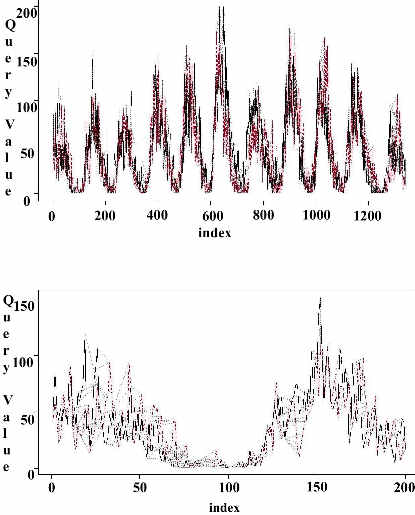}
\caption{Upper panel: DTW of Northern hemispheric Sunspot Number (Quarry data, black colour) and Southern hemispheric Sunspot Number (Reference data, red colour) are mapped for finding the minimum distance between these two sets of data sequences. This mapping is shown more clearly for first 200 sets of data (depicted in the lower panel). Lower panel: DTW of Northern hemispheric Sunspot Number (Quarry data, black colour) and Southern Sunspot Number (Reference data, red colour) are mapped to find finding minimum distance between these two sets (for first 200 data sequences).}
\label{fig:18}
\end{figure}

\subsection{Bootstrap Statistics}

After taking the 10000 samples of sunspot number data generated by bootstrap method, we have computed the cross-correlation values and it is found to be 0.72 with a bias of -0.0002 and the standard error of 0.013. This analysis provided the confidence intervals for the normal representation at 95\% levels as ( 0.6949,  0.7477 )  and ( 0.6956,  0.7481 ) in the basic level and in the BCa representation ( 0.6941,  0.7465 ) and ( 0.6937,  0.7462) at level percentile.

This analysis of cross-correlation of North and South data sets indicates that these two data sets are dependent to some extent on each other. For this reason some small/ mid-term periods may appear commonly to both set of data.
Wave strapping is also done by using bootstrapping method between North and South hemispheric group sunspot area data.
Cross-correlations between North and South group area data have been calculated taking 10,000 samples generated from North and South area data sets as previous and the relevant plots are shown in Figure~\ref{fig:17}.

We have performed a similar statistics on the hemispheric sunspot group area data. The analysis shows that the original cross-correlation value is 0.526 with a bias of 9.181$\times$10$^{-5}$ and standard error is 0.022.   The intervals of estimation of cross-correlation analysis done by different methods gave a value at the 95\% confidence level as (0.483, 0.569) and (0.484, 0.571). The BCa level values at the same 95\% confidence level are (0.482, 0.568) and (0.481, 0.567). The results of this analysis indicates that northern and southern sunspot group area data sets are dependent on each other to some extent.

\begin{figure}[!h]
\centering
\includegraphics[width=0.8\textwidth]{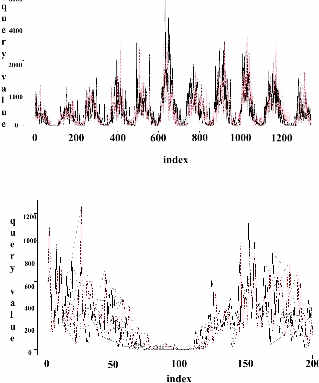} \\
\caption{Upper panel: DTW of Northern hemispheric Sunspot Area (Quarry data, black colour) and Southern hemispheric Sunspot Area (Reference data, red colour) are mapped for finding the minimum distance between these two sets of data sequences. Note: This mapping is shown more clearly for first 200 sets of data (depicted in the lower panel). Lower panel: DTW of Northern hemispheric Sunspot Area (Quarry data, black colour) and Southern hemispheric Sunspot Area (Reference data, red colour) are mapped for finding minimum distance between these two sets (for first 200 data sequences).}
\label{fig:19}
\end{figure}

\subsection{Dynamic Time Warping analysis of sunspot activities}

Dynamic Time Warping (DTW) is a powerful statistical method to compare the similarities between two varying time series which have nearly similar patterns but differ in time i.e., do not synchronize  up perfectly \citep[][etc.]{Keogh2001, Muller2007, Zhang2021}. This technique is utilized to determine the optimal matching between two sequences. Although, this algorithm was initially developed for speech recognition purposes \citep{Myers1981}, but, now a days it is being widely used in many other scientific domains like meteorology, data mining, financial markets etc. DTW algorithm compares the amplitude of first time series at time `t' with amplitude of second time series at time t+1 and t-1 or t+2 and t-2 which makes sure it does not provide low similarity score for two time varying signals of  similar shape with different phases \citep{JEONG20112231}. This technique follows the rule of monotonicity, continuity, with the fact that the first and last points of one signal should be matched with at least the first and last points of the other signal \citep[see:][for details]{Gulzar2018}. It measures a cumulative cost, the DTW score, which exhibits the cost of aligning of two time series in time when their patterns are similar but differ in time \citep{Gorecki2013}. DTW technique computes the distance from the matching of similar elements between two time series with same or different length and this distance is called ``Manhattan distance''. The Manhattan distance between two points x = (x$_{1}$, x$_{2}$, …, x$_{n}$) and y = (y$_{1}$, y$_{2}$, …, y$_{n}$) in n-dimensional space is defined as the sum of the distances in each dimension.

\begin{equation}
    d(x,y) = \sum_{i=1}^{n} \left|x_{i} - y_{i} \right|.
\end{equation}

If two sequences of data sets are same then minimum distance between them is zero. But when the sequences of data are different then minimum distance between them is non-zero and this Manhattan distance increases with the dis-similarity between the data sets \citep{Ratanamahatana2004}.

\citet{Laperre2020} used this method to evaluate the geomagnetic storm index Dst forecast with machine learning technique. Recently, \citet{2022ApJ...927..187S}  have adopted DTW technique to quantify the similarities and dissimilarities between observed and modeled solar wind time series forecasting. Here, we are using this technique to determine the   Manhattan minimum distance for the two time series of sunspot indices and the Figures~\ref{fig:18} to \ref{fig:19} display the relevant DTW plots.

\begin{enumerate}

\item For the North and South Sunspot numbers:  Manhattan minimum distance found = 17029.34 which is quite large and indicates that these two time series are different from one another though some dependency appears.

\item For North and South group Sunspot area: Manhattan minimum distance = 264008.5 which is also quite large and indicates that these two time series are different from one another though some dependency seen in the Wavestrapping analysis.

\end{enumerate}

\section{CONCLUSIONS AND DISCUSSIONS}
\label{sec:dis}

Sunspots are one of the key components of the solar surface magnetic field and its variations in the course of time \citep{2014SSRv..186..251N, 2015LRSP...12....4H}. This index's spatio-temporal evolution helps us to understand the physics behind the solar magnetic field, which is the main driving force of all types of activities inside the Sun.  In this work, we have considered the time series of sunspot numbers and sunspot group area measured from the sun charts at the Kodaikanal Observatory to investigate the presence and Spatio-temporal behavior of the mid-term quasi periodicities as well as QBOs during solar 
cycles 14-24. The search of mid-term periods and their time evolution in the Sun and defining their properties can provide useful information on the global magnetic properties of the Sun \citep{2009A&A...502..981V} like the working behavior of the complex dynamo in different solar cycles. Along with it, the non-constant periodic lengths less than Schwabe cycle ($\sim$~11 years) could enrich the ``solar melody'' \citep{1994seit.conf..221B} with additional components, affecting the space weather /climate and our human activities \citep{2021LRSP...18....4T}. This is the first investigation to detect the periodic and quasi-periodic variations present in this handwritten long-term sunspot data set measured at the KO covering about eleven sunspot cycles 14 to 24. We have applied Wave strapping and Dynamic Time Warping (DTW) methods to determine the statistical relationship between the northern and southern hemispheric sunspot activity data sets. We are providing a summary of the main findings of our study with discussions:

\begin{itemize}
\item It is revealed that both the amplitude and temporal variation of the KO sunspot number and sunspot group area are not the same in the northern and southern hemispheres. The behavior of the ascendant phase, peak time, and decay rate is also different in both hemispheres in several solar cycles. We observed the maximum amplitude of sunspot numbers during solar cycle 19 in both the hemispheres and the whole solar disk. Sunspot number amplitude is minimum during cycle 16 for the northern hemisphere, which is comparable with the strength of cycle 24 in the same hemisphere. But, in the case of the southern hemisphere, cycle 20 is the weakest one. Solar cycle 24 showed the timidest attitude for the whole solar disk data. The dynamical behavior of KO sunspot number has strong similarities with the recently prepared hemispheric sunspot number from different solar proxies \citep{2021A&A...652A..56V} and so may be considered as an alternate proxy of the International Sunspot number for long term hemispheric solar cycle studies.  

\item For the sunspot group area index, both the hemisphere and full-disk data exhibit maximum amplitude during cycle 19. The cycle 15 showed the weakest amplitude in both the hemispheres. For the whole sphere, the amplitude of both cycles 15 and 24 showed a nearly equal weak nature.

\item Double peaks are found around the maximum episode of many solar cycles for both the sunspot number and sunspot group area. However, the nature of the double peak is different in different solar cycles and the opposite hemispheres. This nature is consistent with the temporal evolution of the recently digitized sunspot area as well as the Ca-K plage index data set measured at the KO \citep{2021SoPh..296....2R, 2022ApJ...925...81C}.

\item Sunspot number data measured at Royal Greenwich Observatory (RGO) as well as in the KO indicates that the amplitude of cycles 16, 18, and 20 is smaller than the immediate odd numbered cycle and hence follows the odd-even rule \citep{2021SoPh..296....2R, 2022ApJ...925...81C}. This rule has been violated for the cycle 22-23 pair. Similar behavior is also found for the sunspot group area time series. The full-disk data set of sunspot number and area indices exhibit a weakening nature from solar cycle 21 to 24.

\item The Morlet wavelet analysis method detected several intermediate-term periodicities, including the best known Rieger and Rieger group and QBOs, separately in the northern and southern hemispheres and full disk for both the sunspot number and group sunspot area data set. However, we noticed that these periods' temporal evolution and amplitude are different in the opposite hemispheres in both data sets. We have made a detailed investigation about the spatio-temporal evolution of QBOs in the range of 1.2 -- 2.5 years and observed that these groups of QBOs have some north-south asymmetry in existence in both sunspot numbers and sunspot group area time series. Distinctive peaks around 1.5 years and around 2.2 years were found in all data sets under study. Thus, QBOs $\sim$~2 years were significant in both the sunspot indices measured at KO. But, we have seen that the modulation and appearance of these periods are different in the opposite hemispheres. We have detected the signature of about 5-year quasi-periodicity, which was much prominent in the sunspot number data of the southern hemisphere and full solar disk. This period is prominent and persisted for a long time only in the full solar disk sunspot group area data. We found that all these data sets under investigation have prominent 10 -- 11-year cyclic variations. Along with it, we have also detected the presence of $\sim$~22 years period in both the data sets which reflects the inversion of polar magnetic field. However, we did not find any direct connection between the length of QBOs in the vicinity of $\sim$~2 years and solar cycle strength for the full solar disk data sets.

\item We have applied non stationary ARIMA model to fit the sunspot number and group sunspot area time series and checked the goodness of fit. We have found that the data sets are fitted in ARIMA model and Kolmogorov-Smirnov test states that null hypothesis may be considered true and the standardized residues of the data sets may be considered as white noise.

\item In this investigation, we have utilized two alternative ways of assessing the performance of hemispheric sunspot activities, the so called wavestrapping / bootstrapping and dynamic time warping (DTW) techniques. The wavestrapping methodology indicates that northern and southern hemispheric sunspot data sets are dependent to some extent one on the other data set. The Dynamic Time Warping technique indicates that although the Manhattan minimum distance between the opposite hemispheres are quite large, indicates that these two time series are different from one another though some dependency was observed.

\end{itemize}

In the present work, we have used a complex Morlet wavelet analysis tool to detect the intermediate-term periodicities ($>$ 3 months to $\le$~11 years) in the sunspot number and sunspot group area data sets measured at KO \citep{2015LRSP...12....4H}. About an 11-year periodicity normally dominates sunspot cycles. Along with other periods, the most significant and persistent periods, which several researchers reported, are 130 -- 190 days (Rieger group/type of periods) and 1.3 --  2.5  years \citep[][etc.]{2002A&A...394..701K, 2003ApJ...591..406B, 2005A&A...438.1067K}. Rieger group of periods were detected in different solar and interplanetary parameters like sunspot number/ area \citep{1990ApJ...363..718L, 1992A&A...255..350C, 2009MNRAS.392.1159C,2019SoPh..294..142C, 2014SoPh..289.4365K}; photospheric magnetic flux \citep{2002ApJ...566..505B, 2005A&A...438.1067K}; solar flares and coronal mass ejections \citep{2003MNRAS.345..809L, 2014SoPh..289..649C, 2020SoPh..295..159K}; solar energetic particle events \citep{2005GeoRL..32.2104R, 2006MNRAS.373.1577C, 2009SpWea...7.4008L, 2016SoPh..291.2117R}; north-south asymmetry of sunspot activities \citep[][etc.]{2013ApJ...768..188C, 2021SoPh..296....2R}. These studies indicated that Rieger types of periodicities are not recognized as typical of every sunspot cycle but seem to appear only in different phases of some cycles. This Rieger group of periods was detected in our sunspot activity data sets with intermittent nature, as discussed in Sections~\ref{subsec:psn} and \ref{subsec:psga}.

The quasi-biennial oscillations (QBOs) in the range of 1.2 -- 4 years, although weaker than the main cycle ($\sim$~11 years period), is very robust since it has been notified in many solar activity indicators for the last two decades \citep[e.g., see Table~1 in][]{2014SSRv..186..359B}. The QBO's in the range of 1.3 -- 2 years were also detected in the high latitude ($\ge$~60$^{\circ}$) polar faculae \citep{2020MNRAS.494.4930D} as well as in the frequency shifts of the helioseismic data \citep{2000ApJ...531.1094K, 2013ApJ...765..100S, 2015SoPh..290.3095B}. The temporal variations of the helioseismic proxies are also associated with the changes in the solar internal magnetic fields. Recently, \citet{2021ApJ...920...49I} indicated the presence of QBO-like signals of $\sim$~2 years in different latitudinal bands in the rotation rate residuals at different depths of the deep solar interior. Our wavelet analysis has shown the prominent presence of QBOs in both the hemispheres and whole solar disk data of sunspot number and sunspot group area time series with some intermittency. The QBOs exhibit length variation and appear in the opposite hemispheric data sets at different times. These findings indicate that the QBOs is one of the most prevalent quasi-periodicity in sunspot activity indices. 

The turbulent $\alpha$-dynamos are assumed to be the mechanism responsible for the generation of the QBOs in the range of 1.3 -- 2.5 years \citep{2019A&A...625A.117I}. We have detected the quasi--periodicity of about five years, which was previously detected in the century-long digitized Ca-K plage index \citep{2016Ap&SS.361...54C,  2022ApJ...925...81C} and sunspot area time series \citep{2021SoPh..296....2R} measured at KO. Probably this period which is a harmonic of the 11-year sunspot cycle is closely related to the double peak behavior of the solar cycles \citep{2011ISRAA2011E...2G, 2015LRSP...12....4H}. The 11-year periodicity is the strongest period determined in all the sunspot activity data sets under investigation and is probably connected with the global dynamo mechanism \citep{2020LRSP...17....4C}. The detection of $\sim$~22 years period in both the sunspot indices provides a strong evidence for the existence of a relic magnetic field in the Sun. Further, we have made an effort to determine any correlation between the length of QBOs (1.2 -- 2.5 years span) and the strength of the solar cycle considering both sunspot number and group sunspot area data sets in each solar cycle under study, but, failed to find any such direct connection. In the near future, we aim to study this phenomenon more rigorously considering a long-term data sets of hemispheric sunspot number and sunspot area from different other observatories including KO.

There is no physical model that can explain different aspects of the origin and intermittent behavior of different types of intermediate-term periods found in solar indices. \citet{1993ApJ...409..476B, 2003ApJ...591..406B} studied longitude distribution of major flares for cycles 19--22 and proposed an obliquely rotating structure of wave patterns rotating with a period of 25.5 days about an axis tilted by 40$^{\circ}$ to the solar rotation axis responsible for Rieger type of periods. However, there exists no observational evidence of such a clock-type rotating structure inside the Sun. On the other hand, \citet{2002ApJ...566..505B} linked different mid-term quasi-periods, especially Rieger one with the periodic emergence of magnetic flux from the deep solar interior to the complex active regions in the solar surface. 

After investigating the dynamics near the solar tachocline region, \citet{2010ApJ...709..749Z} proposed Riger type of periods are favored when the field strength is $\le$~10$^{4}$ Gauss, in the upper overshoot layer of the tachocline. A magnetic field with strength $\le$ 10$^{5}$ Gauss near the lower layers of the tachocline leads to oscillations with a period of  $\sim$~2 years \citep{2010ApJ...724L..95Z}. Some researchers argued that the behaviors of different quasi-periodicities are governed by the dynamics of Rossby type of waves in the solar atmosphere \citep{2000ApJ...540.1102L, 2001ApJ...560..466U, 2016ApJ...826...55G, 2018ApJ...862..159D, 2020SpWea..1802109D}. Theoretical studies indicate the possibility of generating Rossby types of waves of various scales in the solar convection zone which splits into low-order Rossby waves to fast and slow magnetic Rossby waves under the influence of magnetic field \citep{2007A&A...470..815Z, 2019ApJ...887....1R}. Helioseismological studies show that Global-scale equatorial Rossby types of waves may be considered as an essential component of internal solar dynamics \citep{2018NatAs...2..568L, 2019A&A...626A...3L}. These types of waves are also observed in bright coronal points \citep{2017NatAs...1E..86M}. It is assumed that Rossby waves in the Sun can partly influence some phenomena in the Earth's atmosphere with periods close to Rieger-type periodicity \citep{2017Ap&SS.362...44S}. 

\citet{2018ApJ...856...32Z} studied the instability of the magnetic field inside the convection zone and indicated that under the condition of reduced gravity, global equatorial fast magneto-Rossby waves match well the about 11-year solar cycle period. On the other hand, \citet{2020MNRAS.497.4376S} argued that periods in the range of 5 -- 6 years might be related to the non-linearities in solar dynamo operation as well as the non-harmonic /asymmetric shape of the sunspot cycles. Some studies have shown that different solar intermediate-term periodicities (including Rieger type and QBOs) could be explained by considering suitable spherical harmonics of magneto-Rossby waves \citep[][etc.]{2005A&A...438.1067K, 2008MNRAS.386.2278D, 2013APh....42...62S, 2013ApJ...768..188C, 2019SoPh..294..142C, 2019ApJ...874..162G, 2020SpWea..1802109D, 2020ApJ...897L..24B}. \citet{2021SSRv..217...15Z} made detailed investigations about the dynamical behavior of different kinds of magnetic Rossby waves and their relationship with different solar periodicities.

We have found double peaks in different solar cycles around the maximum phase, and the gap between these two peaks is called the Gnevyshev gap \citep[GG:][]{1977SoPh...51..175G}. However, this double-peaked nature is not prominent in cycle 19, which had the maximum amplitude. On the other hand, both sunspot number and group sunspot area data sets for recent cycles 23 and 24 showed prominent double-peaked nature. Both these data sets exhibits step like decay during cycle 20. Such nature is also prominent during cycle 18 of the sunspot group area time series. A similar type of complex patterns are also observed during cycle 20 in the low latitude Ca-K plage data sets measured at KO \citep{2022ApJ...925...81C}. This type of dynamical behavior is also detected in the recently constructed long-term hemispheric / full-disk sunspot number by \citet{2021A&A...652A..56V}. 

We have noted that the sunspot activities during different cycles are not symmetric in both the hemispheres and this type of hemispheric imbalance is called ``north-south asymmetry'' \citep{2006A&A...447..735T, 2017A&A...603A.109B, 2021SoPh..296....2R}. The reason for this hemispheric imbalance is under debate, and several mechanisms have been proposed in the literature \citep{2014SSRv..186..251N, 2014A&A...563A..18P}. In addition to that, we also observe that the double peaks, step-like decay, multiple peaks during the rising/descending phase for a given cycle, may occur for a particular hemisphere without having any counterpart of the same in the other hemisphere. The reasons behind the complex nature of some cycles are still under investigation. However, \citet{2018ApJ...853..144D} argued that energy exchange among magnetic fields could generate nonlinear quasi-periodicities, which in turn related to these spikes and multiple peaks during the descending phase in sunspot indices. In this article we have first time adapted and applied the wavestrapping and DTW algorithms for the purpose of evaluating the performance of sunspot time series. Both these two statistical tools indicate that the sunspot data of the opposite hemispheres are dependent on one another which might be the reason behind the appearance of a number of common mid-term periods in wavelet analysis.

For time series, the parametric modeling with any type of mathematical formalism is very useful if the irregular nature of the data can be enclosed by the selected mathematical/ statistical model. Here, we have applied ARIMA model to fit our data sets and noted that both the sunspot data sets can be modeled by higher order ARIMA models. Our statistical significance tests like  Kolmogorov-Smirnov test and the  Anderson-Darling normality test  with the residue of ARIMA model  and random normal data indicates that  both the data sets follows the null hypothesis to some extent  and the  standardized residues may be considered as the white noise. Recently \citet{Abdel-Rahman2018} have investigated some properties of the sunspot number time series observed by National Oceanic and Atmospheric Administration (NOAA) during the period 1991 -- 2017 and indicated that this data set may be fitted with ARIMA (1, 0, 0) model. But, these authors did not study the goodness-of-fit of their model. Our results will be helpful to understand the statistical properties of different kinds of solar data and to forecast the strength of sunspot cycle.

The long-term evolution of the sunspots on the photosphere reflects the process of magnetic dynamo from the interior part of the Sun. The Spatio-temporal evolution, numerical model fitting, statistical properties, and quasi-periodic variations determined from the century-long sunspot number and sunspot group area data sets measured at the KO provide the information to understand different dynamical properties of soar magnetic field during the past sunspot cycles when satellite-borne data were absent. These long term data sets may be utilized to reconstruct solar irradiance in past solar cycles, as well \citep{2016A&A...590A..63D}. A search of all these periodicities, statistical modeling and asymmetry study is expected to provide useful information about the evolution of solar global magnetic and its reflections on the surface layers.  It is assumed that the solar magnetic field is produced by a complex dynamo mechanism that is situated near the base of the solar convection zone. The kinematic dynamo models have shown some success to explain different complex characteristic properties of the solar cycles \citep[e.g.,][]{2017SSRv..210..367C, 2020LRSP...17....4C, 2020A&A...642A..51H}. More investigations about the formation of solar magnetic flux, the behavior of dynamo in the opposite hemispheres, the operating mode of Rossby types of waves, and their variations are required to understand the intrinsic mechanism responsible for these quasi-periodicities.

\acknowledgments
We thank the referees for going through the manuscript and providing insightful comments.
We thank all of the observers at the Kodaikanal Solar Observatory who have observed continuously over a century. We also thank the people who have initiated the digitization program at Kodaikanal Observatory. We thank Mrs. Arockia Mary Anita for helping in data extraction and tabulation. We Thank Javaraiah J for valuable comments on the manuscript.
P. Chowdhury would like to thank Dr. F. Clette and Dr. D.H. Hathaway for valuable discussions during his visits to USA.


\end{document}